\documentclass[conference,anonymous]{IEEEtran}
\IEEEoverridecommandlockouts

\usepackage{cite}
\usepackage{graphicx}
\usepackage{textcomp}
\usepackage{float}
\usepackage{ragged2e}
\usepackage{placeins}
\usepackage[normalem]{ulem}

\usepackage{amsmath}
\usepackage{amssymb}
\usepackage{amsfonts}

\usepackage{booktabs}
\usepackage{tabularx}

\usepackage{xcolor}
\usepackage{soul}

\usepackage{algorithmic}

\usepackage[most]{tcolorbox}

\usepackage{subcaption}
\usepackage{caption}

\usepackage{enumitem}

\def\BibTeX{{\rm B\kern-.05em{\sc i\kern-.025em b}\kern-.08em
 T\kern-.1667em\lower.7ex\hbox{E}\kern-.125emX}}

\tcbset{
 enhanced,
 top=1pt, bottom=1pt,
 left=3pt, right=3pt,
 boxsep=1pt,
 colback=gray!10,
 colframe=gray!80!black,
 before skip=4pt,
 after skip=0pt
}
\author{
\IEEEauthorblockN{
Diksha Goel\IEEEauthorrefmark{1}\IEEEauthorrefmark{2}\thanks{Corresponding author. Email: diksha.goel@data61.csiro.au},
Kristen Moore\IEEEauthorrefmark{1}\IEEEauthorrefmark{2},
Jeff Wang\IEEEauthorrefmark{1}\IEEEauthorrefmark{2},
Minjune Kim\IEEEauthorrefmark{1}\IEEEauthorrefmark{2},
Thanh Thi Nguyen\IEEEauthorrefmark{3}
}

\IEEEauthorblockA{\IEEEauthorrefmark{1}
CSIRO’s Data61, Clayton, Australia}
\IEEEauthorblockA{\IEEEauthorrefmark{2}
Cyber Security Cooperative Research Centre (CSCRC), Joondalup, Australia\\
\{diksha.goel, kristen.moore, jeff.wang, minjune.kim\}@data61.csiro.au}
\IEEEauthorblockA{\IEEEauthorrefmark{3}
Monash University, Melbourne, Australia\\
thanh.nguyen9@monash.edu}
}

\begin{document}
\title{Unveiling the Black Box: A Multi-Layer Framework for Explaining Reinforcement Learning-Based Cyber Agents}

\maketitle

\begin{abstract}
Reinforcement Learning (RL) agents are increasingly used to simulate sophisticated cyberattacks, but their decision-making processes remain opaque, hindering trust, debugging, and defensive preparedness. In high-stakes cybersecurity contexts, explainability is essential for understanding how adversarial strategies are formed and evolve over time. In this paper, we propose a unified, multi-layer explainability framework for RL-based attacker agents that reveals both strategic (Markov Decision Process (MDP)-level) and tactical (policy-level) reasoning. At the MDP-level, we model cyberattacks as a Partially Observable Markov Decision Process (POMDP) to expose exploration-exploitation dynamics and phase-aware behavioural shifts. At the policy-level, we analyse the temporal evolution of Q-values and use Prioritised Experience Replay (PER) to surface critical learning transitions and evolving action preferences. Evaluated across CyberBattleSim environments of increasing complexity, our framework offers interpretable insights into agent behaviour at scale. Unlike previous explainable RL methods, which are {predominantly} post-hoc, domain-specific, or limited in depth, our approach is both agent- and environment-agnostic, {supporting use cases such as red-team simulation, RL policy debugging, phase-aware threat modelling and anticipatory defence planning.} {By transforming black-box learning into interpretable behavioural signals, our framework provides diagnostic insight into agent learning dynamics, with the potential to support developers and defenders in analysing and understanding autonomous cyber threats.}
\end{abstract}


\section{Introduction}

Mapping how attack paths emerge and evolve within a network is critical for understanding systemic vulnerabilities and anticipating adversarial behaviour\cite{Goel2022defending}. To automate this process, recent work increasingly trains reinforcement learning (RL) agents in red-team simulators and cyber-range environments such as CyberBattleSim~\cite{msft:cyberbattlesim} and CybORG~\cite{cage_cyborg_2022}, where autonomous attacker agents learn to explore, pivot, reuse credentials, and escalate privileges through sequential interaction~\cite{terranova:hal-05182437}. These RL agents can uncover complex, multi-step exploitation strategies that are difficult to enumerate manually, enabling systematic stress-testing of enterprise-scale defences \cite{goel2023evolving, goel2023enhancing}.

\textit{However, despite their effectiveness, the decision-making processes of RL-based attacker agents remain largely opaque.} In practice, developers observe aggregate reward curves and final success rates, while the internal learning dynamics governing \emph{why}, \emph{when}, and \emph{how} agents pivot, escalate, or persist, including policy evolution, exploration–exploitation transitions, and convergence or collapse, remain hidden. As a result, it is often unclear whether observed behaviour reflects coherent adversarial reasoning or degenerate learning patterns, such as repeatedly executing low-value actions (e.g., connect(GIT) or connect(PING)) that inflate reward without advancing the attack. This opacity undermines confidence in the realism and reliability of learned behaviours and prevents reliable debugging, benchmarking, and tuning of RL-driven adversaries for trustworthy red-team simulation.

\textit{Explainable Reinforcement Learning (XRL)} seeks to address these challenges by making agent decision processes interpretable, clarifying why particular actions are selected, how policies evolve, and which environmental cues influence behaviour~\cite{wei2021non}. However, interpretability in RL poses fundamentally different challenges than in supervised learning. RL agents learn through continual interaction with uncertain, partially observable environments, producing non-stationary policies whose decision logic evolves throughout training~\cite{dwivedi2023explainable}. As a result, most existing XRL methods remain fundamentally \emph{post-hoc}, focusing on explaining fully trained policies and are typically evaluated in simplified or fully observable settings~\cite{gyevnar2023causal,mathes2023codex,madumal2020explainable,nashed2025causal}.

Within cybersecurity, explainability efforts have further concentrated on narrow supervised tasks, such as malware detection, using feature-attribution techniques (e.g., SHAP, LIME) that provide limited insight into how \emph{sequential attacker strategies} form, adapt, or fail over time~\cite{yu2023airs,foley2023inroads,alabdulkarim2022experiential,sharma2022explainable,nguyen2021evaluation,alenezi2021explainability}. Recent behavioural audits of trained RL agents~\cite{claypoole2025interpreting} offer informative summaries of final policies, yet fail to expose the training-time learning dynamics that generate, stabilise, or destabilise adversarial behaviour. \textit{This gap motivates the need for explainability mechanisms that operate \emph{during learning itself}, rather than only after policy convergence.}

Notably, many of the phenomena that determine behavioural realism and learning stability arise \emph{during training}, not after convergence. In complex, partially observable cyber environments, agents may prematurely commit to narrow tactics, overfit early discoveries, or continue prioritising stale experiences long before such pathologies become visible in aggregate performance metrics. Training-time explainability is therefore not merely descriptive, but a practical diagnostic capability for identifying instability, reward misalignment, and premature exploitation early enough for intervention. Addressing this requires explainability across both high-level strategy formation and low-level policy optimisation.

To address this gap, we introduce a unified \emph{multi-layer explainability framework} for RL-based cyber agents that makes policy formation observable as learning unfolds. Rather than providing static, post-hoc explanations of converged policies, our framework captures how strategies emerge, adapt, and stabilise \emph{during training}, providing a principled diagnostic lens for analysing learning dynamics in adversarial environments.

The framework operates across two complementary levels. 

\noindent 1) At the \emph{strategic} level (Markov Decision Process (MDP)), attacker–environment interaction is modelled as a partially observable Markov decision process (POMDP), enabling interpretable analysis of exploration–exploitation dynamics and phase-wise behavioural transitions. 

\noindent 2) At the \emph{tactical} level (policy), we analyse temporal Q-value evolution and Prioritised Experience Replay (PER) to identify learning inflection points and the consolidation of action preferences. 

Together, these layers characterise how agent behaviour matures over time, revealing dominant strategies, key decision points, and adaptive shifts under uncertainty.

Attacker behaviour in our framework is not governed by fixed heuristics or predefined attack templates, but emerges directly from policy learning through environment interaction. Explainability is derived from training-time signals, including Q-value dynamics, temporal-difference (TD) error trajectories, policy-logit variance, and replay priorities, that expose when and why agents favour particular tactics (e.g., reconnaissance, lateral movement, or privilege escalation), how strategic preferences stabilise or collapse, and under what conditions novel behaviours arise. By grounding explanation in these signals, the framework enables early detection of instability, reward misalignment, and premature exploitation, well before such pathologies manifest in aggregate performance metrics. Unlike prior behavioural audits that characterise only final outcomes, our approach explicitly links internal learning dynamics to externally observable attack behaviour, exposing the causal mechanisms that govern policy formation under uncertainty.

We evaluate the proposed framework in \textit{Microsoft CyberBattleSim}~\cite{msft:cyberbattlesim}, which simulates attacker behaviour in networked environments involving privilege escalation, credential reuse, and lateral movement. Across environments of increasing scale and complexity, the framework reveals how strategic progression and tactical preferences develop throughout training, rather than only in final performance outcomes. Our primary experimental setup instruments a Deep Q-Learning (DQL) agent, a widely used baseline for attacker modelling in RL, where actions are naturally discrete and training proceeds episodically to capture bounded attack campaigns. To assess generality beyond value-based optimisation, we additionally evaluate the framework using the policy-gradient Proximal Policy Optimisation (PPO) algorithm. Collectively, these experiments demonstrate that training-time explainability exposes critical learning dynamics, including replay-induced instability, premature policy collapse, and value-propagation distortions, that remain fundamentally invisible under reward-only evaluation. We further show that na\"ively applying prioritised experience replay (PER), commonly assumed to improve sample efficiency, can degrade learning in large cyber environments, a failure mode not captured by aggregate performance metrics. We further link policy-level collapse to internal confidence–uncertainty signals and externally observable behavioural lock-in.

While our evaluation focuses on attacker agents for red-team simulation, the proposed explainability primitives are algorithm- and environment-agnostic and extend naturally to other sequential cyber decision-makers, including defenders performing containment, deception, or adaptive response. 
Beyond analysis, the framework supports several practical capabilities for developing and validating autonomous cyber agents.

{\textit{First, our framework enables targeted shaping of emergent behaviour.} For instance, red teams or simulator developers may aim to induce specific adversarial styles, e.g., slow, stealthy reconnaissance reminiscent of APTs versus rapid smash-and-grab attacks. By analysing when agents shift from scanning to exploitation or develop preferences for high-impact actions, our framework provides visibility into whether training regimes are producing the intended behavioural profiles. This enables iterative tuning of exploration strategies, reward shaping, or environment complexity to guide learning toward more interpretable and mission-aligned attacker behaviour.}

{\textit{Second, it supports benchmarking and comparison of training configurations.} Rather than relying solely on reward curves, our framework allows developers to track how strategy formation varies across runs. For example, one might ask: Does prioritised experience replay lead to earlier lateral movement? Does PER accelerate learning or instead introduce replay bias that destabilises convergence in large environments? Does a denser reward structure encourage premature escalation? Our multi-layer signals, capturing Q-value evolution, TD-error spikes, and exploration ratios, allow comparative audits of learning dynamics across different agent setups.}

\textit{Third, it enables early diagnosis and control of learning dynamics.}
In complex, partially observable environments such as CyberBattleSim, RL agents may stall in early phases, fail to escalate, or converge on suboptimal behavioural patterns.
Our framework exposes the underlying replay- and optimisation-driven mechanisms governing these dynamics, such as TD-error evolution, replay-priority shifts, entropy contraction, and dominant-action formation, well before such effects become visible in aggregate performance metrics.
This enables earlier diagnosis, targeted intervention, and more reliable shaping of agent behaviour during training and simulation development.

{Taken together, these capabilities position the framework as a diagnostic tool that exposes interpretable learning dynamics. The proposed approach focuses on establishing \textit{technical and behavioural} interpretability by revealing training-time signals that are not observable under conventional reward-based evaluation. These signals provide actionable insights into agent behaviour and learning processes, supporting more informed analysis and intervention during training.}

Our key contributions are as follows:

 \begin{itemize}
 \item A training-time, multi-layer explainability framework for autonomous attacker agents, integrating strategic (MDP-level) and tactical (policy-level) analyses to expose how adversarial strategies emerge, stabilise, or collapse during learning.
 
 \item Empirical insights into adversarial RL learning dynamics, including exploration–exploitation phase transitions, action-dominance formation, replay pathologies under PER, and policy saturation effects that are not observable from reward curves alone.
 
 \item Cross-algorithm validation across value-based (DQL) and policy-gradient (PPO) agents, demonstrating that key behavioural patterns persist despite differing optimisation dynamics.
 
 \item Practical diagnostic utility for red-team simulation developers, enabling debugging, benchmarking, and tuning of autonomous attacker agents to improve realism, stability, and controllability in cyber-range environments.

\end{itemize}

\begin{table*}[t]
\centering
\caption{Comparison of explainable reinforcement 
learning approaches across interpretability 
capabilities, scope, and learning-stage coverage.}
\label{tab:xrl_comparison_detailed}
\tiny
\setlength{\tabcolsep}{3.2pt}

{
\begin{tabular}{p{2cm} p{2.3cm} p{2.2cm} 
p{2.4cm} p{1cm} p{0.7cm} p{0.8cm} p{0.6cm} 
p{1cm} p{1cm} p{0.8cm}}
\toprule
\textbf{Method Category} & 
\textbf{Core Idea} & 
\textbf{Strengths} & 
\textbf{Limitations w.r.t.\ Learning Dynamics} & 
\textbf{Temporal Coverage} & 
\textbf{Training-Time Analysis} & 
\textbf{Explanation Granularity} & 
\textbf{Internal Signal Access} & 
\textbf{Explanation Scope} & 
\textbf{Evaluation Stage} & 
\textbf{Diagnostic Utility} \\
\midrule

\textbf{Post-hoc Local} \newline (SHAP, LIME)
& Explain individual decisions via feature 
attribution or surrogate models.
& Model-agnostic; intuitive local explanations.
& Static; ignores temporal dependencies; does 
not capture policy evolution or training 
dynamics; may be unstable under perturbations.
& None & No & Local & No 
& Decision-level & Post-hoc & Low \\

\midrule

\textbf{Policy-level / Distillation}
& Summarise policies using rules, graphs, 
or distilled models.
& Provide global understanding of 
policy structure.
& Approximate behaviour without explaining 
its formation; no visibility into optimisation 
or temporal learning dynamics.
& Limited & No & Policy-level & No 
& Policy-level & Post-hoc & Low \\

\midrule

\textbf{Trajectory / Behavioural}
& Analyse state--action sequences to 
identify behavioural patterns.
& Capture sequential structure; enable 
behaviour-level interpretation.
& Primarily post-hoc; no direct access to 
internal learning signals; cannot explain 
how behaviour develops over the course 
of learning.
& Limited & No & Behaviour-level & No 
& Behaviour-level & Post-hoc & Moderate \\

\midrule

\textbf{Causal / Counterfactual}
& Use interventions or causal models to 
explain decisions via ``why'' reasoning.
& Strong reasoning over alternatives; 
supports counterfactual analysis.
& Computationally expensive; limited 
scalability; typically post-hoc; does not 
expose RL optimisation dynamics.
& Interventional \newline (not temporal) 
& No & Decision-level & No 
& Decision-level & Post-hoc & Moderate \\

\midrule

\textbf{Reward-based Diagnostics}
& Use reward signals or performance trends 
to infer behaviour.
& Simple; widely applicable; useful for 
coarse monitoring.
& Indirect; does not explain internal 
mechanisms; cannot capture instability 
or learning dynamics.
& Limited & No & Performance-level & No 
& Aggregate & Post-hoc \newline (coarse) 
& Low \\

\midrule

\textbf{Ours (Multi-layer XRL)}
& Jointly models learning dynamics and 
behavioural evolution using Q-values, 
TD-error, replay, and phase-aware 
segmentation.
& Unified multi-layer interpretability 
linking strategy and optimisation; 
explains policy emergence and adaptation 
during training.
& Evaluated in simulation only; 
user-centred validation with practitioners 
not yet conducted; currently focused on 
single-agent settings.
& Strong & Yes & Strategic + Tactical & Yes 
& Multi-level \newline (strategic + tactical) 
& Training-time \newline (continuous) 
& High \\

\bottomrule
\end{tabular}
} 

\end{table*}

\section{Related Work}
\subsection{Reinforcement Learning for Cybersecurity}
\subsubsection{Simulation Environments} Several simulation environments support RL research in cybersecurity, each with distinct emphases. CyberBattleSim \cite{msft:cyberbattlesim} models enterprise networks as graphs to train attacker agents on privilege escalation, lateral movement, and exploitation, with rewards tied to node criticality. NASim \cite{schwartz2019nasim} offers a lightweight, Gym-compatible framework for penetration testing under partial observability. In contrast, CybORG \cite{cage_cyborg_2022} 
and FARLAND \cite{molina2021network} focus on defender training, integrating emulation (CybORG) or scalable probabilistic models (FARLAND) to support adaptive blue team agents and complex attack scenarios. \\

\subsubsection{RL-based Cyber Defence and Adversarial Simulation}
Recent work has explored both defensive and offensive RL agents. On the defensive front, Thompson et al. \cite{thompson2024entity} used entity-based RL with transformer policies to enable generalisation in dynamic networks. Goel et al. \cite{goel2024optimizing} co-trained attacker and defender policies in a Stackelberg game for Active Directory protection. { Wiebe et al. \cite{wiebe2023learning} employed cooperative MARL to enable distributed defenders to coordinate joint responses in complex host-based attack scenarios.} On the offensive side, Sultana et al. \cite{sultana2021autonomous} trained deep RL agents for multi-layered cyberattack strategies across network, service, and application layers. Other approaches simulate red team behaviour using RL to support defender training \cite{andrew2022developing, bierbrauer2023autonomous}.

\textit{Unlike these efforts, which aim to optimise RL agent performance, our approach focuses on interpreting agent behaviour, providing a principled, multi-layer framework for understanding attacker decision-making in adversarial settings.}

{
\subsection{Explainable Reinforcement Learning 
(XRL)}

\subsubsection{General XRL Techniques}

Explainable reinforcement learning (XRL) has 
evolved into a broad set of approaches that 
enhance transparency across different stages 
of learning and levels of abstraction. However, 
as we discuss below, existing methods share a 
common limitation: they predominantly operate 
post-hoc and do not expose the internal 
optimisation dynamics that govern how policies 
form and evolve during training. Existing 
methods can be systematically categorised into 
five groups: (i) post-hoc local explanation 
methods, (ii) policy-level and intrinsic 
explanation methods, (iii) trajectory- and 
behaviour-level explanation methods,  (iv) 
causal and counterfactual explanation methods, and (v)  reward-based diagnostic 
methods. 
Table~\ref{tab:xrl_comparison_detailed} 
provides a structured comparative overview, 
highlighting their limitations with respect 
to training-time learning dynamics and 
contrasting them with the capabilities of 
our framework.

\noindent\textbf{Post-hoc local explanation 
methods} interpret individual decisions after 
training using feature attribution or local 
surrogate models, such as SHAP~\cite{alenezi2021explainability} 
and LIME~\cite{nguyen2021evaluation}. While 
model-agnostic, these approaches provide 
static, instance-level explanations that 
ignore temporal dependencies and policy 
evolution. Their robustness has also been 
challenged, as perturbation-based methods 
can be manipulated to yield misleading 
explanations~\cite{slack2020fooling}, 
prompting recent work on evaluating 
explanation fidelity~\cite{bello2025level}. 
Consequently, they offer limited insight 
into sequential decision-making and fail 
to capture how behaviour emerges during 
learning.

\noindent\textbf{Policy-level and intrinsic 
explanation methods} aim to embed 
interpretability into the learning process 
or provide global summaries of learned 
policies. Alabdulkarim et 
al.~\cite{alabdulkarim2022experiential} 
model action influence during training, 
while Topin et 
al.~\cite{topin2019generation} construct 
Abstract Policy Graphs to capture policy 
structure, and Guo et 
al.~\cite{guo2021edge} identify temporally 
critical steps contributing to final 
rewards. Although these approaches provide 
higher-level insights, they primarily 
compress or approximate policies and do 
not expose the underlying optimisation 
dynamics or explain how behaviours 
develop over time.

\noindent\textbf{Trajectory- and 
behaviour-level explanation methods} 
analyse sequences of state--action 
interactions to characterise higher-level 
behavioural patterns. Mathes et 
al.~\cite{mathes2023codex} propose CODEX 
for clustering behavioural trajectories, 
Deshmukh et 
al.~\cite{deshmukh2023explaining} 
attribute decisions to influential training 
trajectories via sequence embeddings, and 
Takagi et al.~\cite{takagi2024abstracted} 
introduce abstracted trajectory 
visualisations for improved 
interpretability. While these methods 
capture sequential structure, they remain 
largely post-hoc and do not reveal the 
internal learning signals that drive 
behavioural evolution.

\noindent\textbf{Causal and counterfactual 
explanation methods} provide 
intervention-based reasoning to answer 
questions such as ``why'' and ``why not.'' 
Gyevnar et al.~\cite{gyevnar2023causal} 
propose a causal framework for multi-agent 
settings, while Madumal et 
al.~\cite{madumal2020explainable} and 
Nashed et al.~\cite{nashed2025causal} 
leverage structural causal models for 
contrastive explanations. Despite their 
expressive power, these approaches are 
computationally demanding, challenging to 
scale to complex environments, and 
typically do not capture training-time 
optimisation dynamics.

\noindent\textbf{Reward-based diagnostic 
methods} use aggregate reward signals and 
performance trends as proxies for agent 
behaviour. While simple and widely 
applicable, these approaches are indirect, they cannot access internal learning 
mechanisms and therefore cannot capture 
instability, replay bias, or premature 
policy collapse \cite{li2017deep}. As our experiments 
demonstrate, reward curves can actively 
mask underlying learning pathologies 
rather than reveal them.

\noindent\textbf{Gap and Motivation.} 
Despite substantial progress, existing 
XRL approaches provide limited support 
for analysing how policies evolve during 
training. Most methods focus on post-hoc 
or static explanations and do not capture 
the internal optimisation signals that 
drive behavioural adaptation. This 
limitation is particularly critical in 
adversarial cybersecurity settings, where 
understanding when and why an agent 
transitions from reconnaissance to 
exploitation, or exhibits instability 
due to replay bias or premature 
convergence, is essential for ensuring 
realistic simulation and reliable 
evaluation.

To address this gap, we introduce a 
unified, multi-layer explainability 
framework that bridges three key 
limitations in prior work. First, it 
addresses the \textit{temporal gap} by 
enabling training-time analysis rather 
than relying on post-hoc explanations. 
Second, it resolves the 
\textit{granularity gap} by jointly 
modelling strategic (MDP-level) 
progression and tactical (policy-level) 
decision-making. Third, it overcomes the 
\textit{signal gap} by exposing internal 
learning signals, Q-value evolution, 
TD-error dynamics, and replay effects, 
alongside observable behavioural 
outcomes. Together, these capabilities 
enable a principled and mechanistic 
understanding of how adversarial policies 
emerge, stabilise, and adapt over time 
in complex, partially observable cyber 
environments.
}

\begin{figure*}[t!]
 \centering

 \begin{subfigure}[b]{0.32\textwidth}
 \centering
 \includegraphics[width=\textwidth]{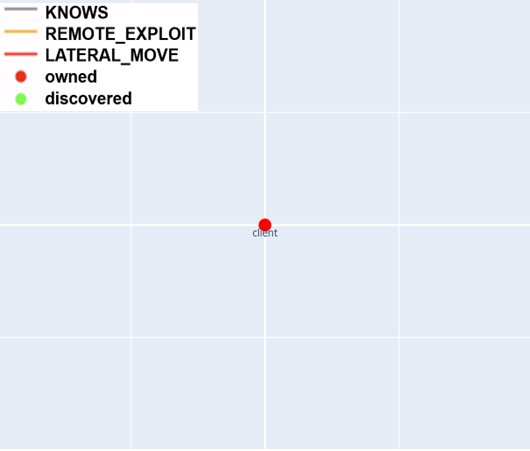}
 \caption{\scriptsize Early stage: Initial foothold (Client node owned)}
 \end{subfigure}
 \hspace{0.15em}
 \begin{subfigure}[b]{0.32\textwidth}
 \centering
 \includegraphics[width=\textwidth]{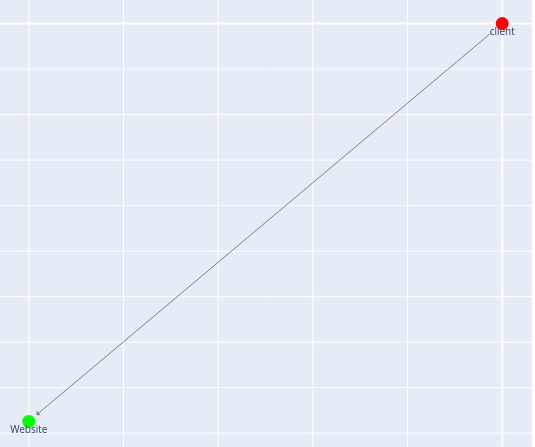}
 \caption{\scriptsize Early stage: Website discovered}
 \end{subfigure}
 \hspace{0.15em}
 \begin{subfigure}[b]{0.32\textwidth}
 \centering
 \includegraphics[width=\textwidth]{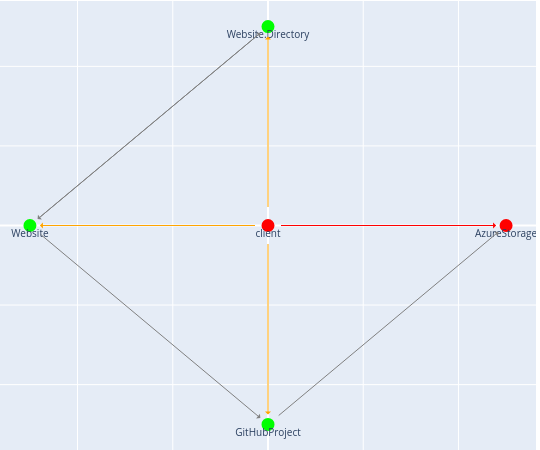}
 \caption{\scriptsize Mid-stage: Expanded discovery and node compromise}
 \end{subfigure}

 \vspace{0.3cm}

 \begin{subfigure}[b]{0.32\textwidth}
 \centering
 \includegraphics[width=\textwidth]{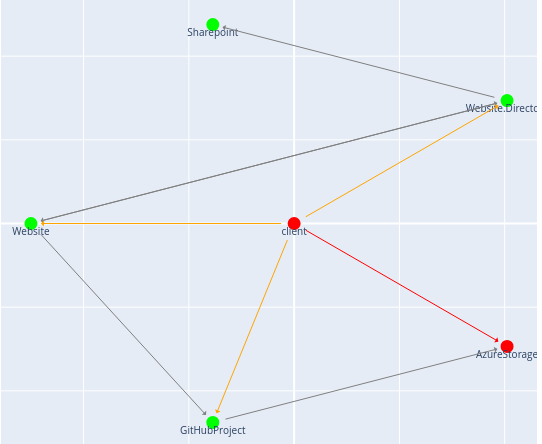}
 \caption{\scriptsize Mid-stage: Remote exploits and lateral movement}
 \end{subfigure}
 \hspace{0.15em}
 \begin{subfigure}[b]{0.32\textwidth}
 \centering
 \includegraphics[width=\textwidth]{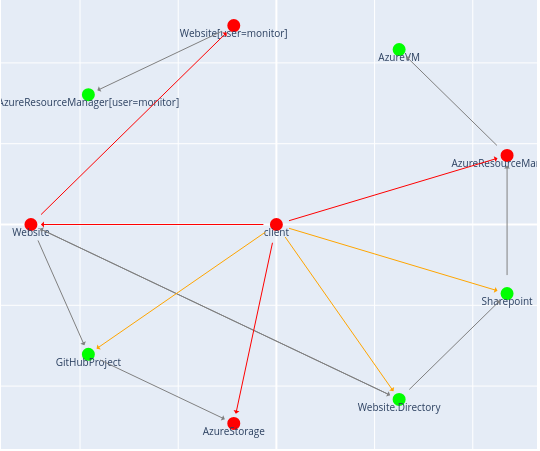}
 \caption{\scriptsize Late-stage: Full visibility with partial control}
 \end{subfigure}
 \hspace{0.15em}
 \begin{subfigure}[b]{0.32\textwidth}
 \centering
 \includegraphics[width=\textwidth]{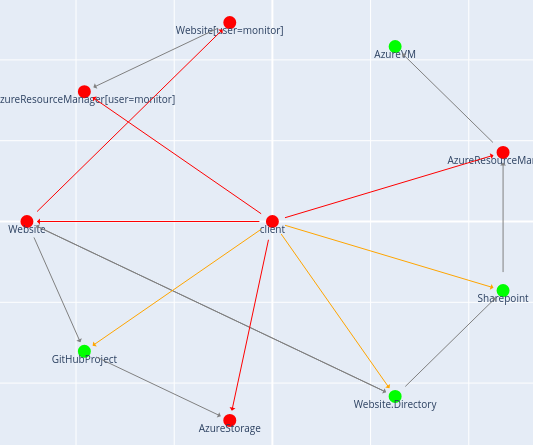}
 \caption{\scriptsize Late-stage: Full visibility and expanded control}
 \end{subfigure}

\caption{\small{
{Progressive discovery and control escalation by the RL-based attacker agent in the ToyCTF environment.
\textbf{Figure (a–b)} In the early-stages, the agent initially controls only the Client node and discovers the adjacent Website node. 
\textbf{Figure (c–d)} In the mid-stage, the agent uncovers additional nodes and expands its control via remote exploits and lateral movements.
\textbf{Figure (e–f)} In the late-stages, the agent attains full visibility and compromises additional critical nodes.
}}}
\label{fig:toyctf-exploration}
\end{figure*}

\subsubsection{XRL Techniques in Cybersecurity}
XRL in cybersecurity has largely targeted static classification tasks (e.g., malware detection \cite{yu2023airs}, intrusion analysis \cite{sharma2022explainable}) rather than sequential decision-making.
Yu et al. \cite{yu2023airs} incorporated explainability into malware mutation agents, while Sharma et al. \cite{sharma2022explainable} explored adversarial XRL for classifier sensitivity. For instance, they used adversarial perturbations to identify minimal input changes affecting classification outcomes, {but such methods fail to capture the sequential, adaptive nature of cyber operations.}
Foley et al. \cite{foley2023inroads} proposed a post-hoc XRL framework for defender agents in CybORG using SHAP, feature ablations, and state visualisations to improve situational awareness. However, their attacker agents follow fixed, hierarchical rules, limiting insight into real-time red-team adaptation.
\vspace{0.04in}

\textit{By contrast, our work centres the attacker in the explanatory process, introducing a general, environment-agnostic XRL framework that captures behaviour across both strategic (MDP-level) and tactical (policy-level) layers. Unlike existing approaches that are {predominantly post-hoc}, {limited to static tasks}, or {tied to specific environments or agents}, our framework supports both {training-time} and {post-training} analysis. This dual capability captures the temporal evolution of strategies as they emerge and adapt under uncertainty, enabling defenders to anticipate adversarial tactics and developers to refine agent training, while delivering actionable, interpretable insights that remain opaque in existing approaches.}

\section{Problem Formulation}
Enterprise networks are increasingly targeted by lateral movement attacks, where adversaries navigate interconnected systems to escalate privileges and compromise critical assets. In Microsoft’s \textit{CyberBattleSim}, such networks are modelled as directed graphs $G = (V, E)$, where nodes $v \in V$ represent machines (e.g., workstations, servers, domain controllers) and edges $e \in E$ represent communication links. Each node has an associated reward $r(v)$, reflecting its operational importance. An attacker, modelled as an RL agent, seeks to maximise cumulative reward $R = \sum_{v \in V_C} r(v)$, where \( V_C \subseteq V \) denotes the set of compromised nodes. The agent operates under POMDP, reflecting the uncertainty and limited visibility typical of real-world cyberattacks. While RL agents can learn effective attack policies in this setting, their decision-making processes remain opaque. Traditional explainability methods, such as static feature attribution and post-hoc visualisations, fall short of capturing the sequential, adaptive nature of adversarial behaviour. As a result, defenders lack insight into \textit{when}, \textit{why}, and \textit{how} attackers pivot laterally, escalate privileges, or shift tactics, insights that are critical for proactive defence and trust in autonomous systems.

\section{CyberBattleSim Platform} 
CyberBattleSim is an open-source simulation platform designed to model adversarial cyber operations within enterprise network environments. Networks are represented as directed graphs, where nodes encapsulate system configurations such as operating systems, services, known vulnerabilities, firewall rules, and edges represent connectivity. 
The environment is inherently attacker-centric: it trains RL agents to autonomously discover and exploit vulnerabilities through local privilege escalation, remote exploitation, and credential-based lateral movement. Unlike CybORG \cite{cage_cyborg_2022}, which emphasises defender modelling and blue-team coordination, CyberBattleSim provides a focused setting for evaluating autonomous red-team strategies. CyberBattleSim does not model an explicit, adaptive defender agent; instead, defensive behaviour is encoded implicitly through static network configurations, access control rules, and vulnerability placement.

Agents begin with limited visibility (Figure \ref{fig:toyctf-exploration}) and explore to uncover system properties and reachable nodes. 
The reward model incentivises strategic behaviour, assigning higher rewards to the compromise of mission-critical nodes (e.g., 100 for a domain controller vs. 10 for a workstation). Successfully breaching a node may also reveal previously hidden segments of the network, enabling progressive expansion of the attack agent's surface. CyberBattleSim provides a structured yet non-trivial attacker action space and partially observable environment dynamics, making it suitable for studying how autonomous attacker behaviour and learning dynamics can be interpreted in sequential decision-making settings.

\subsection{CyberBattleSim Environments}

\noindent\textbf{\textit{CyberBattle Capture-the-Flag (CTF) Environment.}} 
This environment uses a loosely connected hub-and-spoke topology, with nodes representing enterprise services like web applications, GitHub, Azure storage, SharePoint, and cloud VMs. 
Each node exposes multiple context-dependent exploit paths (e.g., scanning content, mining Git history, credential reuse), requiring chained reasoning to progress. Supported actions include \texttt{SearchEdgeHistory}, \texttt{ScanPageContent}, \texttt{NavigateWebDirectory}, \texttt{AccessDataWithSASToken}, and local and remote privilege escalation actions, with vulnerabilities embedded through exposed tokens, weak authentication, and misconfigurations. The environment's non-linear structure supports analysis of both strategic variation and tactical adaptation.

Figure~\ref{fig:toyctf-exploration} illustrates the agent’s discovery and control escalation in the ToyCTF environment. 
Early episodes show limited visibility and control of only the client node (Figure~\ref{fig:toyctf-exploration}(a)). 
As training progresses (Figures~\ref{fig:toyctf-exploration}(b)–(d)), the agent uncovers adjacent services and additional assets (e.g., Website, GitHubProject, AzureStorage), leveraging lateral movement and remote exploits. 
In later stages (Figures~\ref{fig:toyctf-exploration}(e)–(f)), the agent attains broad visibility and compromises critical assets such as AzureResourceManager and AzureVM, demonstrating the transition from initial access to network-wide dominance and highlighting ToyCTF’s suitability for analysing enterprise-scale attack progression. \\

\noindent\textbf{\textit{CyberBattleChain Environment.}} 
CyberBattleChain simulates a linear, fixed-sequence network topology alternating between Linux and Windows nodes: \texttt{start → (Linux → Windows)\textsuperscript{n} → Linux[Flag]}. Progression requires exploiting platform-specific vulnerabilities and reusing credentials revealed upon node compromise (e.g., 
SSH credentials from Windows to the subsequent Linux node). 
Embedded traps lure agents into suboptimal paths, requiring strategic discernment to avoid costly dead ends. The environment ends with a high-value target (``flag" node), making it ideal for evaluating long-horizon planning, sequential decision-making, and explainability in high-stakes attacker behaviour.

\section{Multi-Layer Explainability Framework for RL-Based Attackers}
To address the opacity of RL agents in adversarial environments, we propose a \textit{unified, multi-layer explainability framework} that reveals both strategic intent and tactical decision-making over time. Grounded in core explainable AI principles, such as transparency, temporal insight, and behavioural attribution, our framework operates at two complementary levels:
\begin{enumerate}
 \item Strategic-Level (MDP): Models attacker progression through exploration–exploitation dynamics and phase transitions over the attack lifecycle.
 \item Tactical-Level (Policy): Traces evolving action preferences and learning signals using temporal Q-value analysis and Prioritised Experience Replay (PER).
\end{enumerate}
{This layered design enables analysis of not just what actions were taken, but why and when they emerged, providing interpretable behavioural signals that may assist further human analysis. }
 
\subsection{Strategic-Level (MDP) Explainability}
We formalise attacker behaviour as a \textit{Partially Observable Markov Decision Process}, defined by the tuple \( (S, A, \Omega, O, T, R, \gamma) \), where \( S \) is the set of true (hidden) states, \( A \) the action space, \( \Omega \) the set of possible observations, \( O: S \times A \rightarrow \mathcal{P}(\Omega) \) is the observation function (a probability distribution over observations), \( T: S \times A \rightarrow \mathcal{P}(S) \) the state transition function, \( R: S \times A \rightarrow \mathbb{R} \) the reward function, and \( \gamma \in [0, 1] \) the discount factor. This formulation captures uncertainty and limited visibility adversaries face in real networks, and allows us to analyse strategic adaptations over time.\\

\subsubsection{Exploration-Exploitation Dynamics}

\noindent We examine how different exploration strategies shape the agent's learning and behaviour:

\begin{enumerate}
\item \textbf{Early Exploitation Strategy:} This strategy models adversaries who aim for fast wins. The agent begins with a low exploration rate (\( \epsilon_{\text{low}} \)) up to step \( T_{\text{exploit}} \), after which the exploration rate gradually increases, prioritising rapid exploitation before defences adapt \cite{tokic2010adaptive}. We define the exploration rate \( \epsilon_t \) at time step \( t \) as:

 \begin{equation*}
 \epsilon_t =
 \begin{cases}
 \epsilon_{\text{low}}, & t \leq T_{\text{exploit}} \\
 \epsilon_{\text{low}} + \frac{(\epsilon_{\text{high}} - \epsilon_{\text{low}})(t - T_{\text{exploit}})}{T - T_{\text{exploit}}}, & t > T_{\text{exploit}}
 \end{cases}
 \end{equation*}

\item \textbf{Standard Exploration Strategy:} This strategy reflects more reconnaissance-heavy adversaries. The agent begins with high exploration (\( \epsilon_{\text{high}} \)) until step \( T_{\text{explore}} \), after which exploration gradually decays in favour of exploiting the learned policy. Such agents aim for long-term advantage by systematically mapping vulnerabilities and refining their actions for maximum impact.

 \begin{equation*}
 \epsilon_t =
 \begin{cases}
 \epsilon_{\text{high}}, & t \leq T_{\text{explore}} \\
 \epsilon_{\text{high}} - \frac{(\epsilon_{\text{high}} - \epsilon_{\text{low}})(t - T_{\text{explore}})}{T - T_{\text{explore}}}, & t > T_{\text{explore}}
 \end{cases}
 \end{equation*}
\end{enumerate}
where, \( T \) denotes the total number of training steps; \( t \) is the current training step, \( T_{\text{exploit}} \) and \( T_{\text{explore}} \) specify the step thresholds that mark the end of the initial low- and high-exploration phases, respectively, for each strategy.

Comparing these schedules reveals how different attacker profiles prioritise speed, breadth, or depth in their learning trajectories and strategic development.\\

\subsubsection{Phase Transitions and Behaviour Evolution}
To further capture strategic adaptation, we segment attacker behaviour into \textit{early} and \textit{late} attack phases based on the network compromise ratio ($C_t$):

\begin{equation*}
C_t = \frac{|\text{Compromised nodes at time } t|}{|N|}
\end{equation*}

\noindent where $C_t$ represents the proportion of compromised nodes, and $|N|$ denotes the total number of nodes in the network.

\begin{enumerate}
 \item \textbf{Early Phase:} In the early stages of an attack, strong defences and limited visibility are expected to constrain the agent's progress, prompting a cautious, reconnaissance-driven approach. This phase is typically characterised by uncertainty, minimal network compromise, and broad probing, as the agent gathers information and identifies potential vulnerabilities. To capture this behaviour, our framework isolates transitions from early phases of the attack, where only a small fraction of the network has been compromised, enabling analysis of how strategies form under high uncertainty ($C_t < \text{Threshold}$ and $\text{Threshold} \in (0,1)$).

\item \textbf{Late Phase:} As more nodes are compromised and visibility improves, the agent is expected to transition to a more aggressive phase of the attack. With increased access and diminishing defensive barriers, this phase is typically associated with behaviours such as privilege escalation, lateral movement, and targeting of high-value assets. To capture this shift, our framework isolates transitions where the network compromise ratio exceeds a defined threshold, enabling analysis of how attacker strategies evolve from broad exploration to more focused, high-impact exploitation ($C_t > \text{Threshold}$). \\
\end{enumerate}

\subsection{Tactical-Level (Policy) Explainability}
To complement the strategic lens provided by MDP-level modelling, our framework incorporates a tactical layer focused on how the agent's preferences evolve during learning. 
This layer provides insights into how specific actions become favoured over time, and which experiences most influence policy updates. We achieve this through two techniques: (1) tracking temporal Q-value evolution and (2) leveraging Prioritised Experience Replay. 
Together these approaches support temporal attribution and highlight learning dynamics behind critical decisions, key goals in explainable AI.\\

\subsubsection{Temporal Q-Value Evolution and Action Preference}

\noindent We track the evolution of Q-values over time to analyse how the agent's assessment of state-action pairs changes with experience. In RL settings, Q-values begin largely undifferentiated due to sparse early feedback, and progressively evolve to reflect long-term reward expectations. Our framework leverages this temporal progression to surface how tactical preferences may emerge over time, offering insight into the agent's evolving prioritisation of specific actions. The Q-values are updated via Q-learning rule \cite{watkins1992qlearning}:

\begin{equation*}
Q(s,a) \leftarrow Q(s,a) + \alpha \left[ r + \gamma \max_{a'} Q(s',a') - Q(s,a) \right]
\end{equation*}

where \( \alpha \in (0, 1] \) is the learning rate controlling the update magnitude, 
\( \gamma \) is the discount factor, \( r \) is the immediate reward, and \( s' \) is the next state.

Over time, the Q-values converge toward the expected return defined by the Bellman expectation equation under the learned policy \cite{bellman1957dynamic}:

\begin{equation*}
Q(s,a) = \mathbb{E}\left[\sum_{t=0}^{\infty} \gamma^t r_{t}\right]
\end{equation*}

where $Q(s,a)$ denotes the expected cumulative reward for taking action $a$ in state $s$, $\gamma$ is the discount factor determining the importance of future rewards, and $r_t$ is the reward received at time step $t$.

Tracking Q-value progression allows us to infer \emph{when} an agent's behaviour transitions from broad exploration (e.g., probing for vulnerabilities) to targeted exploitation (e.g., privilege escalation or lateral movement toward critical assets). This \emph{temporal Q-value analysis} supports temporally grounded explanations of evolving intent and highlights emerging tactical focus within agent's policy.\\

\subsubsection{Policy Interpretation via Prioritised Experience Replay}
Prioritised Experience Replay complements Q-value evolution by revealing what drives learning. It amplifies the influence of high-impact transitions, those with large Temporal-Difference (TD) errors, focusing the agent’s attention on experiences that most significantly shape policy updates.

\paragraph{Prioritised Experience Replay.}
Each experience tuple $
\tau_i = (s_i, a_i, r_i, s'_i, d_i)$ is stored in a replay buffer \(\mathcal{D}\), where $d_i \in \{0, 1\}$ is a binary done flag indicating whether the transition leads to episode termination. Rather than sampling these transitions uniformly, PER prioritises transitions with higher learning potential and assigns a priority \(p_i\) based on the transition’s TD error $\delta_i$ \cite{schaul2016prioritized} as:
 
\[
\delta_i = r_i + \gamma \max_{a'} Q (s'_i,a') \;-\; Q (s_i,a_i)
\] 
Transitions with large \(\delta_i\) (\emph{e.g.}, unexpected outcomes, high-value exploits) receive higher priority: 
\[
p_i = |\delta_i| + \epsilon_0,\quad 
P(\tau_i) = \frac{p_i^\alpha}{\sum_{j \in \mathcal{D}} p_j^\alpha},
\]
where \(p_i\) is the priority; \(\epsilon_0\) avoids zero priority; \(\alpha\) controls prioritisation; \(P(\tau_i)\) is the sampling probability; \(\mathcal{D}\) is the replay buffer. Importance-sampling weights are applied to ensure unbiased learning updates.

By surfacing the transitions that most influence learning, such as early successes in credential acquisition or late-stage access to high-value nodes, PER reveals the moments of highest explanatory value in the agent's trajectory. These inflection points help clarify how the agent's tactics evolve in response to new information and highlight where strategic shifts are most likely to occur. 

\begin{tcolorbox}
By integrating strategic-level progression with tactical decision analysis, our framework provides a comprehensive and temporally grounded explanation of RL-based attacker behaviour. It elucidates not only the actions taken by the agent, but also the underlying rationale, timing, and policy evolution driving those decisions. This layered interpretability enhances transparency, supports informed defensive planning, and enables the development of phase-aware countermeasures aligned with the agent’s behavioural dynamics and learning patterns. Moreover, its generalisable design allows seamless adaptation to diverse environments and attacker profiles, making it a practical tool for analysing and refining autonomous cyber operations in both research and operational contexts. \end{tcolorbox}

\begin{figure*}[t!]
 \centering
 \includegraphics[width=0.40\paperwidth]{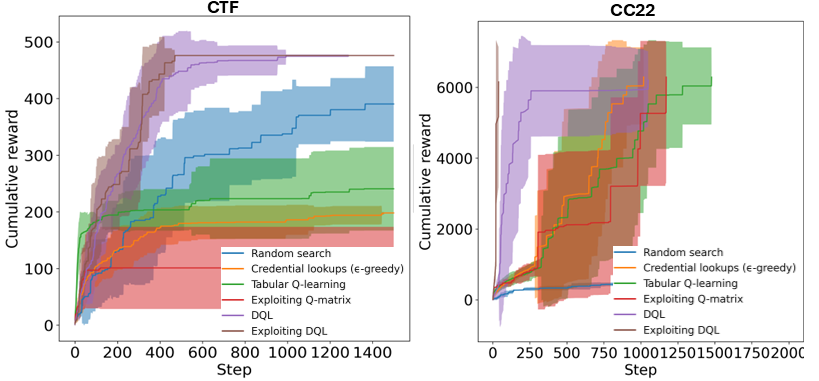}
 \vspace{0.15in}
 \includegraphics[width=0.4\paperwidth]{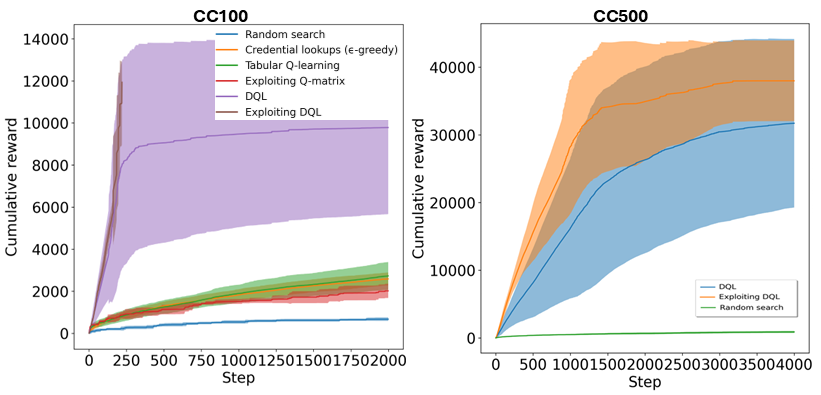}
\caption{\small
Cumulative rewards comparison of attacker policies across environments. The shaded region represents the standard deviation of cumulative rewards across training steps for each agent. (Results indicate that while Exploiting-DQL attains the highest rewards by leveraging a fixed, trained DQL policy, standard DQL is the most effective core agent, providing high cumulative rewards with stable and scalable learning across all environments).
}
 \label{fig:Setup1}
\end{figure*}

\section{Experimental Evaluation}\label{experiments}

We first evaluate our framework across five experimental setups designed to highlight key explainability dimensions over raw performance: 
(1) baseline policy selection, 
(2) exploration--exploitation behaviour, 
(3) phase-aware behavioural shifts, 
(4) temporal Q-value evolution, and 
(5) strategic learning signals via PER. 
{These experiments are designed to reveal how and why attacker agents adapt over time, surfacing interpretable patterns such as evolving intent, phase transitions, action prioritisation, and critical learning events. }

We then extend this evaluation with four additional studies (Setups~6--9) that deepen the analysis along algorithmic and behavioural dimensions, including cross-algorithm generalisability, policy-level dynamics of confidence and uncertainty, joint policy--performance explainability, and behavioural lock-in through dominant action evolution. 
Together, these experiments explain not only \emph{what} decisions the agent makes, but \emph{how}, \emph{when}, and \emph{why} its learning dynamics emerge, stabilise, and ultimately collapse, yielding a coherent mechanistic account of attacker behaviour beyond surface-level performance metrics.

\vspace{0.1in}

{
\noindent \textbf{Evaluation of Interpretability (Functional Diagnostic Utility).}
We evaluate interpretability based on the framework’s ability to expose diagnostically meaningful learning behaviours during training. Rather than relying solely on aggregate metrics (e.g., cumulative reward), we assess whether the proposed explanations reveal temporally grounded signals that enable systematic analysis of agent behaviour. An explanation is considered effective if it supports identification of key learning phenomena, including:
(i) phase transitions between exploration and exploitation,
(ii) emergence of dominant actions and tactical preferences,
(iii) instability in learning dynamics (e.g., replay-induced oscillations or delayed value propagation), and
(iv) convergence characteristics such as stagnation or policy collapse.

These behaviours are captured through signals such as Q-value trajectories, temporal-difference error evolution, replay-priority dynamics, and phase-aware segmentation, linking internal optimisation processes to observable behavioural patterns. 
{These signals directly support debugging efficiency by enabling developers to identify when learning becomes unstable, detect policy collapse or replay-induced bias, and trace how alternative strategies emerge following environmental changes, thereby reducing reliance on trial-and-error tuning and providing actionable insight into policy behaviour during training.}
}\\

\begin{table}[t!]
\centering
\caption{{Quantitative signals used to evaluate interpretability of learning dynamics.}}
\label{table}
\begin{tabular}{l l}
\toprule
{\textbf{Phenomenon}} & {\textbf{Quantitative Signal}} \\
\midrule
{Phase transition} & {Reward slope, entropy change} \\
{Dominant actions} & {Q-value / logit separation} \\
{Instability}      & {TD-error magnitude and variance} \\
{Adaptation}       & {TD-error spikes + recovery rate} \\
{Convergence}      & {Q-value stabilisation, entropy decay} \\
\bottomrule
\end{tabular}
\end{table}

{
\noindent \textbf{Quantitative Characterisation of Interpretability Signals.}
While the proposed framework exposes interpretable learning dynamics through temporal signals such as Q-values and TD-error, we further formalise these signals to provide a structured and measurable understanding of how policies evolve during training. These signals serve as quantitative interpretability metrics that enable systematic and measurable  evaluation of learning dynamics across training. They provide measurable criteria for identifying phase transitions, instability, adaptation, and convergence behaviour.

\begin{enumerate}
    \item \textit{Phase Transition Detection.}
    Transitions between exploration and exploitation are identified through coordinated changes in reward progression, reductions in policy entropy, and shifts in the distribution of Q-values over time.

    \item \textit{Dominant Action Emergence.}
    The formation of stable action preferences is captured through increasing separation between the highest-valued action and its alternatives (in Q-values or logits).

    \item \textit{Learning Instability.}
    Instability is quantified using the magnitude and temporal variation of TD-error. Sustained elevation indicates ongoing adaptation under persistent disruption, while sharp spikes correspond to abrupt environmental changes.

    \item \textit{Policy Adaptation.}
    Adaptation is characterised by transient spikes in TD-error followed by rapid recovery, reflecting the agent’s ability to reorganise value estimates and identify alternative strategies after disruption.

    \item \textit{Policy Convergence or Collapse.}
    Convergence is characterised by stabilisation of Q-values and entropy reduction, whereas collapse is indicated by excessive concentration of action probabilities and reduced policy diversity.
\end{enumerate}

Together, these signals provide a clear and structured view of how learning progresses, allowing us to distinguish between stable convergence, continuous adaptation, and structural reorganisation, and to directly relate internal learning dynamics to observable agent behaviour. 
Table~\ref{table} summarises these quantitative metrics, linking each interpretability phenomenon to measurable signals used in our analysis.
}

\vspace{0.1in}

\noindent \textbf{Environment Description.}
We evaluated our framework using \textit{CyberBattleSim}\cite{msft:cyberbattlesim} across two core environments: (1) \textit{CTF}, a 12-node network designed for targeted attack scenarios, and (2) \textit{CyberBattleChain}, with three scalable variants, CC22, CC100, and CC500, representing networks of 22, 100, and 500 nodes, respectively. Table \ref{tab:network_config} summarises these configurations, where “Size” indicates the number of exploitable nodes and “Max Nodes” represents the full network, including non-exploitable components such as firewalls and auxiliary infrastructure.

\vspace{0.1in}
\begin{table}[t!]
 \centering
\caption{Summary of CyberBattleSim environments, showing exploitable nodes (Size) and total nodes (Max Nodes).}
\label{tab:network_config}
 \begin{tabular}{p{2.5cm}p{1.5cm}p{2cm}}
 \toprule
 \textbf{Environment} & \textbf{Size} & \textbf{Max Nodes} \\
 \midrule
 CTF & 9 & 12 \\
 CC22 & 12 & 22 \\
 CC100 & 70 & 100 \\
 CC500 & 350 & 500 \\
 \bottomrule
 \end{tabular}
\end{table}

\noindent \textbf{Training Parameters.} 
We scaled the architecture of the Deep Q-Network to the environment complexity: 3 hidden layers were used for smaller scenarios (CTF, CC22), and 5 layers for larger ones (CC100, CC500). Agents used a replay buffer of 20000 transitions, a target network update frequency of 10 episodes, a learning rate of 0.005 with the Adam optimiser, and batch size of 128. The discount factor was set to $\gamma = 0.95$. Unless stated otherwise, exploration followed a standard $\epsilon$-greedy policy, decaying from $\epsilon = 0.90$ to $\epsilon = 0.10$ over 5000 steps. All experiments are implemented in PyTorch and run on a high-performance cluster with 1 CPU and 256 GB RAM.

\begin{figure*}[t!]
 \centering
 \begin{subfigure}[b]{0.24\textwidth}
 \centering
 \includegraphics[width=\textwidth, trim=0 0 0 0, clip]{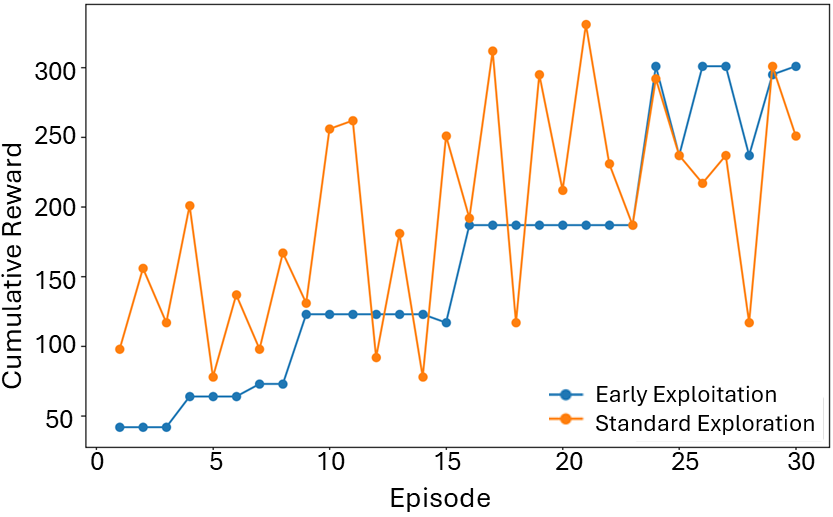}
 \caption{CTF}
 \label{fig:bar_qvals_70}
 \end{subfigure}
 \hspace{0.06em}
\begin{subfigure}[b]{0.24\textwidth}
 \centering
 \includegraphics[height=2.67cm, width= 4.4cm]{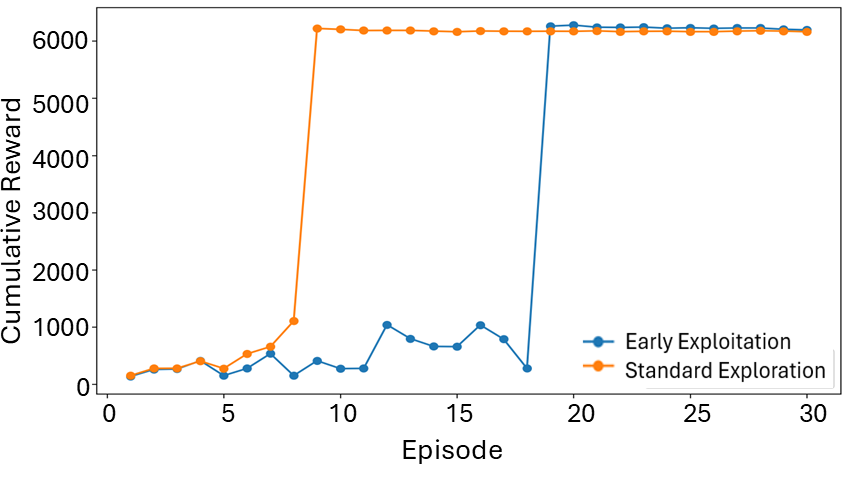}
 \caption{CC22}
 \label{fig:bar_qvals_1500}
\end{subfigure}
 \hspace{0.06em}
 \begin{subfigure}[b]{0.24\textwidth}
 \centering
 \includegraphics[width=\textwidth, trim=0 0 0 0, clip]{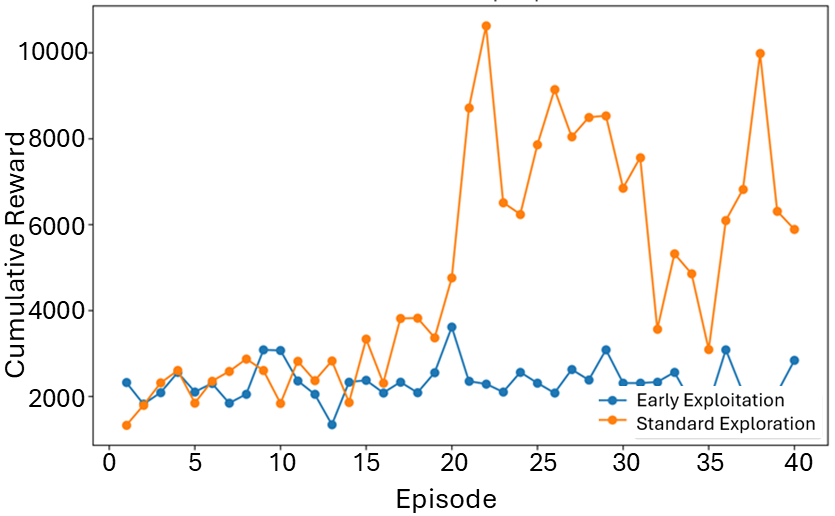}
 \caption{CC100}
 \label{fig:bar_qvals_1500_2}
 \end{subfigure}
 \hspace{0.06em}
 \begin{subfigure}[b]{0.24\textwidth}
 \centering
 \includegraphics[width=\textwidth, trim=0 0 0 0, clip]{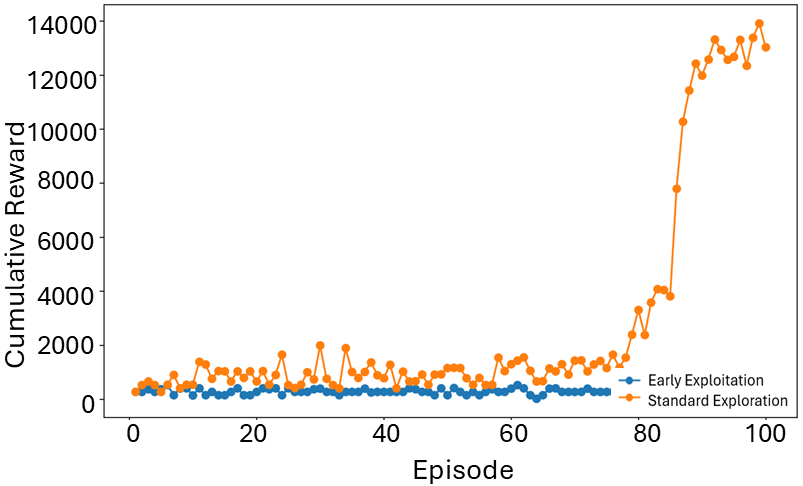}
 \caption{CC500}
 \label{fig:bar_qvals_1500_3}
 \end{subfigure}

\caption{\small{Impact of Exploration Strategies on cumulative rewards across environments {(Results show that Standard Exploration achieves broader state-space coverage, yielding higher rewards and facilitating more robust policy convergence).}}}
\label{fig:rewardMDP1}
\end{figure*}

\begin{figure*}[t!]
 \centering
 \begin{subfigure}[b]{0.24\textwidth}
 \centering
 \includegraphics[width=\textwidth, trim=0 0 0 0, clip]{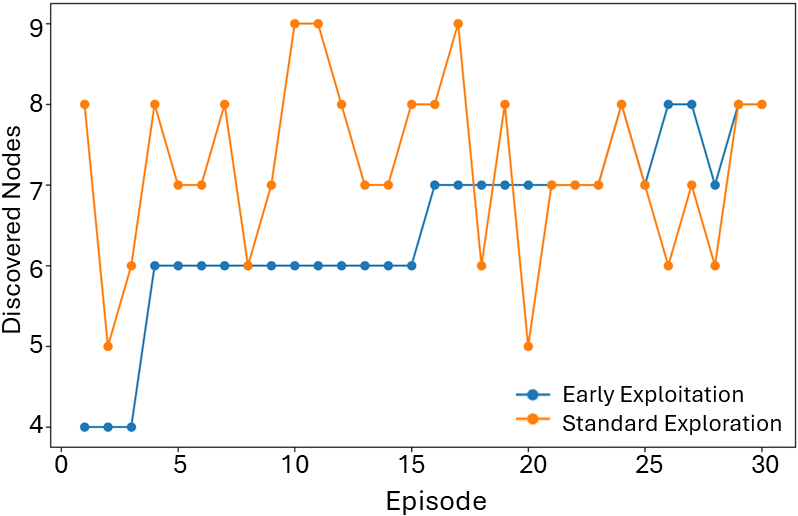}
 \caption{CTF}
 \label{fig:bar_qvals_70_d}
 \end{subfigure}
 \hspace{0.06em}
 \begin{subfigure}[b]{0.24\textwidth}
 \centering
 \includegraphics[width=\textwidth, trim=0 0 0 0, clip]{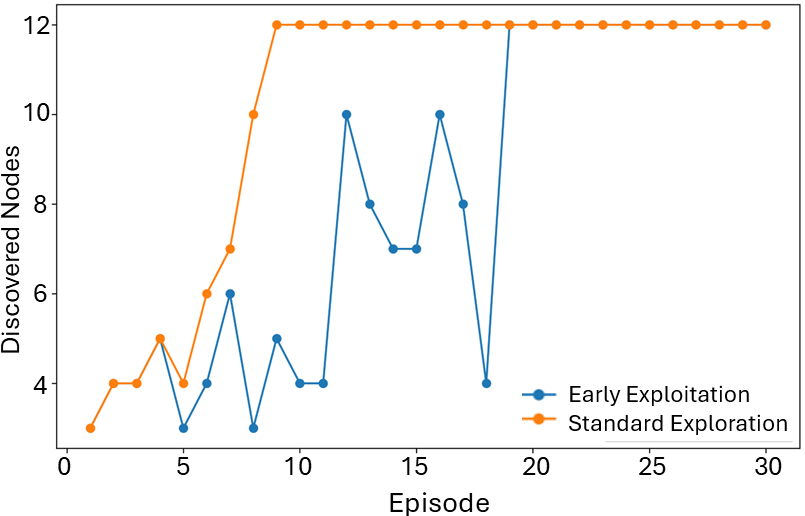}
 \caption{CC22}
 \label{fig:bar_qvals_1500_d1}
 \end{subfigure}
 \hspace{0.06em}
 \begin{subfigure}[b]{0.24\textwidth}
 \centering
 \includegraphics[width=\textwidth, trim=0 0 0 0, clip]{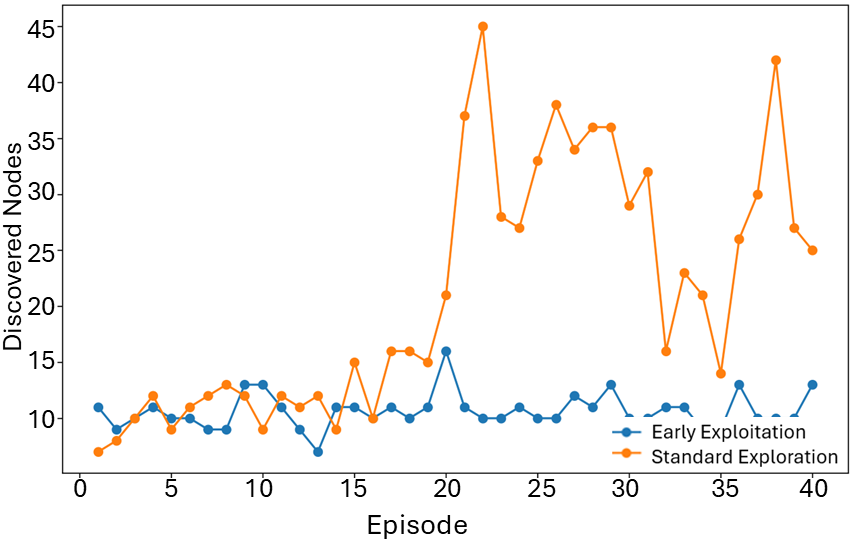}
 \caption{CC100}
 \label{fig:bar_qvals_1500_d2}
 \end{subfigure}
 \hspace{0.06em}
 \begin{subfigure}[b]{0.24\textwidth}
 \centering
 \includegraphics[width=\textwidth, trim=0 0 0 0, clip]{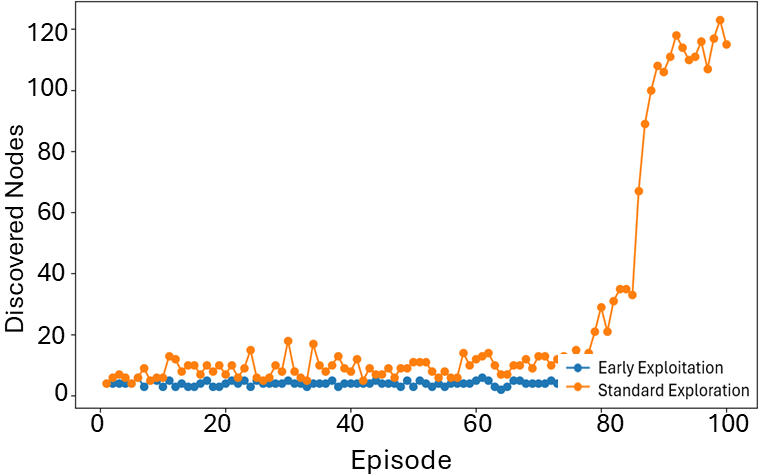}
 \caption{CC500}
 \label{fig:bar_qvals_1500_d3}
 \end{subfigure}

\caption{\small{Impact of Exploration Strategies on node discovery rate across environments 
{(Results show that Standard Exploration consistently enables faster and broader discovery, facilitating more informed policy learning).}}}
 \label{fig:discoveredMDP1}
\end{figure*}

\subsection{Setup 1: Baseline Policy Selection}\label{s1}
To ensure that the interpretability results are grounded in a stable, scalable attacker policy, we benchmarked six RL-based agents: Random Search, Credential Lookup ($\epsilon$-greedy), Tabular Q-Learning, Exploiting Q-Matrix, Deep Q-Learning (DQL), and Exploiting DQL. These agents were evaluated across CTF, CC22, CC100, and CC500 using cumulative reward as the primary metric.

\begin{figure*}[t!]
 \centering
 \begin{subfigure}[b]{0.24\textwidth} 
 \centering
 \includegraphics[width=\textwidth, trim=0 0 0 0, clip]{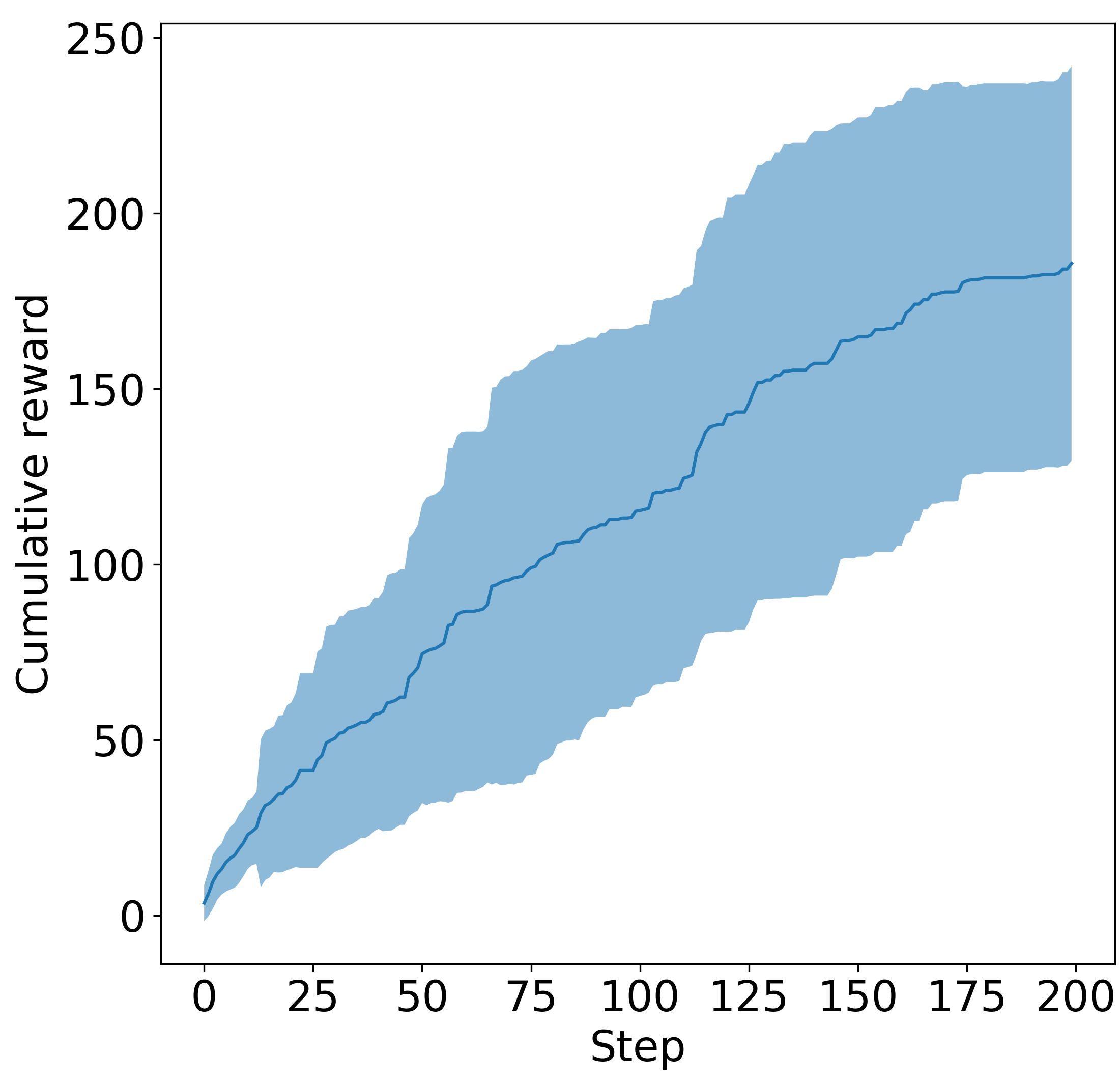}
 \captionsetup{font=scriptsize}
 \caption{CTF: Early Phase}
 \label{fig:ctf}
 \end{subfigure}
 \hfill
 \begin{subfigure}[b]{0.24\textwidth}
 \centering
 \includegraphics[width=\textwidth, trim=0 0 0 0, clip]{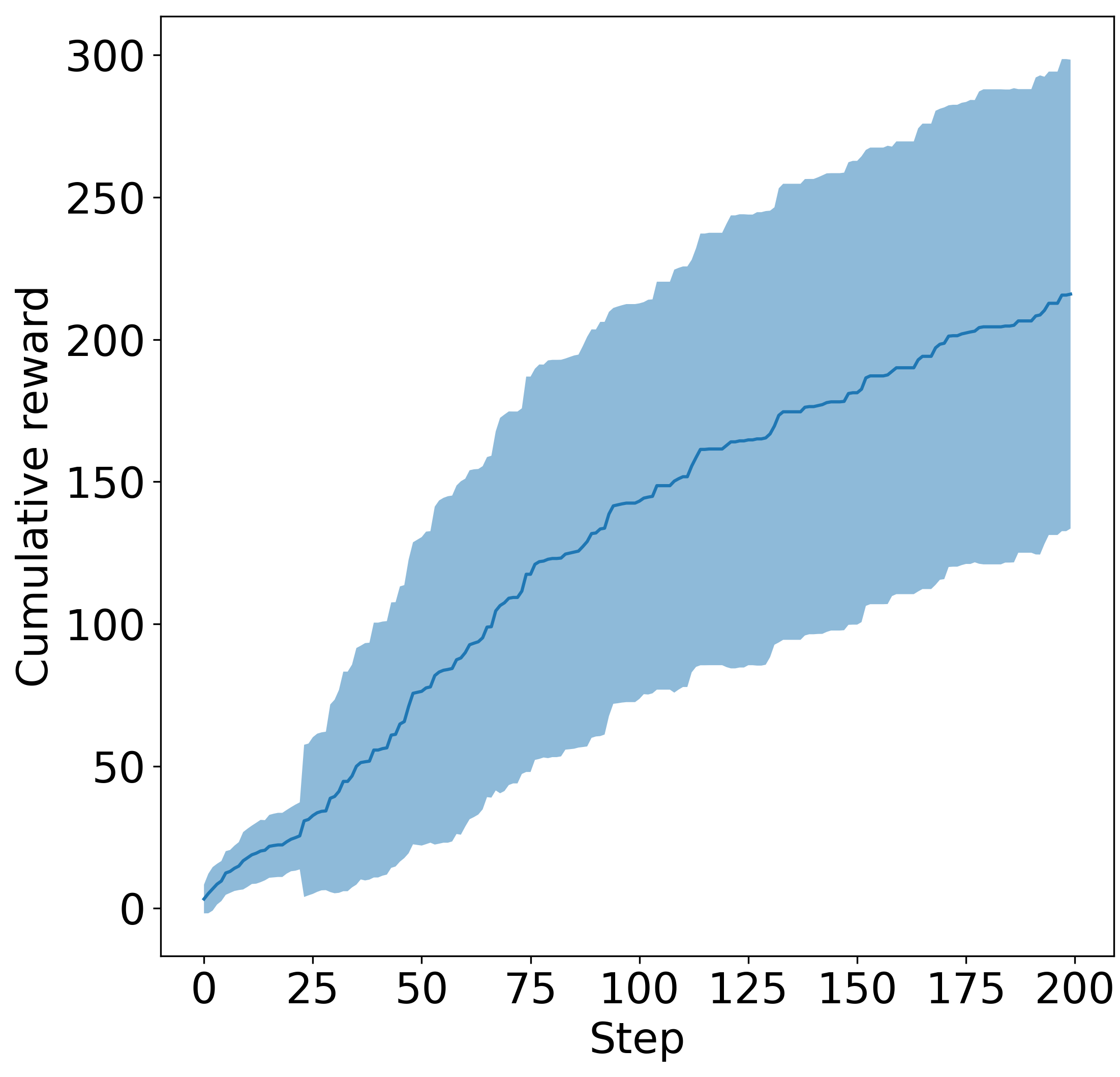}
 \captionsetup{font=scriptsize}
 \caption{CTF: Late Phase}
 \label{fig:cc22}
 \end{subfigure}
 \hfill
 \begin{subfigure}[b]{0.24\textwidth}
 \centering
 \includegraphics[width=\textwidth, trim=0 0 0 0, clip]{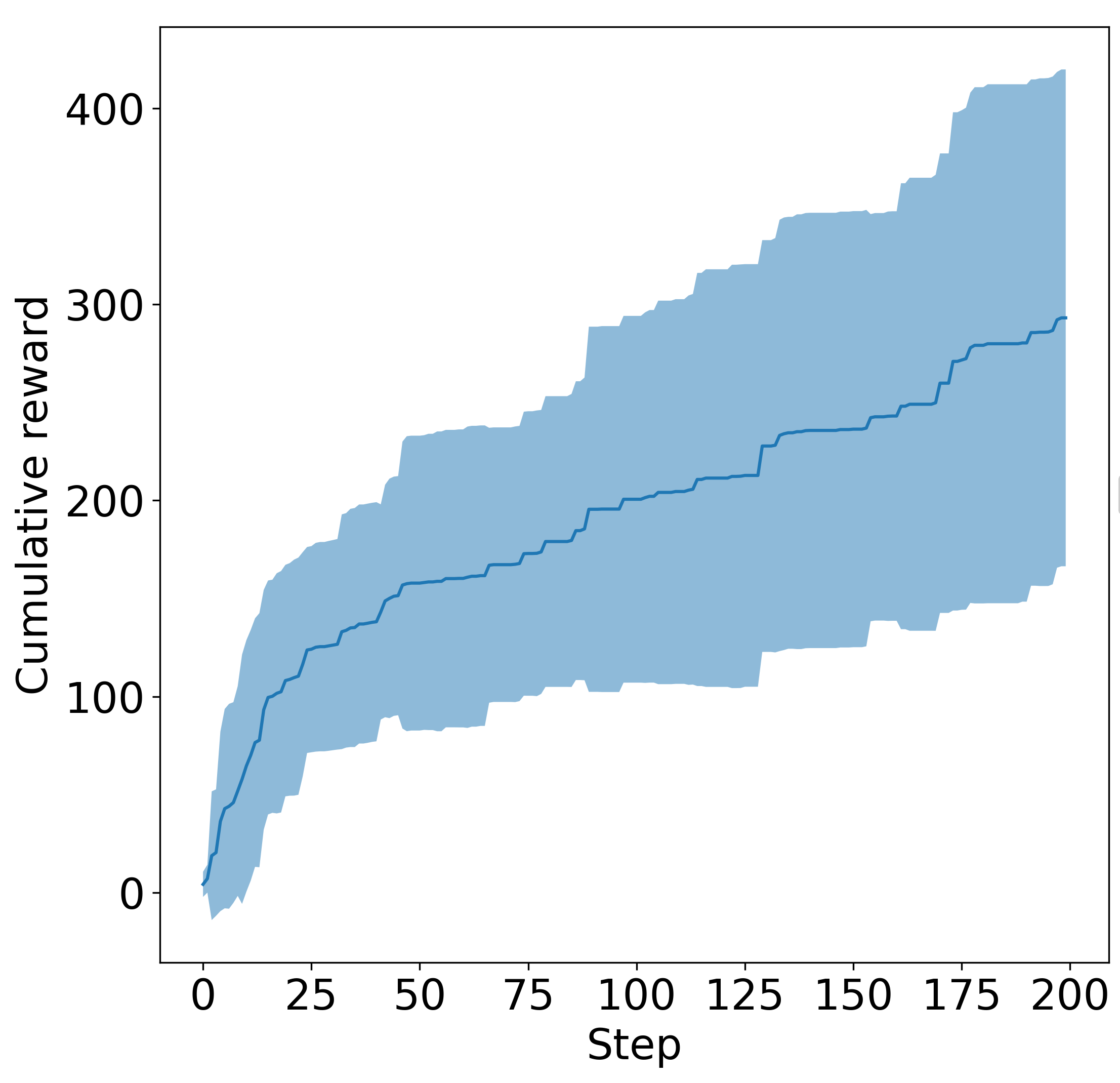}
 \captionsetup{font=scriptsize}
 \caption{CC22: Early Phase}
 \label{fig:cc100}
 \end{subfigure}
 \hfill
 \begin{subfigure}[b]{0.24\textwidth}
 \centering
 \includegraphics[width=\textwidth, trim=0 0 0 0, clip]{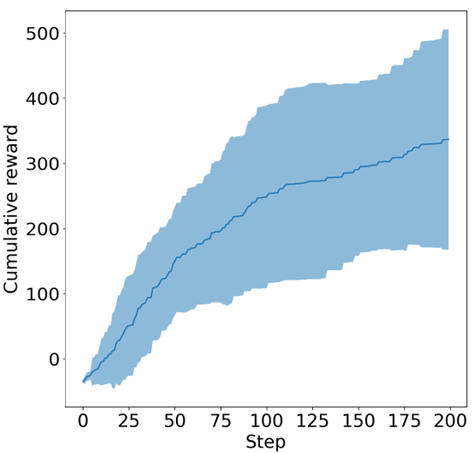}
 \captionsetup{font=scriptsize}
 \caption{CC22: Late Phase}
 \label{fig:cc500}
 \end{subfigure}
 \begin{subfigure}[b]{0.24\textwidth} 
 \centering
 \includegraphics[width=\textwidth, trim=0 0 0 0, clip]{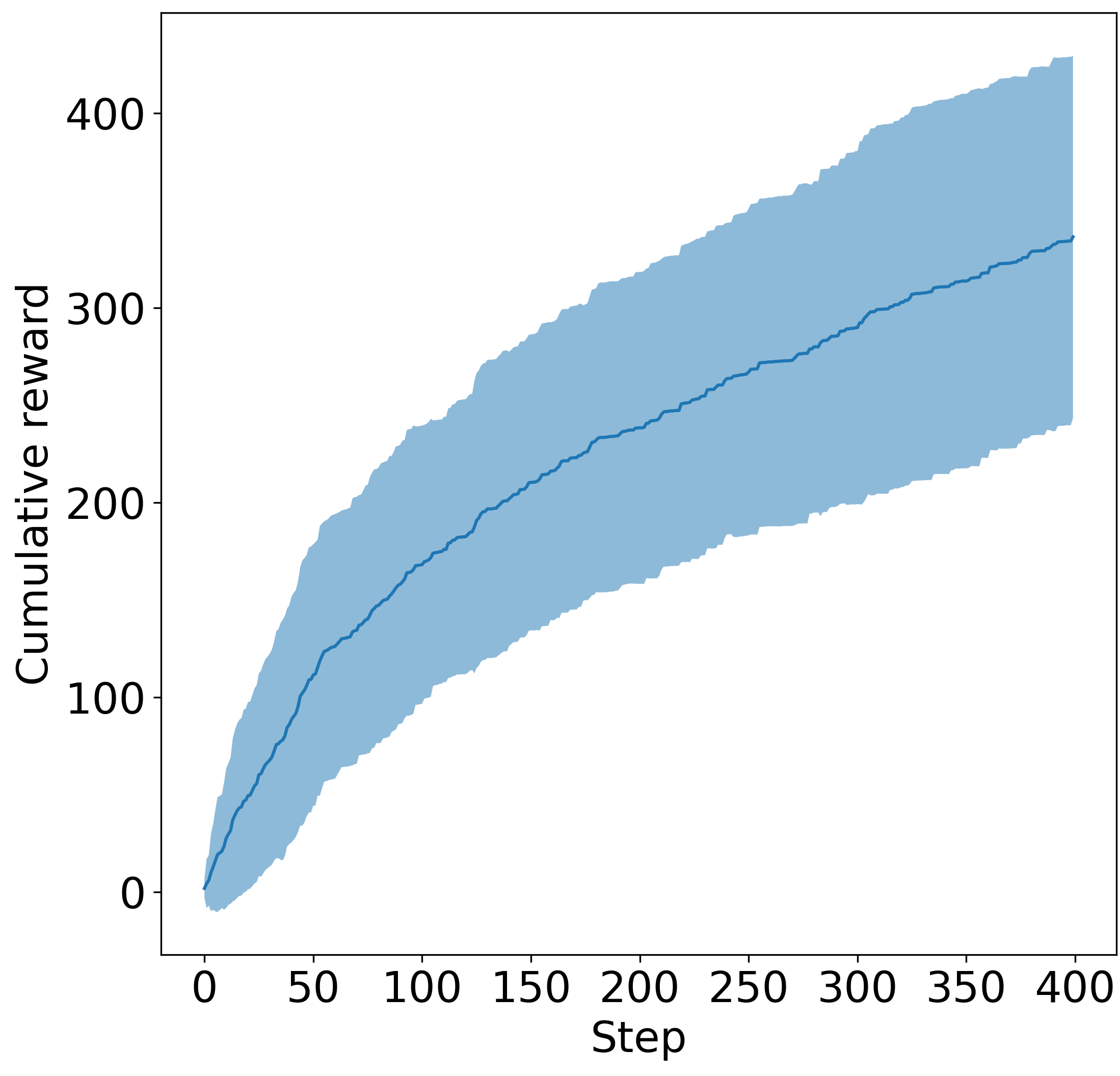}
 \captionsetup{font=scriptsize}
 \caption{CC100: Early Phase}
 \label{fig:ctf}
 \end{subfigure}
 \hfill
 \begin{subfigure}[b]{0.24\textwidth}
 \centering
 \includegraphics[width=\textwidth, trim=0 0 0 0, clip]{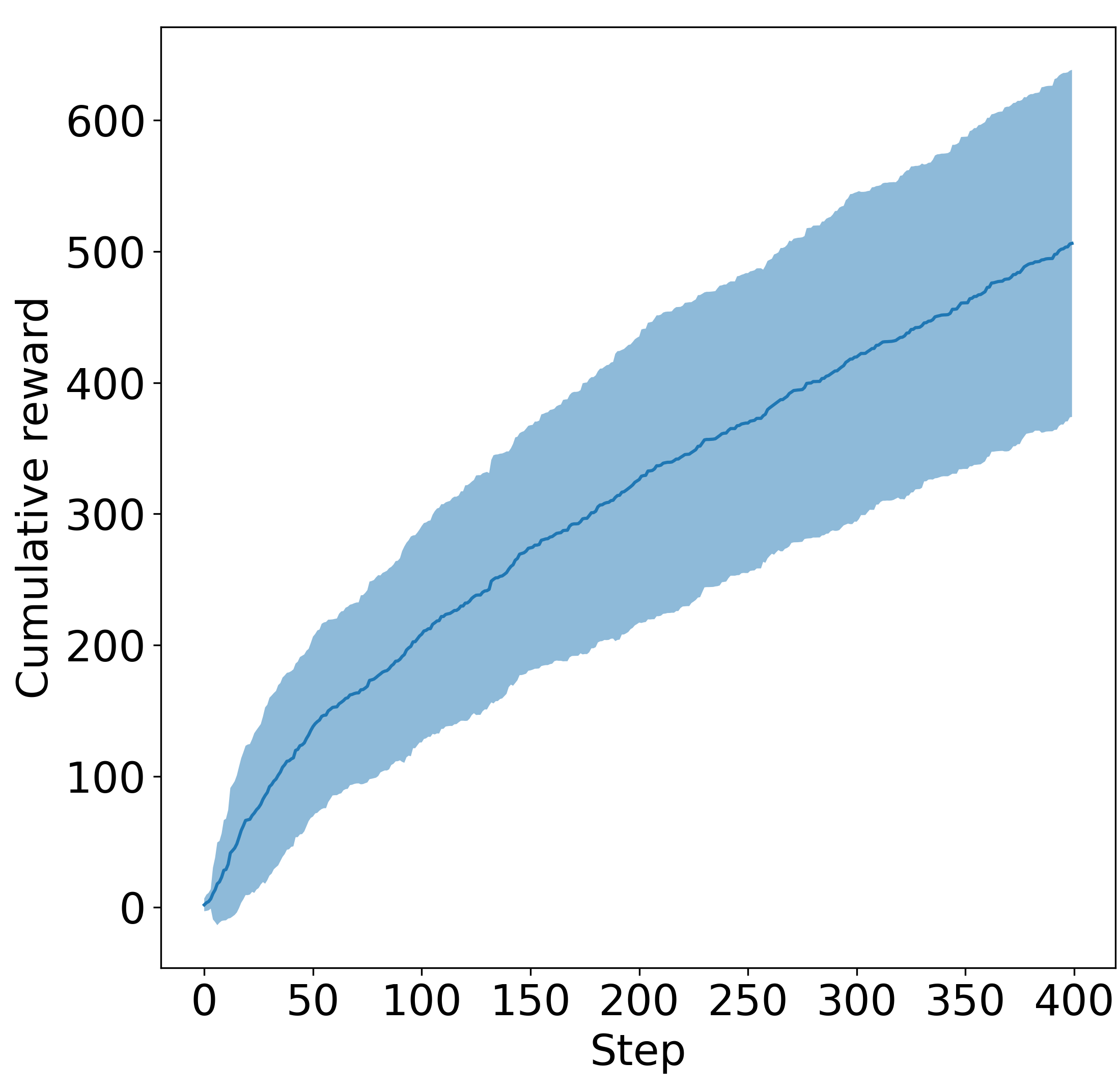}
 \captionsetup{font=scriptsize}
 \caption{CC100: Late Phase}
 \label{fig:cc22}
 \end{subfigure}
 \hfill
 \begin{subfigure}[b]{0.24\textwidth}
 \centering
 \includegraphics[width=\textwidth, trim=0 0 0 0, clip]{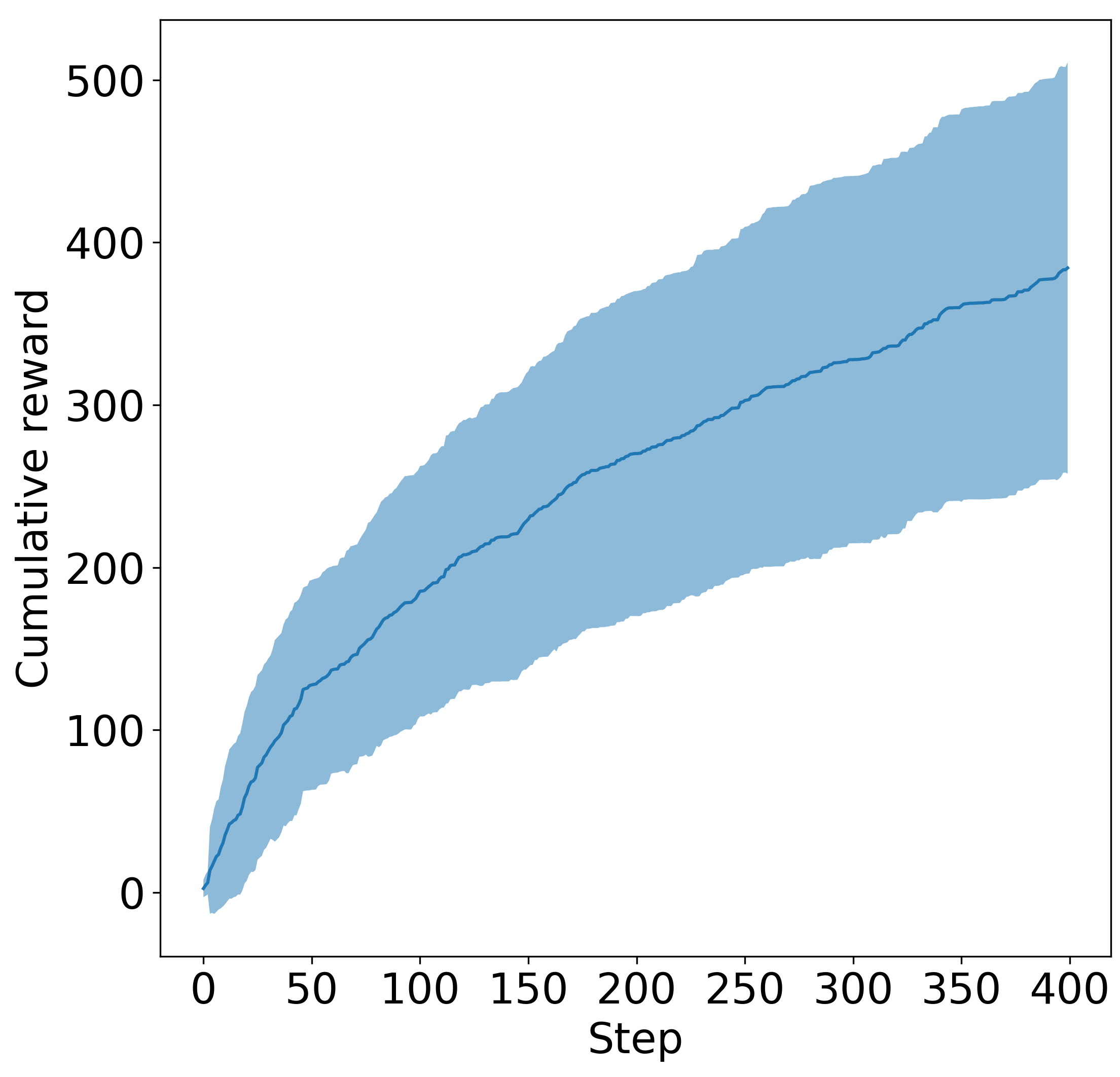}
 \captionsetup{font=scriptsize}
 \caption{CC500: Early Phase}
 \label{fig:cc100}
 \end{subfigure}
 \hfill
 \begin{subfigure}[b]{0.24\textwidth}
 \centering
 \includegraphics[width=\textwidth, trim=0 0 0 0, clip]{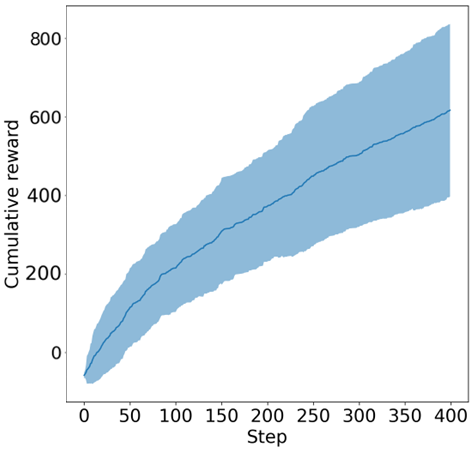}
 \captionsetup{font=scriptsize}
 \caption{CC500: Late Phase}
 \label{fig:cc500}
 \end{subfigure}
\caption{\small{Cumulative rewards comparison in early vs. late attack phases across environments (Results show that agents achieve higher rewards in the late phase, indicating a progression from exploratory behaviour to exploitation-focused strategies as training advances).}} \label{fig:MDPlate}
\end{figure*}

\noindent \textit{\textbf{Results.}} 
We trained each policy with a consistent optimisation setup, adjusting the number of episodes and iterations per episode to reflect the complexity of each environment. Specifically, the agent was trained for 20 episodes (1500 iterations) in CTF, 20 episodes (2000 iterations) in CC22, 100 episodes (2000 iterations) in CC100, and 200 episodes (4000 iterations) in CC500 environment.

Figure~\ref{fig:Setup1} summarises performance across environments and the results show that DQL significantly outperformed the baseline agents in terms of stability, scalability, and cumulative reward. Classical methods such as Tabular Q-Learning and Credential Lookup struggled in larger environments (CC100, with no viable results in CC500). Random search underperformed in all structured settings.
 While the Exploiting DQL variant achieved slightly higher cumulative rewards, it is not a standalone method; it relies on a fully trained DQL agent and simply exploits the final policy without any further training or adaptation. As such, it does not reflect the ongoing decision-making process that our explainability framework aims to explain. We therefore selected DQL as the attacker policy for all subsequent experiments. We emphasise that the results in Figure \ref{fig:Setup1} serve to establish a realistic setting to demonstrate the utility of our multi-layer explainability framework.

\begin{tcolorbox}
Our benchmarking shows DQL as the most effective and scalable attacker policy across all environments. While the Exploiting DQL variant achieves marginally higher rewards, it leverages a trained DQL policy without further learning, making it less representative of adaptive behaviour. Classical approaches like Random Search, Tabular Q-Learning, and Credential Lookup fail to scale, particularly in larger networks (CC100 and CC500). We therefore adopt DQL for all subsequent experiments, ensuring that our explainability framework is evaluated on a stable and representative attacker policy.
\end{tcolorbox}

\subsection{Setup 2: Exploration-Exploitation Dynamics}\label{s2}
To analyse how exploration-exploitation strategies shape agent behaviour, we compare two strategies: Early Exploitation and Standard Exploration and track their impact on cumulative reward and node discovery rate across environments. This reveals whether the agent behaves opportunistically, methodically, or inefficiently over time.

\noindent \textit{\textbf{Results.}} For Early Exploitation, we initialised the exploration rate at $\epsilon = 0.001$ and increased it additively by 0.005 per step until reaching $\epsilon = 0.9$. In contrast, Standard Exploration began with $\epsilon = 0.9$ and decayed multiplicatively to $\epsilon = 0.01$ using a decay factor of 0.95. Training was conducted for 30 episodes (200 iterations per episode) in CTF and CC22, 40 episodes (300 iterations per episode) in CC100, and 100 episodes (300 iterations per episode) in CC500.
Figure \ref{fig:rewardMDP1} shows {the impact of exploration strategies on cumulative rewards across environments.} In smaller topologies (CTF and CC22), \textit{Standard Exploration} quickly outperformed \textit{Early Exploitation}, benefitting from more aggressive probing. 
In larger environments (CC100 and CC500), this advantage widened considerably: Early Exploitation frequently stagnated at lower cumulative rewards, while Standard Exploration achieved sustained improvement and a late-stage reward surge in CC500 after broad, systematic probing. 
These patterns are echoed in the node discovery rates in Figure \ref{fig:discoveredMDP1} where Standard Exploration uncovered more nodes earlier, while Early Exploitation remained constrained.\\

\begin{tcolorbox}
Our findings demonstrate that the choice of exploration strategy fundamentally drives both the efficiency of learning and the tactical profile of the agent. {Standard Exploration} promotes broader state-space coverage and faster reward accumulation, enabling the agent to form well-informed policies through deliberate reconnaissance and systematic probing. In contrast, {Early Exploitation} narrows the exploration horizon, leading to opportunistic but less comprehensive behaviour that limits performance in larger and more complex environments such as CC100 and CC500. By exposing these strategy-driven behavioural signatures, our framework provides interpretable evidence of how exploration dynamics influence policy formation, enabling rigorous analysis of attacker decision-making and implications for scalable, realistic adversarial simulations.
\end{tcolorbox}

\begin{figure*}[t!]
\centering
\begin{subfigure}[t]{0.48\textwidth}
\centering
\includegraphics[width=\textwidth]{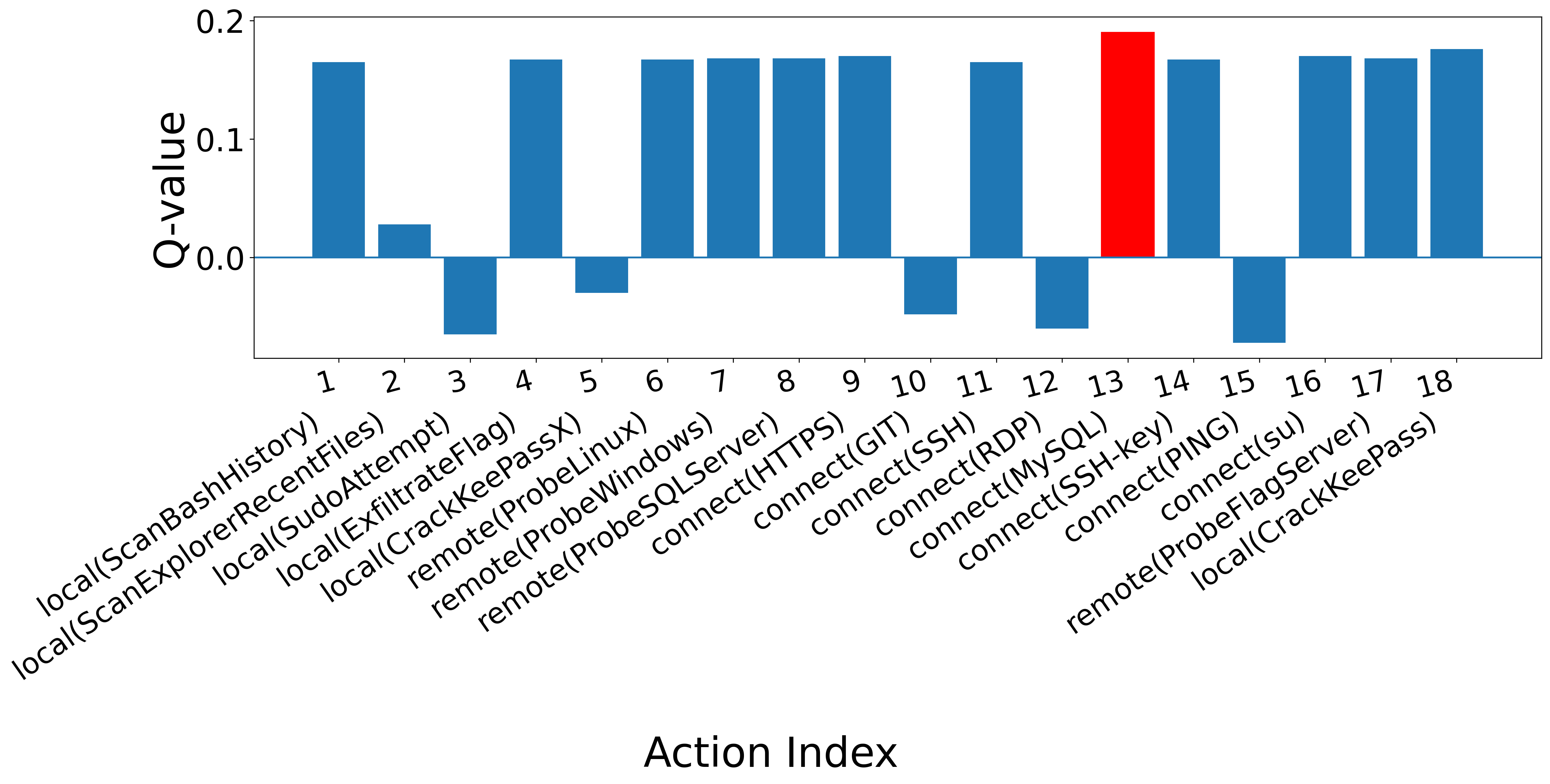}
\caption{CTF: Episode 10}
\label{fig:ctf1}
\end{subfigure}
\hfill
\begin{subfigure}[t]{0.48\textwidth}
\centering
\includegraphics[width=\textwidth]{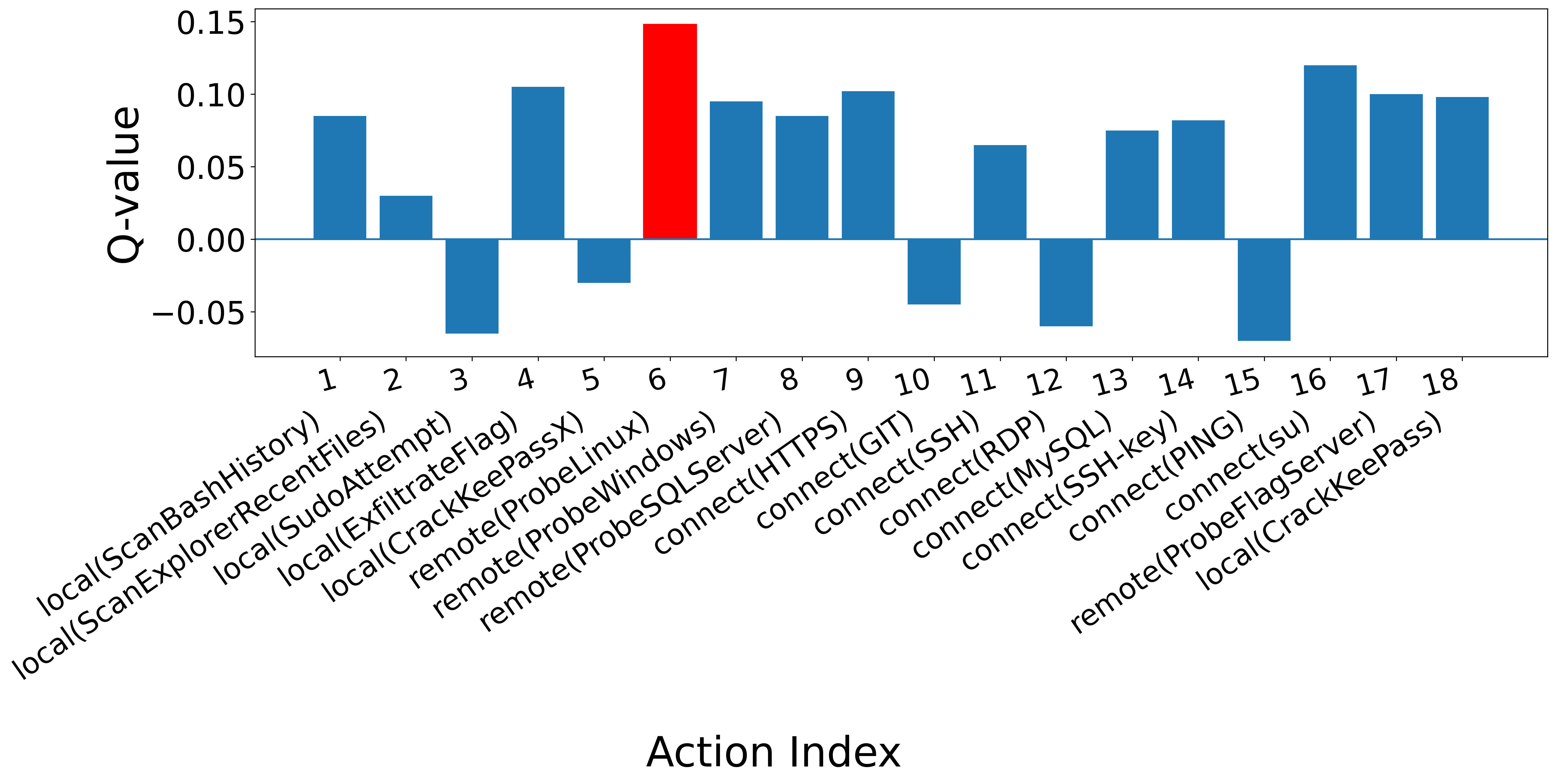}
\caption{CTF: Episode 25}
\label{fig:ctf2}
\end{subfigure}
\vspace{0.6em}
\begin{subfigure}[t]{0.48\textwidth}
\centering
\includegraphics[width=\textwidth]{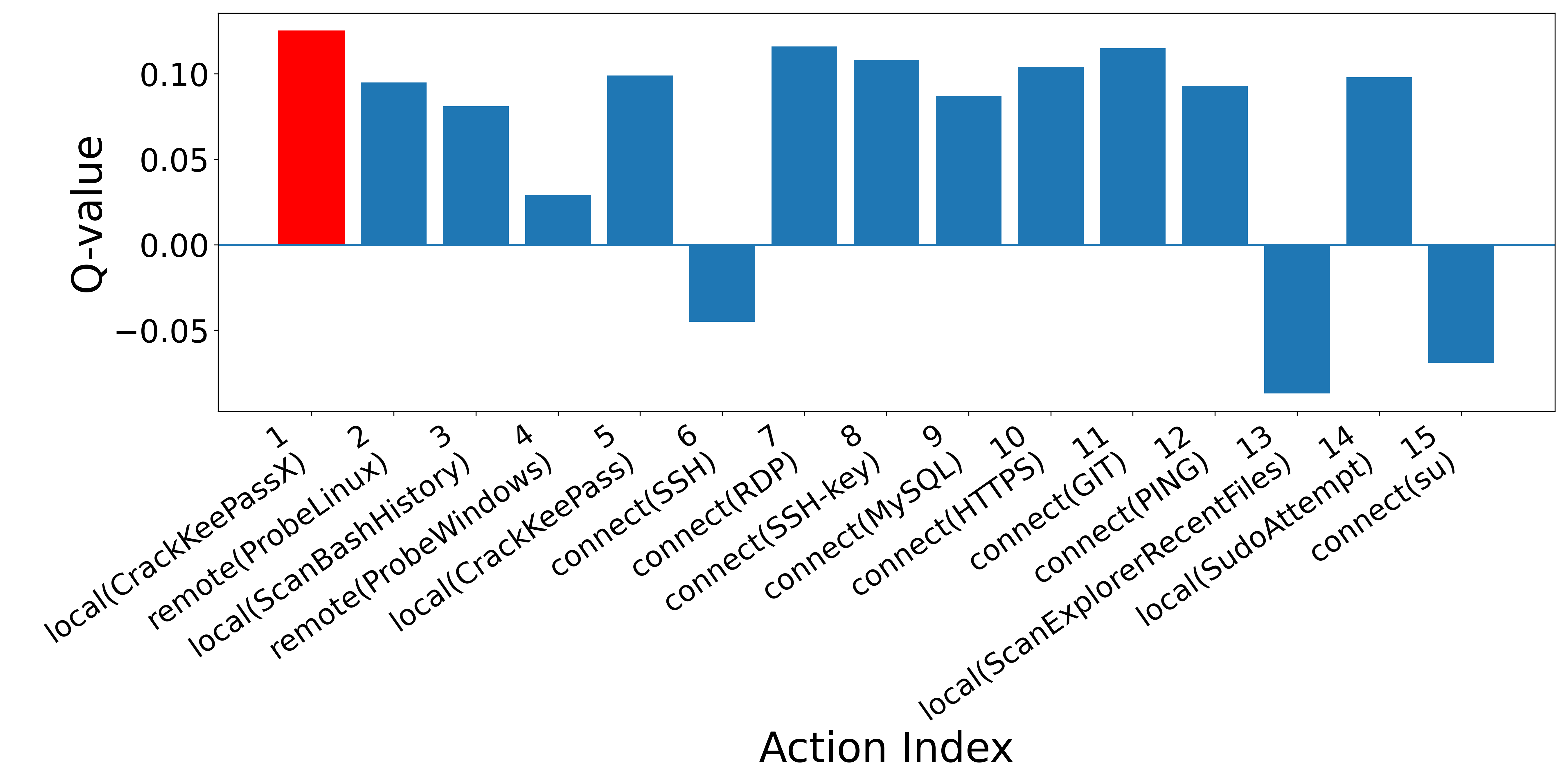}
\caption{CC22: Episode 20}
\label{fig:cc22_1}
\end{subfigure}
\hfill
\begin{subfigure}[t]{0.48\textwidth}
\centering
\includegraphics[width=\textwidth]{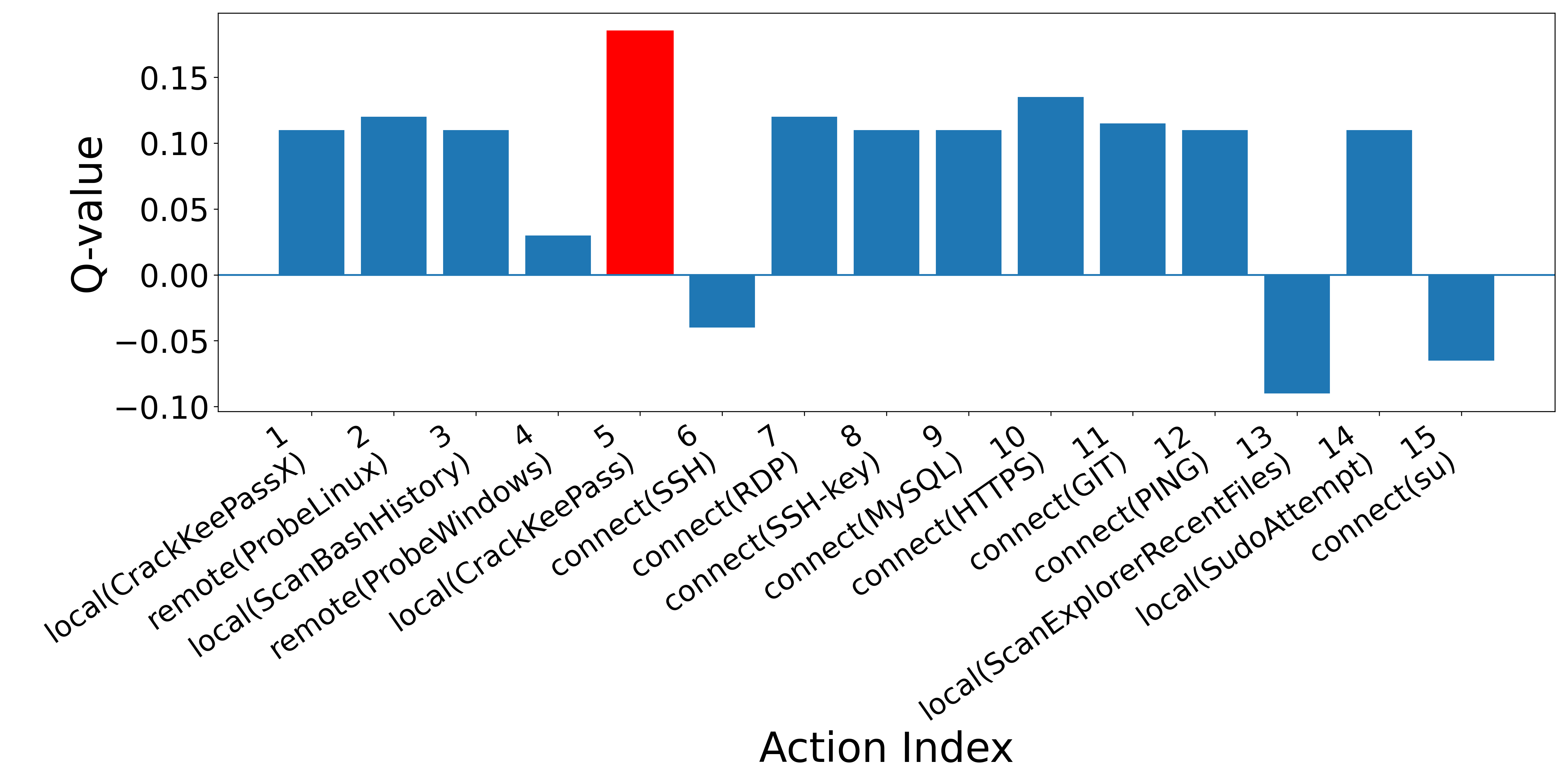}
\caption{CC22: Episode 35}
\label{fig:cc22_2}
\end{subfigure}
\vspace{0.6em}
\begin{subfigure}[t]{0.48\textwidth}
\centering
 \includegraphics[width=\textwidth]{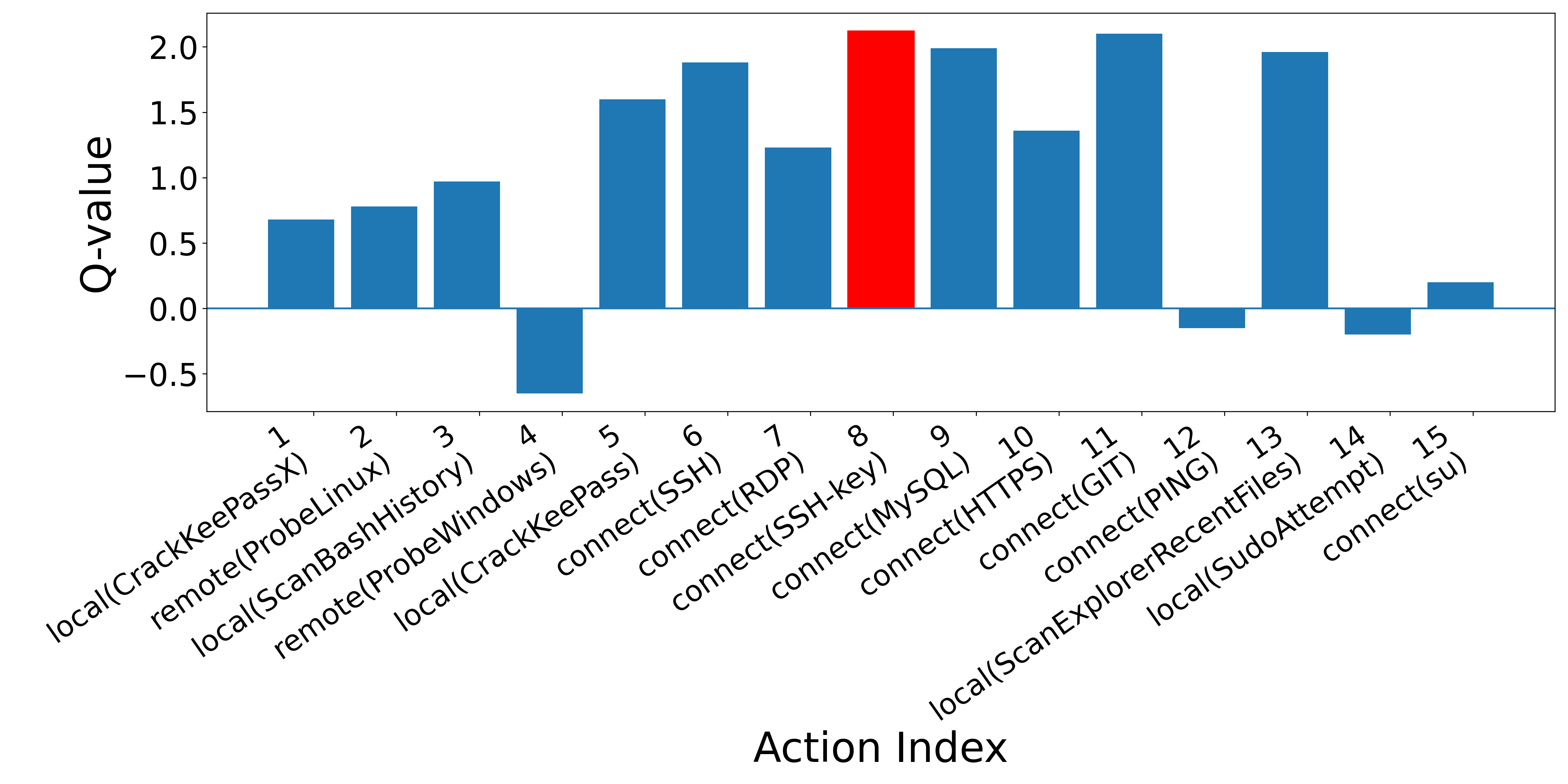}
\caption{CC100: Episode 10}
\label{fig:cc100_1}
\end{subfigure}
\hfill
\begin{subfigure}[t]{0.48\textwidth}
\centering
 \includegraphics[width=\textwidth]{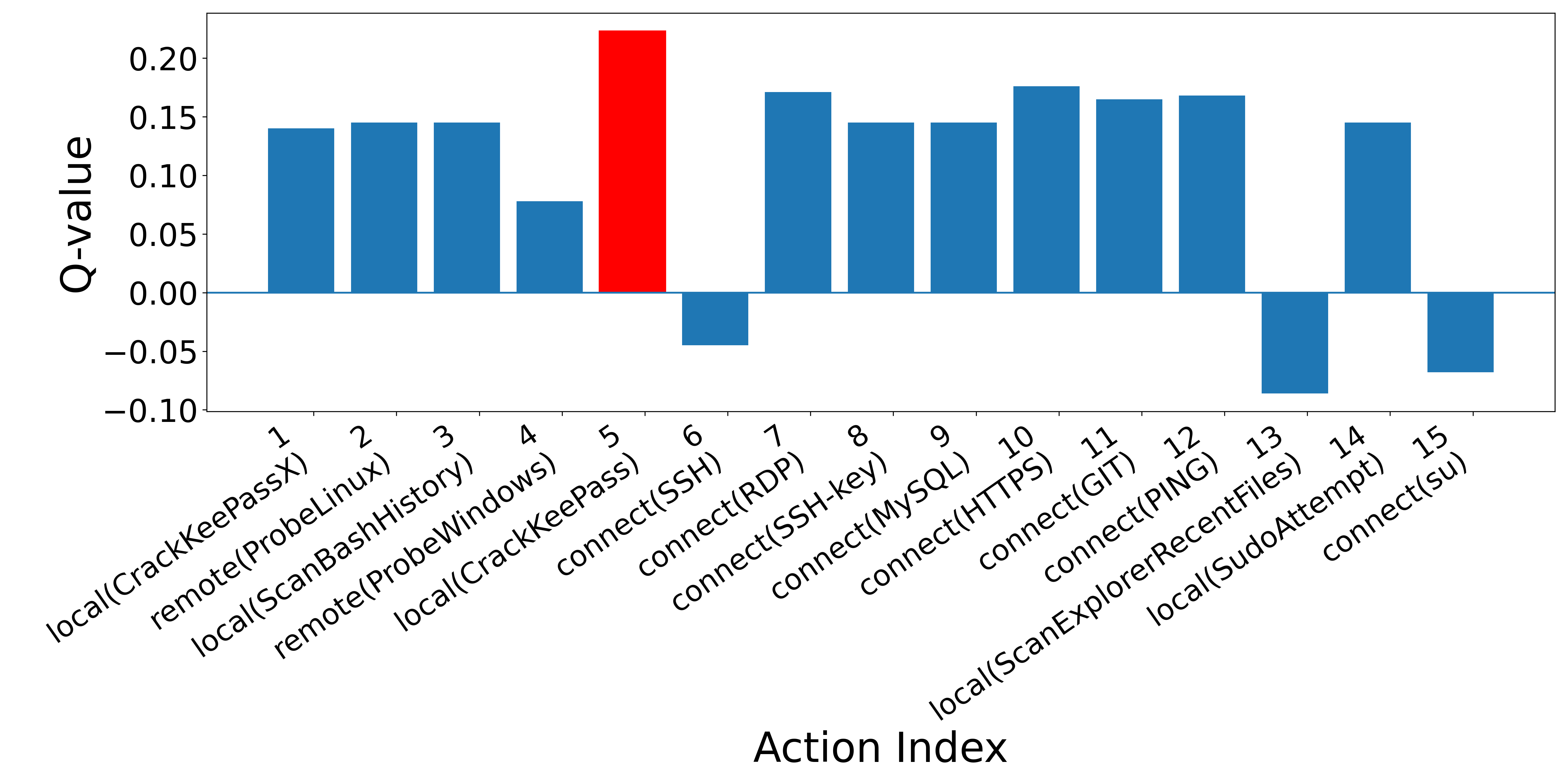}
\caption{CC100: Episode 35}
\label{fig:cc100_2}
\end{subfigure}

\vspace{0.6em}

\begin{subfigure}[t]{0.48\textwidth}
\centering
 \includegraphics[width=\textwidth]{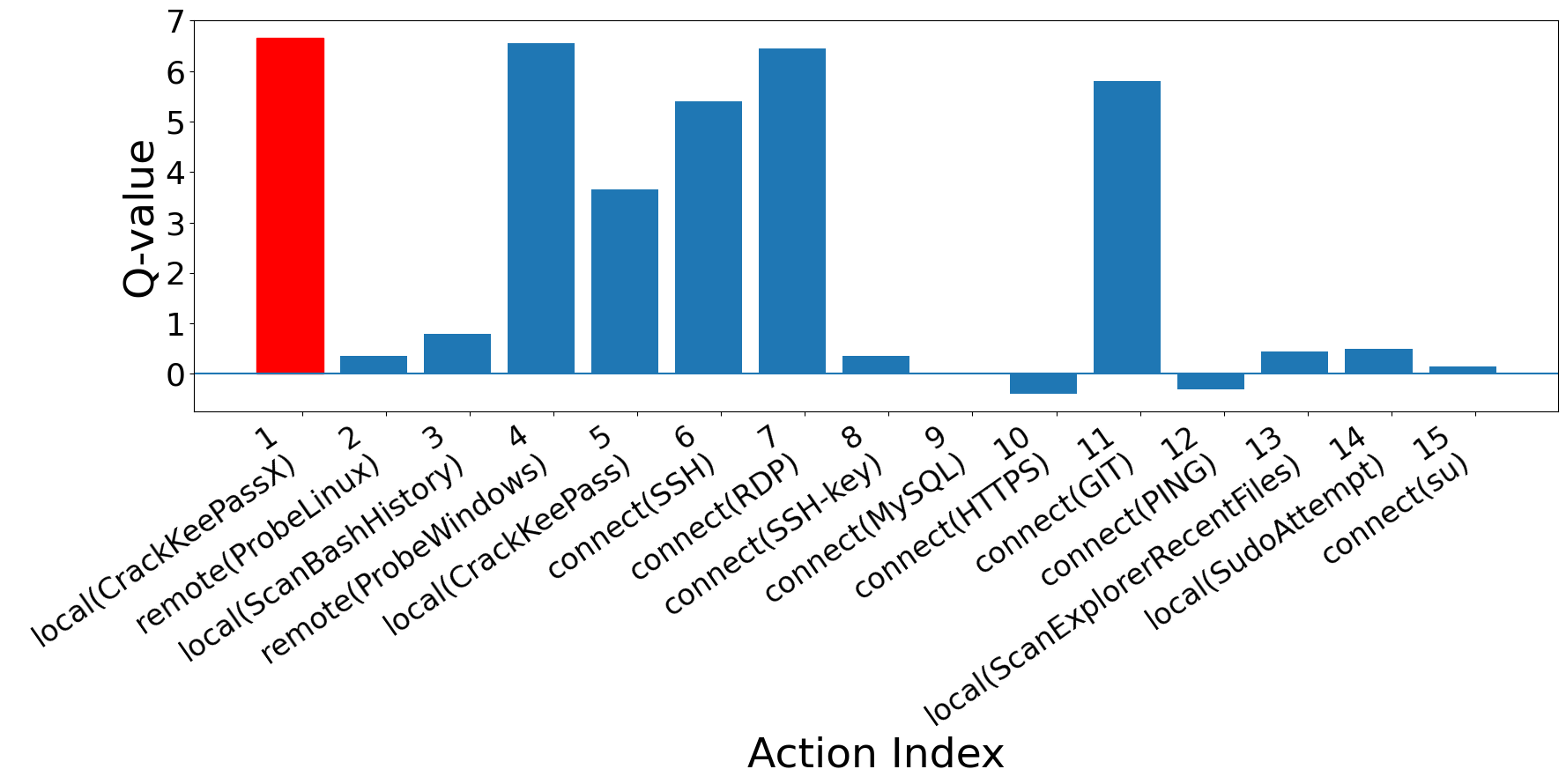}
\caption{CC500: Episode 5}
\label{fig:cc500_1}
\end{subfigure}
\hfill
\begin{subfigure}[t]{0.48\textwidth}
\centering
 \includegraphics[width=\textwidth]{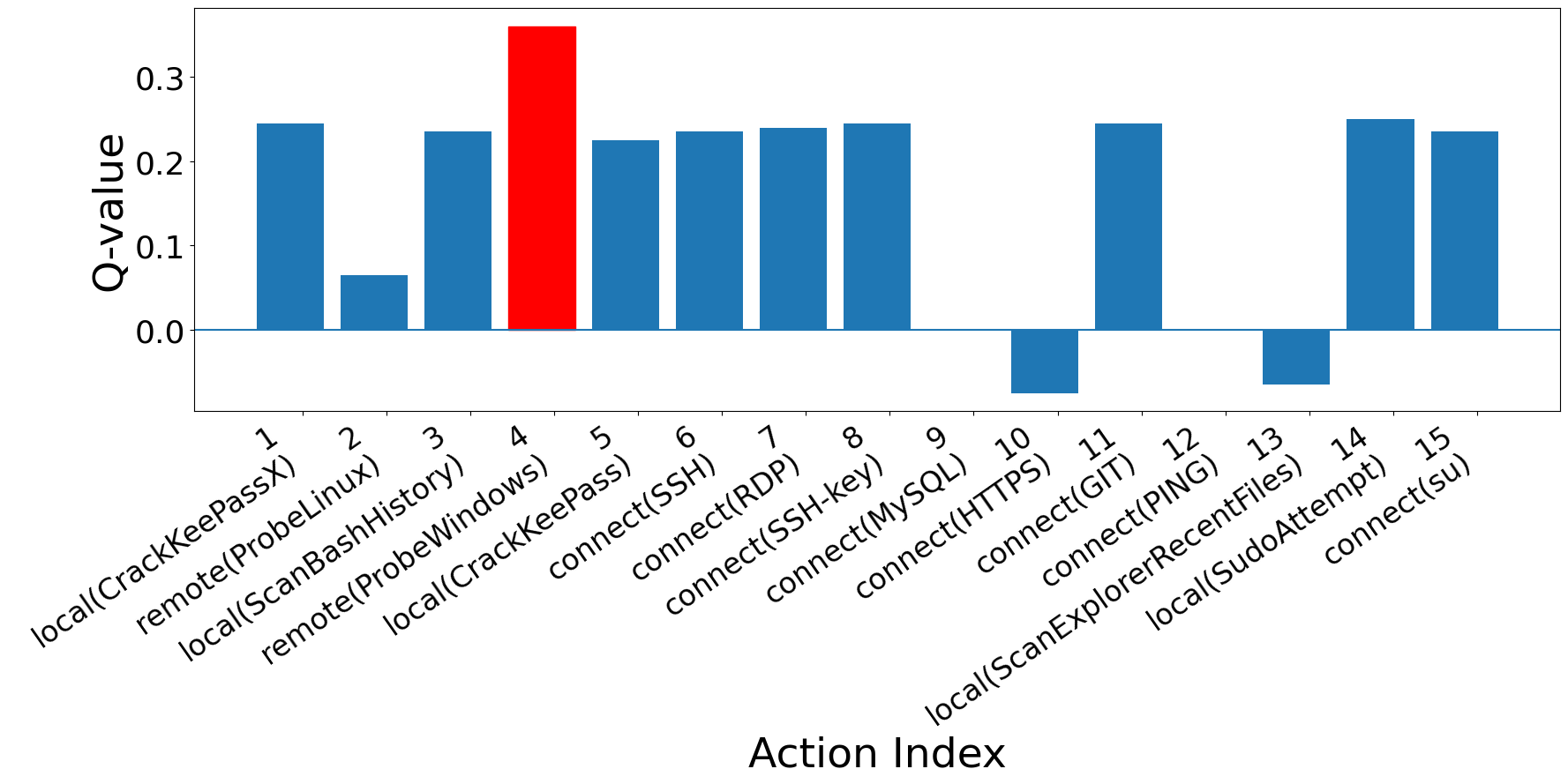}
\caption{CC500: Episode 40}
\label{fig:cc500_2}
\end{subfigure}

\caption{\small{{Emergence of action preferences via state-aggregated Q-values across episodes. The highlighted dominant action marks the shift from exploration to high-value, environment-specific tactics. DQL’s sharp Q-value gradients reflect discrete exploitation bursts as the agent locks onto effective attack sequences. Action indices for CyberBattleChain (1–15) and CTF (1–18) correspond to the x-axis and are listed in Table \ref{tab:action_ladder} and Table \ref{tab:abstract-actions}, respectively.}}}
 \label{fig:PL1_actions}
\end{figure*}

 \subsection{Setup 3: Phase-Aware Behaviour Evolution}\label{s3}
This setup supports phase-aware explainability by segmenting attacker behaviour into early and late stages using the compromise ratio $C_t$, defined as the proportion of compromised nodes, to highlight when strategic shifts occur.

\noindent \textit{\textbf{Results.}} Agents were trained for 20 episodes (200 iterations) in the CTF and CC22 environments, 100 episodes (400 iterations) in CC100, and 200 episodes (400 iterations) in CC500. The \textit{early phase} is defined as $C_t < 0.5$, while the \textit{late phase} corresponds to $C_t \geq 0.5$. {We select $\tau = 0.5$ as a balanced 
midpoint separating low-compromise 
exploratory behaviour from high-compromise 
exploitative behaviour. As shown in 
Table~\ref{tab:threshold_sensitivity}, 
varying $\tau$ affects the magnitude of 
the reward gap between phases but not 
its direction, the late phase 
consistently achieves higher cumulative 
reward than the early phase across all 
environments and threshold values. Lower 
thresholds assign more reward-rich 
transitions to the late phase, widening 
the gap, while higher thresholds reduce 
it by including more of the trajectory 
in the early phase. The midpoint 
$\tau = 0.5$ yields a stable and 
interpretable phase boundary that 
balances these effects, with consistent 
trends observed across environments of 
varying sizes and structural 
characteristics.} Figure~\ref{fig:MDPlate} shows that early-phase rewards remain modest due to limited visibility and cautious probing. In contrast, late-phase curves exhibit higher rewards and steeper growth, reflecting increased access, confidence, and exploitation capability.

In CyberBattleChain environments, the attack path is fixed, with each compromised node granting access to the next. Behavioural variation therefore reflects how efficiently the agent escalates privileges, selects exploits, and avoids traps, rather than differences in path selection. The observed late-phase deceleration in these environments likely reflects structural constraints rather than a fundamental strategic shift. By contrast, the CTF environment presents a more open topology, where phase-aware analysis reveals how strategies evolve across multiple possible paths, including which services and nodes are prioritised once situational awareness improves. Such insights can assist defenders in identifying high-risk assets and anticipating the timing of escalation.

The $C_t = 0.5$ threshold serves as an interpretable behavioural inflection point rather than a hard operational rule. Below this threshold, agents predominantly engage in reconnaissance-driven exploration under high uncertainty, whereas beyond it, expanded visibility and access enable more targeted, high-impact exploitation. This phase-based segmentation provides a principled and consistent basis for analysing how attacker behaviour evolves over the course of training.

\begin{tcolorbox}
Our framework simulates an adaptive attacker that improves across training episodes via policy (parameter) updates, without using explicit episodic memory. Segmenting attacker behaviour into early and late phases using the network compromise ratio ($C_t$) reveals how strategies evolve during training. In open environments like CTF, agents shift from cautious probing to targeted exploitation as access expands, whereas in constrained settings such as CyberBattleChain, behaviour plateaus earlier due to structural limits on strategic diversity. This phase-aware analysis exposes when and how escalation occurs and helps defenders assess how network structure shapes adversarial learning. The resulting phase-specific reward curves also provide diagnostic value, enabling deeper evaluation of policy generalisation, convergence dynamics, and environmental impacts on learned tactics. {Equivalent PPO experiments confirm these trends and the persistence of structural transitions across learning paradigms (see §\ref{s7})}.
\end{tcolorbox}

\begin{table}[t!]
\centering
\caption{{Sensitivity analysis of phase segmentation across compromise ratio thresholds. 
We vary the transition threshold $\tau \in \{0.3, 0.5, 0.7\}$ and report mean cumulative reward aggregated over transitions belonging to 
the early ($C_t < \tau$) and late ($C_t \geq \tau$) phases.}}
\label{tab:threshold_sensitivity}
{
\begin{tabular}{p{1.5cm}p{1.5cm}p{1.5cm}p{1cm}}
\toprule
\textbf{Env} & \textbf{$\tau$} & \textbf{Early Phase} & \textbf{Late Phase} \\
\midrule
CTF   & 0.3 & 146.82 & 223.41 \\
CTF   & 0.5 & 185.30 & 215.75 \\
CTF   & 0.7 & 203.94 & 216.73 \\
\midrule
CC22  & 0.3 & 229.63 & 361.87 \\
CC22  & 0.5 & 290.47 & 335.43 \\
CC22  & 0.7 & 321.48 & 341.92 \\
\midrule
CC100 & 0.3 & 271.34 & 528.63 \\
CC100 & 0.5 & 335.36 & 500.87 \\
CC100 & 0.7 & 391.52 & 458.37 \\
\midrule
CC500 & 0.3 & 321.43 & 649.82 \\
CC500 & 0.5 & 385.29 & 610.61 \\
CC500 & 0.7 & 448.37 & 564.91 \\
\bottomrule
\end{tabular}
}
\end{table}

\begin{table*}[t!]
\centering
\caption{\small{Action indices and descriptions for Chain environments (CC22, CC100, CC500). These indices (1-15) correspond to the x-axis in Figure \ref{fig:PL1_actions} and are used to interpret the emergence of action preferences.}}
\label{tab:action_ladder}
\renewcommand{\arraystretch}{1.2}
\footnotesize
\begin{tabularx}{\textwidth}{@{}c >{\raggedright\arraybackslash}p{5cm} X@{}}
\toprule
\textbf{Index} & \textbf{Abstract Action} & \textbf{Description} \\
\midrule
1 & \texttt{local(CrackKeePassX)} & Local attempt to crack KeepPassX and retrieve RDP credentials on Linux nodes. \\
2 & \texttt{remote(ProbeLinux)} & Remote probe to confirm whether a target host is running Linux. \\
3 & \texttt{local(ScanBashHistory)} & Local scan of bash history to discover references to other machines. \\
4 & \texttt{remote(ProbeWindows)} & Remote probe to determine if a target host is running Windows. \\
5 & \texttt{local(CrackKeePass)} & Local attempt to crack KeepPass on Windows nodes to obtain SSH credentials. \\
6 & \texttt{connect(SSH)} & Network connection to a target over the SSH service using known credentials. \\
7 & \texttt{connect(RDP)} & Network connection to a target over the RDP service using known credentials. \\
8 & \texttt{connect(SSH-key)} & SSH connection attempt using a leaked or cached SSH key. \\
9 & \texttt{connect(MySQL)} & Connection to a MySQL service on the target host. \\
10 & \texttt{connect(HTTPS)} & HTTPS service connection, typically for basic enumeration. \\
11 & \texttt{connect(GIT)} & Connection to a Git service port on the target. \\
12 & \texttt{connect(PING)} & ICMP ping action to check reachability of the target host. \\
13 & \texttt{local(ScanExplorerRecentFiles)} & Local scan of Windows Explorer recent files (trap on Linux systems). \\
14 & \texttt{local(SudoAttempt)} & Local attempt to escalate privileges via \texttt{sudo} on Linux. \\
15 & \texttt{connect(su)} & Connection or command using \texttt{su} credentials on the target host. \\
\bottomrule
\end{tabularx}
\end{table*}

\subsection{Setup 4: Temporal Q-Value Analysis}\label{s4}

In this setup, we track how action preferences evolve over training, revealing shifts in tactical focus. We compute state-aggregated Q-values per action across episodes to visualise which actions the agent increasingly favours as it learns.

\noindent \textit{\textbf{Results.}} We trained the attacker for 40 episodes (400 iterations), logging Q-values for all visited state-action pairs. Figure \ref{fig:PL1_actions} shows Q-value distributions per action at different training stages across environments, with Tables \ref{tab:action_ladder} and \ref{tab:abstract-actions} defining the mapping from action indices to semantic descriptions for the Chain and CTF environments, respectively. {
The mapping between action indices and their semantic descriptions is not manually constructed. Each action index corresponds directly to a predefined action in the CyberBattleSim environment, and its semantic label (e.g., \texttt{remote(ProbeLinux)}, \texttt{connect(GIT)}) is automatically retrieved from the environment’s action space specification. The Tables \ref{tab:action_ladder} and \ref{tab:abstract-actions} are included to further explain the meaning of each action and support easier understanding of the results. }

In early episodes, Q-values are relatively low and undifferentiated, indicating limited experience and high uncertainty. As training progresses, a preference hierarchy emerges: a small number of actions become consistently favored, with one action (highlighted in red) attaining the highest mean Q-value across states, indicating stronger expected long-term value.

\noindent \textbf{Learning Trajectories by Environment:}
\begin{itemize}[leftmargin=0pt,label=\textbullet]
 \item {\textbf{CTF (Figure \ref{fig:PL1_actions}(a), \ref{fig:PL1_actions}(b)):} A clear pattern emerges where several actions, including \texttt{ICMP ping}, \texttt{Attempt sudo privilege escalation}, \texttt{Connect to Git service}, \texttt{RDP connection with cached creds}, and \texttt{Crack KeepPassX vault (Linux)} (action 3, 5, 10, 12, 15), consistently receive negative Q-values. This suggests these options are ineffective or yield poor long-term value under the current reward and exploration strategy. Moreover, the agent converges on \texttt{remote(ProbeLinux)} (action 6) as its highest-valued action by episode 25, indicating a deliberate pivot toward OS fingerprinting and reconnaissance. This reflects an interpretable, interpretable shift toward information-gathering tactics even in smaller environments.}

 \item {\textbf{CC22 (Figure \ref{fig:PL1_actions}(c), \ref{fig:PL1_actions}(d)):} Early exploration favours \texttt{local(CrackKeePassX)}, \texttt{connect(RDP)}, and \texttt{connect(GIT)} (actions 1, 7, 11), consistent with lateral movement and credential reuse. By episode 35, the agent settles on \texttt{local(CrackKeePass)} (action 5), marking a transition to reliable, local credential cracking once the network is partially mapped.}

 \item {\textbf{CC100 (Figure \ref{fig:PL1_actions}(e), \ref{fig:PL1_actions}(f)):} The agent begins with broad exploration, including \texttt{connect(SSH-key)}, \texttt{connect(MySQL)}, \texttt{connect(GIT)}, and \texttt{local(ScanExplorerRecentFiles)} (actions 8, 9, 11, 13), and converges on the \texttt{local(CrackKeePass)} action by episode 35. This again signals the emergence of credential-based exploitation as a robust tactic once credentials become available.}

 \item {\textbf{CC500 (Figure \ref{fig:PL1_actions}(g), \ref{fig:PL1_actions}(h)):} Unlike smaller environments, CC500 favours reconnaissance. The agent begins with trials of \texttt{local(CrackKeePassX)}, \texttt{remote(ProbeWindows)}, and \texttt{connect(RDP)} (action 1, 4, 7), but by episode 40, \texttt{remote(ProbeWindows)} (action 4) dominates. This pivot reflects a visibility-first strategy in large networks, where reconnaissance precedes coordinated exploitation. For cyber range developers/red team trainers, this validates that training regime fosters realistic, stepwise escalation rather than brittle or impulsive policies.}

\end{itemize}

\begin{tcolorbox} 
These results highlight the agent’s progression from broad exploration to targeted exploitation of high-reward paths, with environment scale shaping the final strategy. In smaller environments (CC22, CC100), the agent converges on local exploits such as, \texttt{local(CrackKeePass)} while in larger networks like CC500, it prioritises reconnaissance actions such as \texttt{remote(ProbeWindows)} (action 4), reflecting a preference for scalable reconnaissance before deeper compromise. By tracking the evolution of Q-values over time, our framework surfaces these strategy-dependent patterns, enabling developers to diagnose brittle convergence, identify underutilised actions, and assess policy stability. For cybersecurity teams engaged in red/blue-team exercises or AI-driven threat simulation, these insights provide a clear view into how autonomous agents learn and prioritise tactics, informing training design, service hardening, and the strategic timing of decoy deployments.
\end{tcolorbox}

\subsection{Setup 5: Strategic Decision Attribution via PER}\label{s5}
This experiment demonstrates how Prioritised Experience Replay surfaces the most impactful (high-TD-error) transitions, acting as an attention mechanism that steers the agent toward strategically significant experiences.\\

\begin{table}[t!]
\centering
\caption{\small{Action indices and descriptions for CTF environment. These indices (1-18) correspond to the x-axis in Figure \ref{fig:PL1_actions} and are used to interpret the emergence of action preferences.}}
\renewcommand{\arraystretch}{1.3} 
\scriptsize
\setlength{\tabcolsep}{6pt} 
\begin{tabular}{p{0.4cm}p{2.9cm}p{3.8cm}}
\hline
\textbf{Index} & \textbf{Abstract Action} & \textbf{Description} \\
\hline
1 & local(ScanBashHistory) & Scan bash history for host references \\
2 & local(ScanExplorerRecentFiles) & Scan Windows recent-files list \\
3 & local(SudoAttempt) & Attempt sudo privilege escalation \\
4 & local(ExfiltrateFlag) & Read local flag file \\
5 & local(CrackKeePassX) & Crack KeepPassX vault (Linux) \\
6 & remote(ProbeLinux) & Probe to confirm Linux OS \\
7 & remote(ProbeWindows) & Probe to confirm Windows OS \\
8 & remote(ProbeSQLServer) & Probe for SQL-Server service \\
9 & connect(HTTPS) & Connect to HTTPS service \\
10 & connect(GIT) & Connect to Git service \\
11 & connect(SSH) & SSH connection with cached creds \\
12 & connect(RDP) & RDP connection with cached creds \\
13 & connect(MySQL) & Connect to MySQL service \\
14 & connect(SSH-key) & SSH using leaked key \\
15 & connect(PING) & ICMP ping \\
16 & connect(su) & Command/connection via su creds \\
17 & remote(ProbeFlagServer) & Probe the flag server \\
18 & local(CrackKeePass) & Crack KeepPass vault (Windows) \\
\hline
\end{tabular}
\label{tab:abstract-actions}
\end{table}

\begin{figure*}[t!]
 \centering
 \begin{subfigure}[b]{0.24\textwidth}
 \centering
 \includegraphics[width=\textwidth, trim=0 0 0 0, clip]{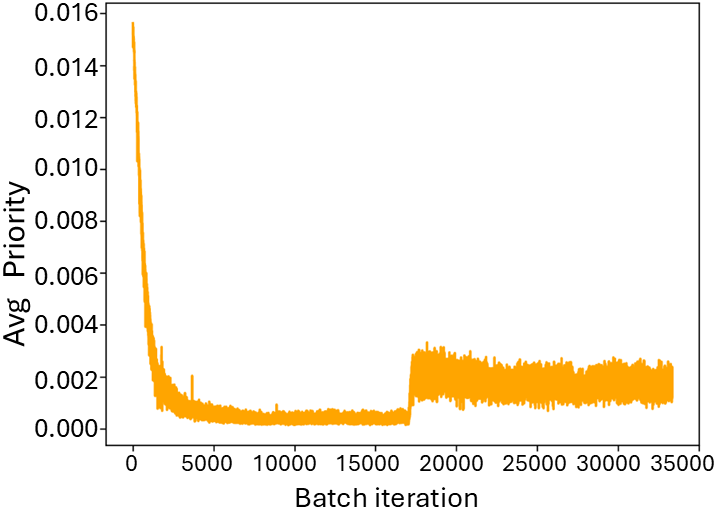}
 \caption{CTF}
 \label{fig:ctf}
 \end{subfigure}
 \hspace{0.06em}
 \begin{subfigure}[b]{0.24\textwidth}
 \centering
 \includegraphics[width=\textwidth, trim=0 0 0 0, clip]{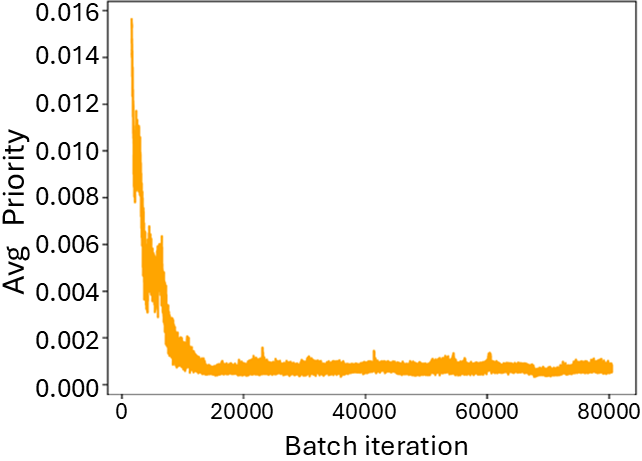}
 \caption{CC22}
 \label{fig:cc22}
 \end{subfigure}
 \hspace{0.06em}
 \begin{subfigure}[b]{0.24\textwidth}
 \centering
 \includegraphics[width=\textwidth, trim=0 0 0 0, clip]{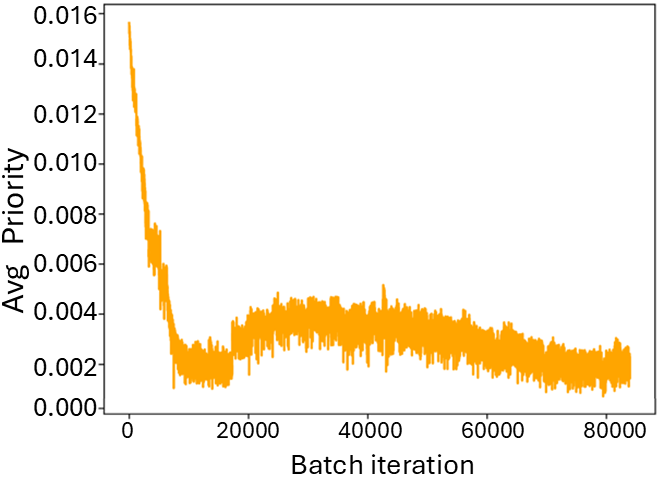}
 \caption{CC100}
 \label{fig:cc100}
 \end{subfigure}
 \hspace{0.06em}
 \begin{subfigure}[b]{0.24\textwidth}
 \centering
 \includegraphics[width=\textwidth, trim=0 0 0 0, clip]{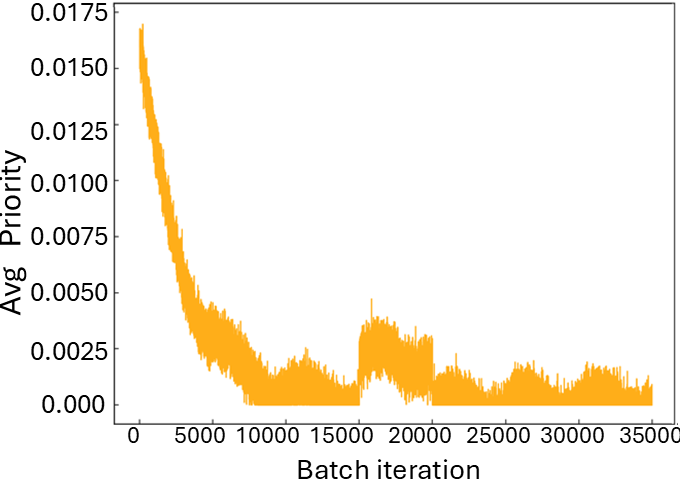}
 \caption{CC500}
 \label{fig:cc500}
 \end{subfigure}

 \caption{{\small{Temporal evolution of average TD-error (PER priority) across environments. Smaller environments (CTF, CC22) stabilise quickly, whereas larger networks (CC100, CC500) exhibit prolonged surges in TD-error, signalling persistent replay of unstable transitions. These spikes correspond to the performance degradation discussed in Section \ref{s5} (“Unexpected PER Behaviour”).}}}
 \label{fig:PL2_priority}
\end{figure*}

\begin{figure*}[t!]
 \centering
 \begin{subfigure}[b]{0.24\textwidth}
 \centering
 \includegraphics[width=\textwidth, trim=0 0 0 0, clip]{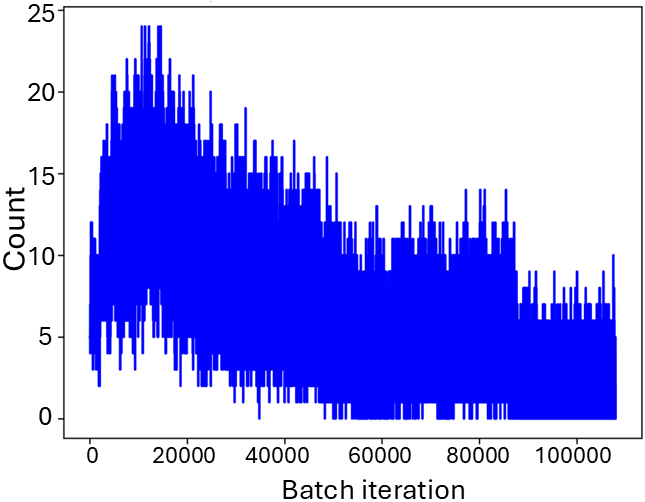}
 \caption{CTF}
 \label{fig:ctf}
 \end{subfigure}
 \hspace{0.06em}
 \begin{subfigure}[b]{0.24\textwidth}
 \centering
 \includegraphics[width=\textwidth, trim=0 0 0 0, clip]{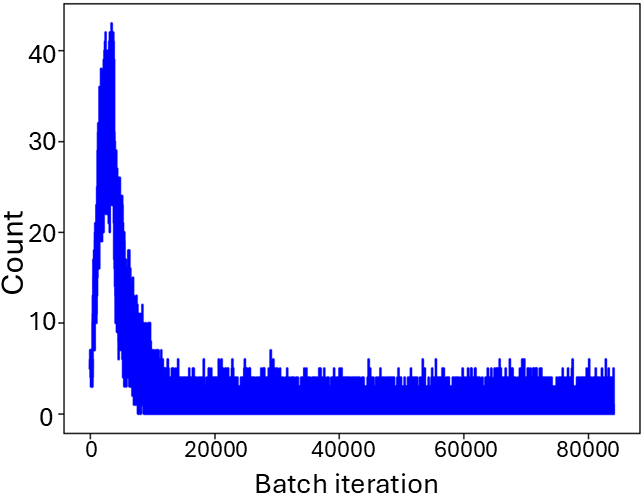}
 \caption{CC22}
 \label{fig:cc22}
 \end{subfigure}
 \hspace{0.06em}
 \begin{subfigure}[b]{0.24\textwidth}
 \centering
 \includegraphics[width=\textwidth, trim=0 0 0 0, clip]{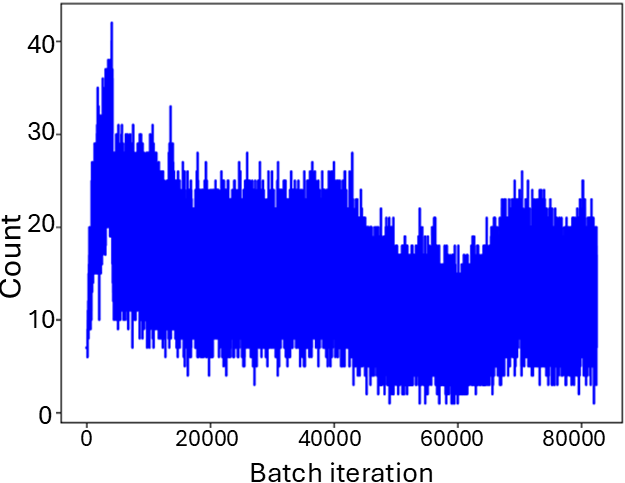}
 \caption{CC100}
 \label{fig:cc100}
 \end{subfigure}
 \hspace{0.06em}
 \begin{subfigure}[b]{0.24\textwidth}
 \centering
 \includegraphics[width=\textwidth, trim=0 0 0 0, clip]{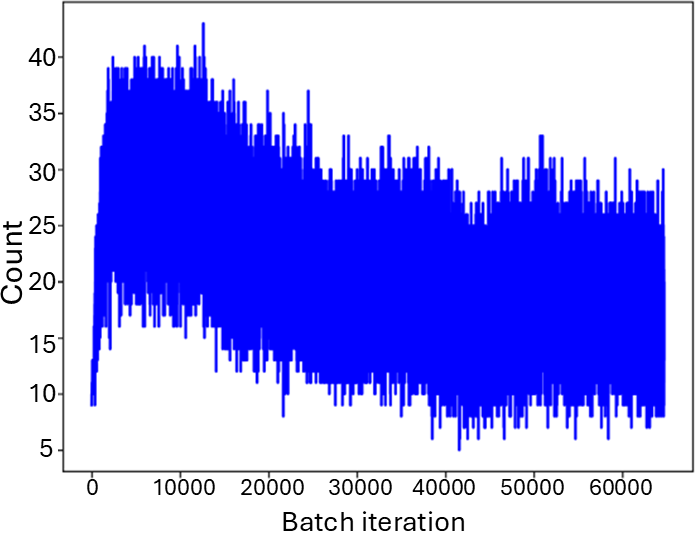}
 \caption{CC500}
 \label{fig:cc500}
 \end{subfigure}

 \caption{\small{Number of Key High-Priority States Across Environments. {In smaller networks (CTF, CC22), the agent quickly converges to a small set of key states, indicating early policy convergence. In contrast, larger networks (CC100, CC500) exhibit broader, persistent distributions, reflecting prolonged exploration and slower convergence due to higher state-space complexity.}}}
 \label{fig:PL2_states}
\end{figure*}

\vspace{0.1in}

\noindent \textit{\textbf{Results.}} We trained the agent for 20 episodes (1500 iterations each) in CTF, 20 episodes (2000 iterations each) in CC22, 100 episodes (2000 iterations each) in CC100, and 200 episodes (4000 iterations each) in CC500. PER was configured with $\alpha = 0.6$, $\beta = 0.7$, and $\epsilon_0 = 0.01$, following standard practice in the PER literature, where $\alpha$ controls the strength of prioritisation, $\beta$ compensates for the bias introduced by non-uniform sampling via importance sampling, and $\epsilon_0$ prevents zero sampling probability \cite{schaul2016prioritized}. We filtered transitions with rewards above 2.0 to isolate high-impact experiences such as privilege escalation. Figure \ref{fig:PL2_priority} shows {the evolution of }average transition priority over time. All environments exhibit early TD-error spikes from policy instability and exploration. Smaller networks (CTF, CC22) stabilise quickly, while larger ones (CC100, CC500) show sustained fluctuation, indicating slower convergence and repeated reprioritisation of experience samples. In CC500, a sequence of mid-training spikes {between approximately 9K and 18K batch iterations reflects intermittent bursts of partial policy reorganisation. These surges likely correspond to successive discoveries of new pivoting or credential paths, followed by unstable replay of high-TD transitions. The irregular amplitude and recurrence of these spikes suggest that, beyond initial learning gains, PER began over-emphasising transient experiences, amplifying noise and delaying convergence.} 
Figure \ref{fig:PL2_states} tracks the number of distinct high-priority states across training. 
CTF and CC22 converge to compact sets of key states, implying repeated exploitation of an optimal state set. In contrast, larger networks (CC100, CC500) maintain broader, persistent distributions, reflecting ongoing refinement and diverse learning trajectories in larger state spaces.\\

\begin{tcolorbox} 
\textbf{Interpretation and Utility.} PER visualisations reveal \emph{when} and \emph{where} attacker agents learn most rapidly during training. 
{In small networks, contraction of the high-priority state set signifies convergence toward a stable attack strategy.
In larger environments, intermittent TD-error surges mark critical phases of strategic exploration, information that developers can use to assess convergence timing, reward stability, and exploration schedules.} 
For defenders, while PER signals are not observable in deployed agents, { analysing these training-time patterns helps identify the states, assets, or pathways most likely to be targeted or re-used by adaptive agents, insights that can inform the design and placement of decoys, monitoring hooks, or segmentation policies in simulated or operational settings.}
\end{tcolorbox}

\vspace{0.5in}

\begin{figure*}[t!]
\centering
\begin{subfigure}[b]{0.49\textwidth}
\centering
\includegraphics[width=\textwidth, trim=0 0 0 0, clip]{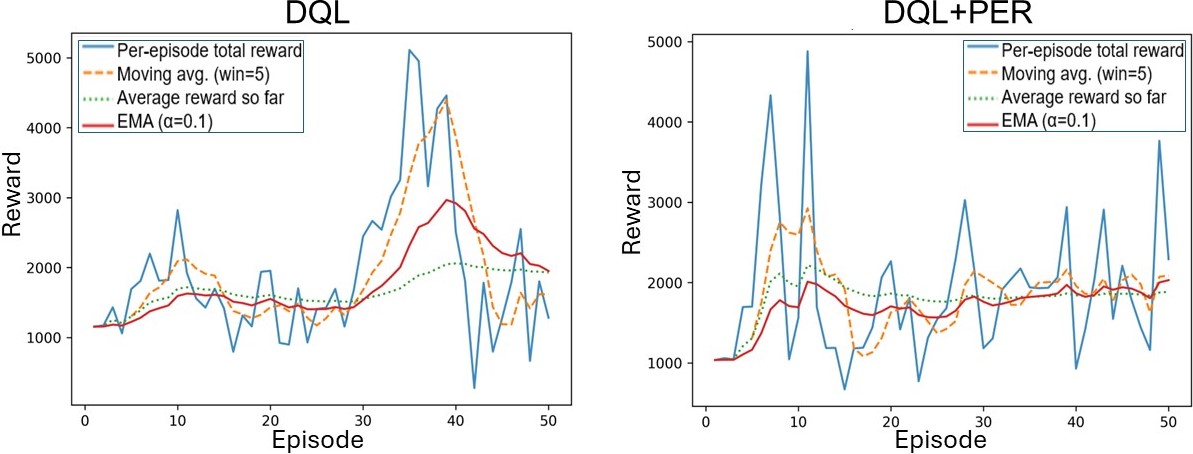}
\caption{CC100}
\label{fig:CC100PER}
\end{subfigure}
\hspace{0.06em}
 \begin{subfigure}[b]{0.49\textwidth}
 \centering
 \includegraphics[width=\textwidth, trim=0 0 0 0, clip]{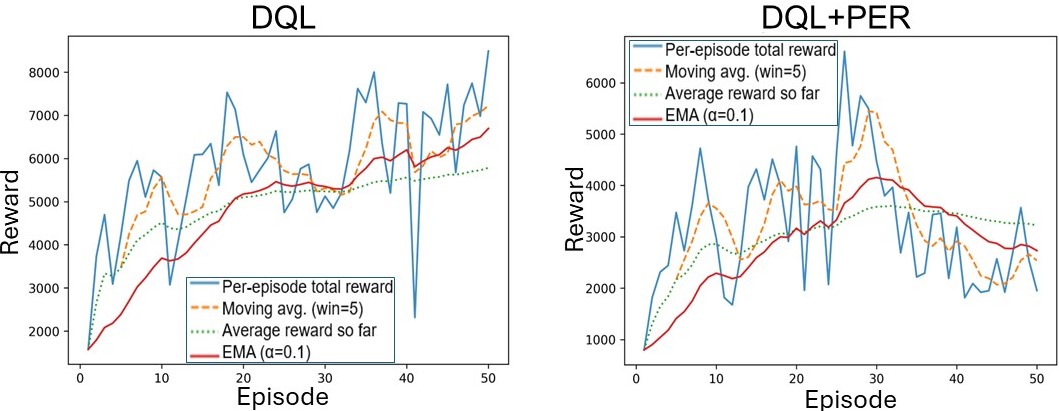}
 \caption{CC500}
 \label{fig:CC100PER}
 \end{subfigure}
\caption{\small{Reward progression of DQL and DQL+PER in CC100 and CC500 environments. DQL achieves higher cumulative rewards and steadier learning, while DQL+PER underperforms due to its bias toward high-TD-error transitions, which oversamples transient experiences and slows convergence.}
}
 \label{fig:DQLVSPER}
\end{figure*}

\noindent \textbf{Explainability as a Diagnostic Lens: Revealing PER Failure Modes.}
PER is widely adopted to improve sample efficiency by amplifying transitions with high TD error. While often assumed to accelerate convergence, its behaviour in large, partially observable cyber environments remains poorly understood. Using our training-time explainability framework, we identify a diagnostic failure mode of PER that is not observable from aggregate reward trajectories alone. As shown in Figure~\ref{fig:DQLVSPER}, DQL augmented with PER (DQL+PER) exhibits less stable learning and inferior final rewards than standard DQL in both CC100 and CC500 environments.

\begin{figure}[t!]
\centering
\includegraphics[width=\columnwidth, trim=0 0 0 0, clip]{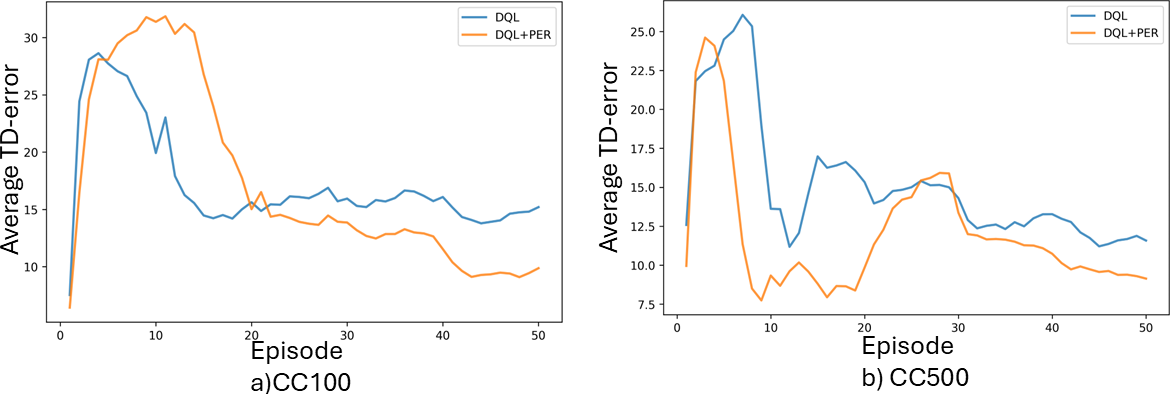}
\caption{\small{Average TD-error of DQL and DQL+PER in CC100 and CC500. DQL learns more stably, whereas PER introduces early instability and biased convergence, leading to inferior rewards.}}
\label{fig:errorVSepisode}
\end{figure}

Reward curves show the symptom; internal signals expose the mechanism. Explainability signals reveal a replay-driven distortion in the learning process that fundamentally alters value propagation at scale. Figure~\ref{fig:errorVSepisode} exposes the internal dynamics responsible for this behaviour. In CC100, DQL+PER sustains elevated TD-error for a prolonged period relative to standard DQL, indicating repeated replay of high-error but weakly informative transitions that slow value stabilisation and delay convergence. In CC500, while both agents exhibit large TD-errors during early exploration, standard DQL subsequently stabilises, whereas DQL+PER shows slower decay and persistent TD-error fluctuations across later episodes. This pattern reflects misaligned replay priorities and inconsistent value propagation as the state space expands.

Inspection of PER-priority traces and temporal Q-value dynamics further indicates that many high-TD-error transitions sampled in later training stages correspond to unstable, short-lived behaviours rather than enduring strategic improvements. Continued oversampling of such transitions after their relevance has decayed injects noise into the replay buffer, disrupts value estimation, and impedes policy stabilisation. This constitutes a negative result enabled by training-time explainability: under reward-only evaluation, PER appears to underperform without clear explanation, whereas internal diagnostics reveal that the degradation arises from replay-driven distortions rather than insufficient exploration, inadequate training duration, or poor hyperparameter choices.

These results highlight a fundamental limitation of naïvely applying PER at scale. Prioritisation schemes must be co-designed with environment complexity and learning dynamics, as their effectiveness is inherently context-dependent rather than universally guaranteed. More broadly, this analysis demonstrates how training-time explainability enables the discovery of optimisation pathologies that remain opaque under performance-centric evaluation, reinforcing the necessity of internal diagnostic signals when developing and validating autonomous cyber agents.

\begin{figure*}[t!]
 \centering
 \begin{subfigure}[b]{0.24\textwidth} 
 \centering
 \includegraphics[width=\textwidth, trim=0 0 0 0, clip]{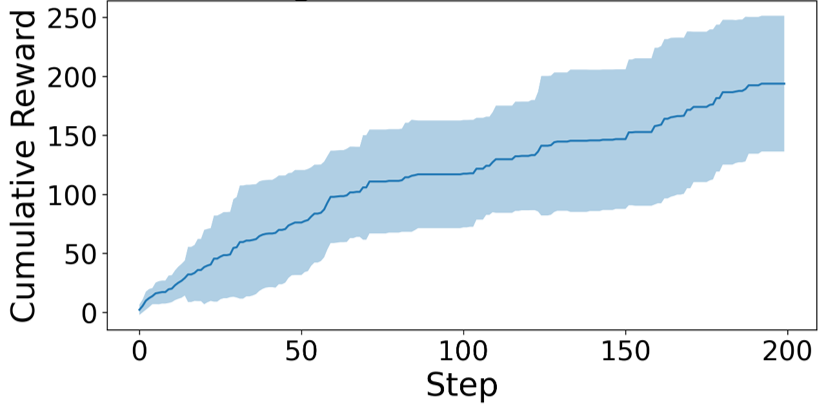}
 \captionsetup{font=scriptsize}
 \caption{CTF: Early Phase}
 \label{fig:ctf}
 \end{subfigure}
 \hfill
 \begin{subfigure}[b]{0.24\textwidth}
 \centering
 \includegraphics[width=\textwidth, trim=0 0 0 0, clip]{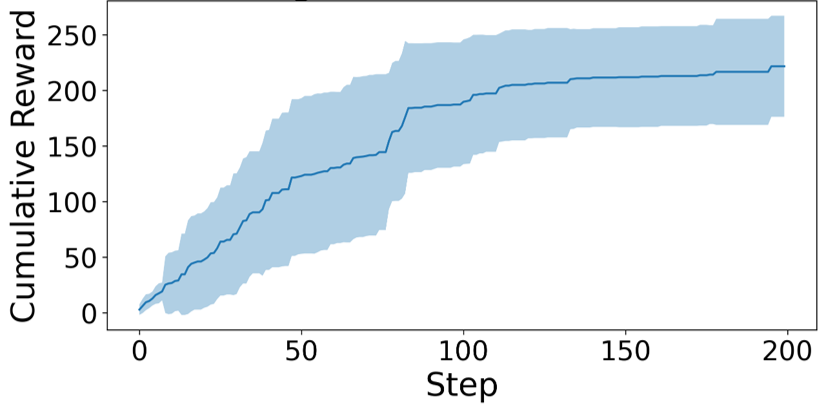}
 \captionsetup{font=scriptsize}
 \caption{CTF: Late Phase}
 \label{fig:cc22}
 \end{subfigure}
 \hfill
 \begin{subfigure}[b]{0.24\textwidth}
 \centering
 \includegraphics[width=\textwidth, trim=0 0 0 0, clip]{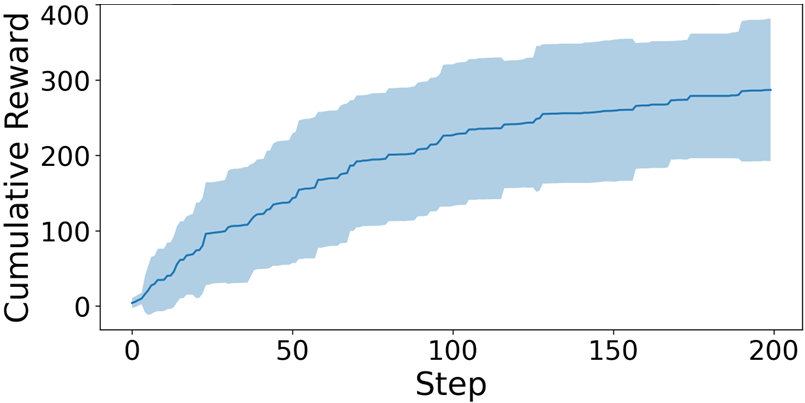}
 \captionsetup{font=scriptsize}
 \caption{CC22: Early Phase}
 \label{fig:cc100}
 \end{subfigure}
 \hfill
 \begin{subfigure}[b]{0.24\textwidth}
 \centering
 \includegraphics[width=\textwidth, trim=0 0 0 0, clip]{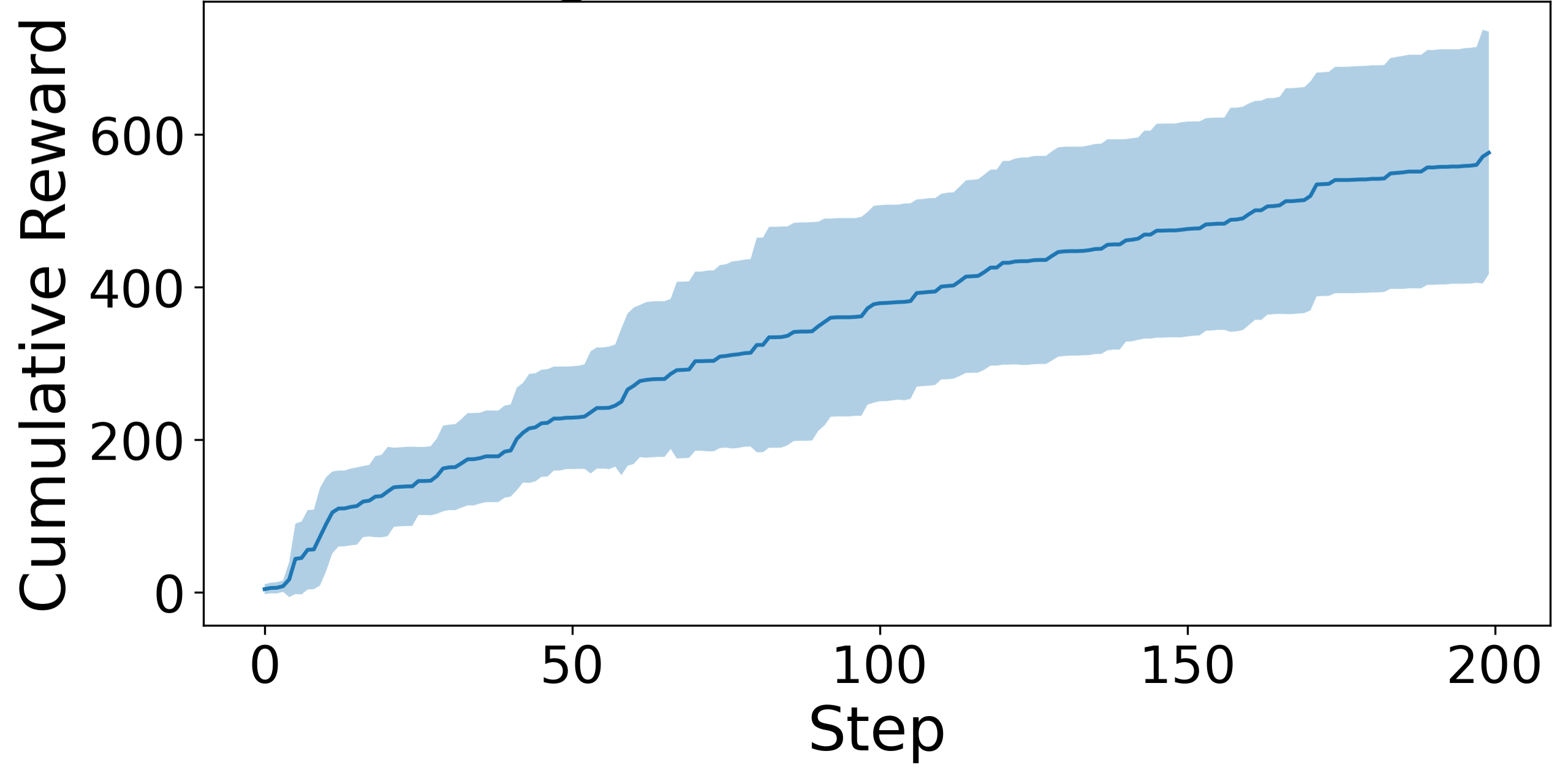}
 \captionsetup{font=scriptsize}
 \caption{CC22: Late Phase}
 \label{fig:cc500}
 \end{subfigure}
 \begin{subfigure}[b]{0.24\textwidth} 
 \centering
 \includegraphics[width=\textwidth, trim=0 0 0 0, clip]{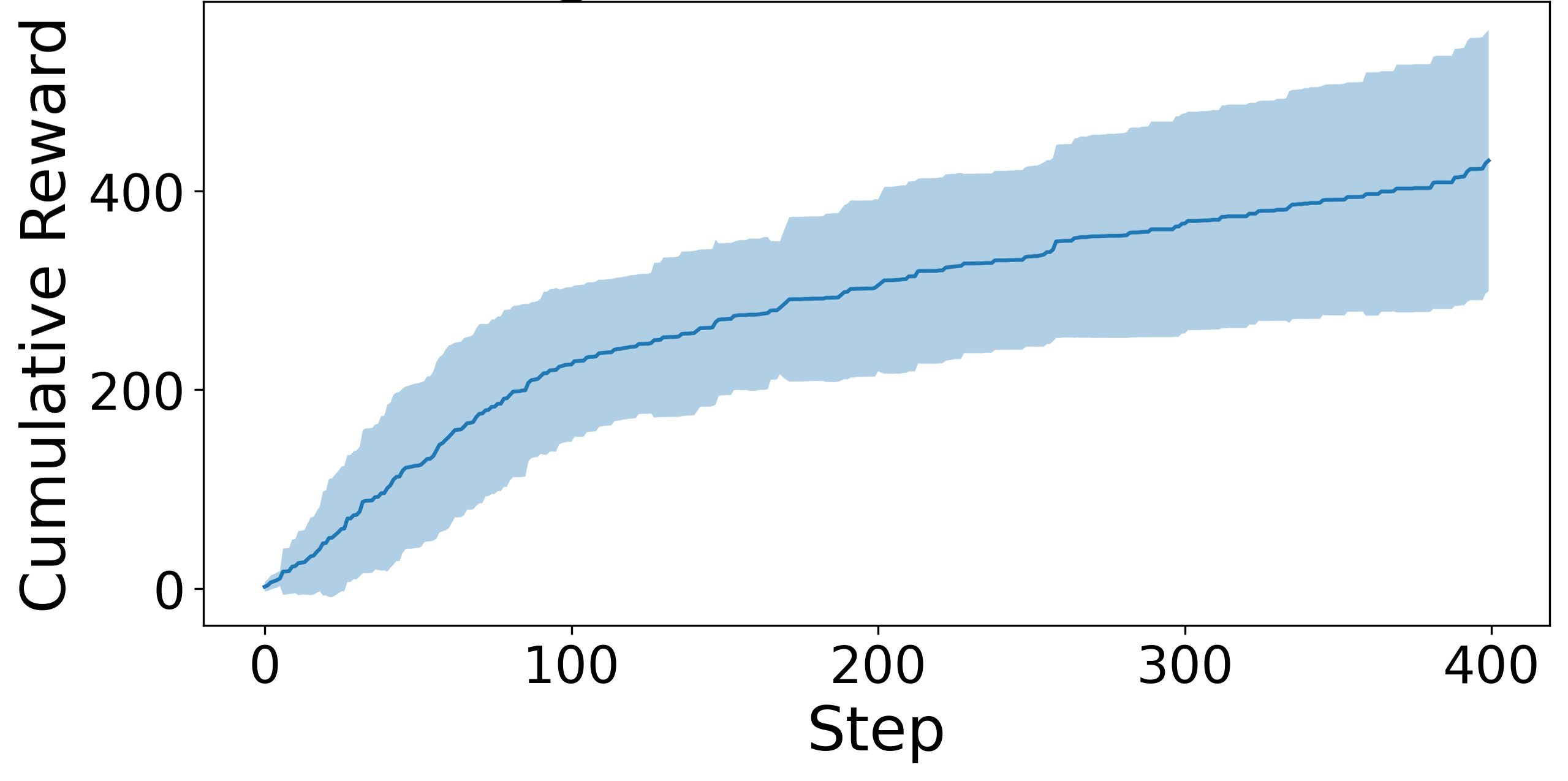}
 \captionsetup{font=scriptsize}
 \caption{CC100: Early Phase}
 \label{fig:ctf}
 \end{subfigure}
 \hfill
 \begin{subfigure}[b]{0.24\textwidth}
 \centering
 \includegraphics[width=\textwidth, trim=0 0 0 0, clip]{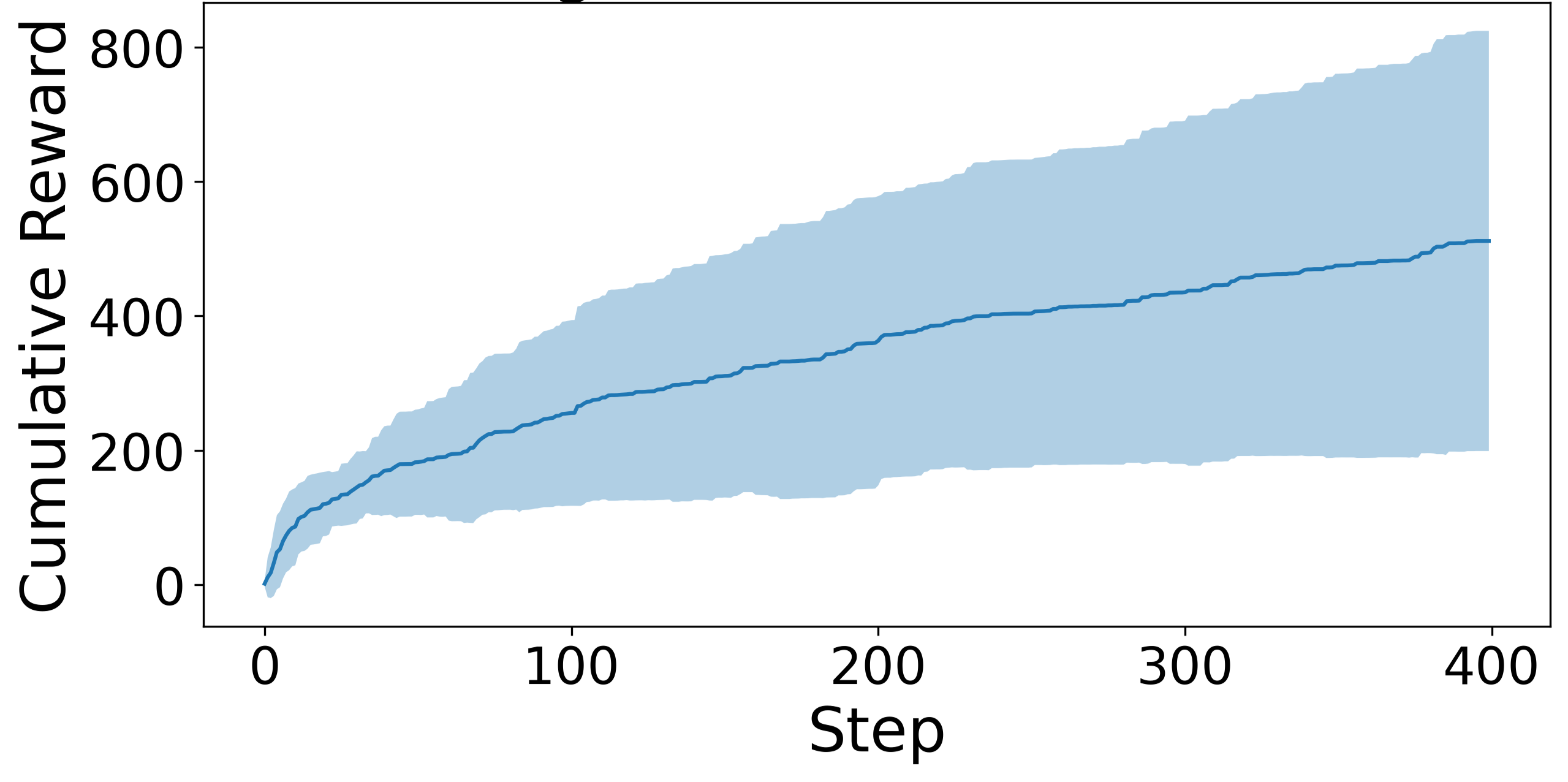}
 \captionsetup{font=scriptsize}
 \caption{CC100: Late Phase}
 \label{fig:cc22}
 \end{subfigure}
 \hfill
 \begin{subfigure}[b]{0.24\textwidth}
 \centering
 \includegraphics[width=\textwidth, trim=0 0 0 0, clip]{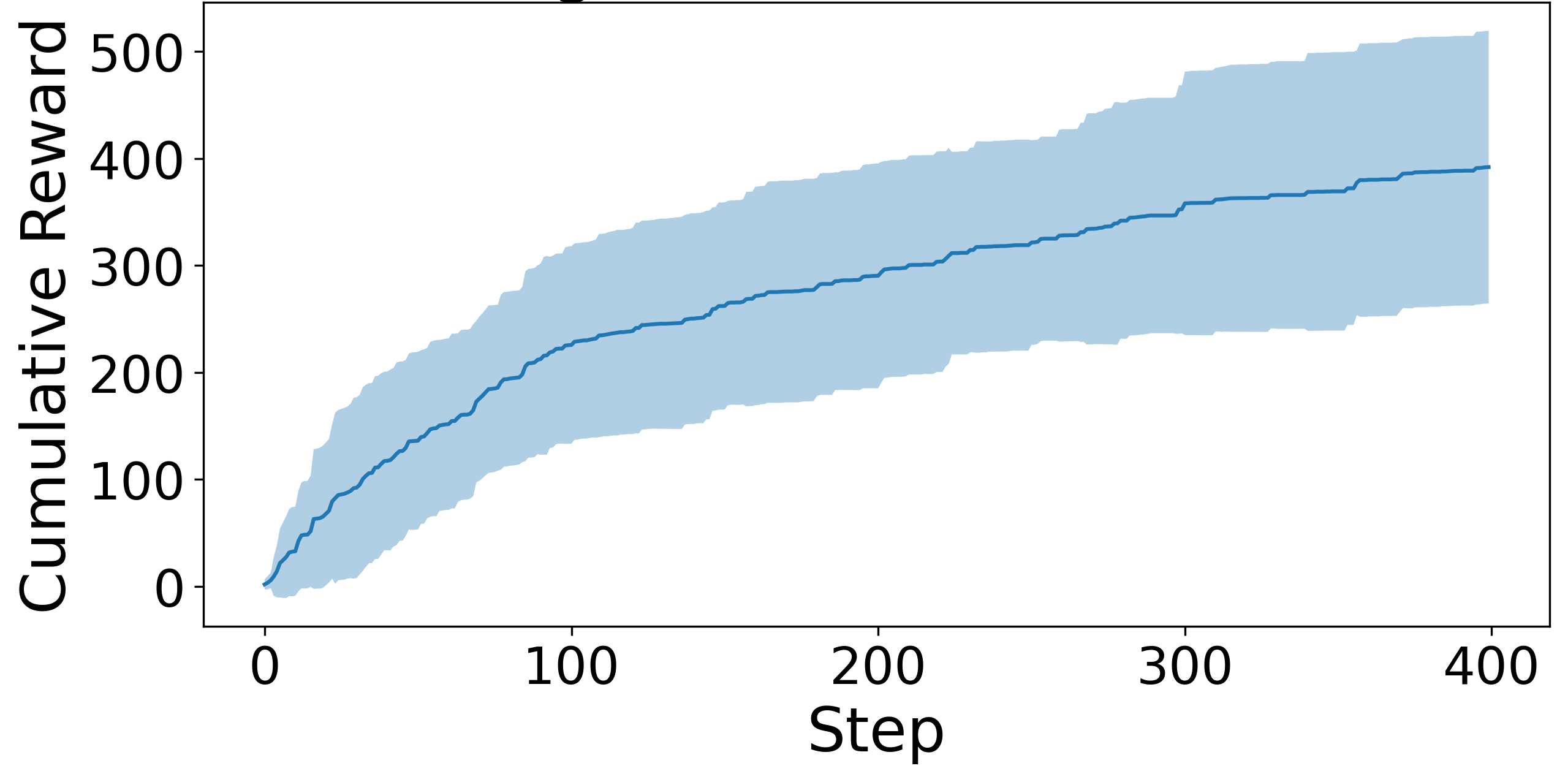}
 \captionsetup{font=scriptsize}
 \caption{CC500: Early Phase}
 \label{fig:cc100}
 \end{subfigure}
 \hfill
 \begin{subfigure}[b]{0.24\textwidth}
 \centering
 \includegraphics[width=\textwidth, trim=0 0 0 0, clip]{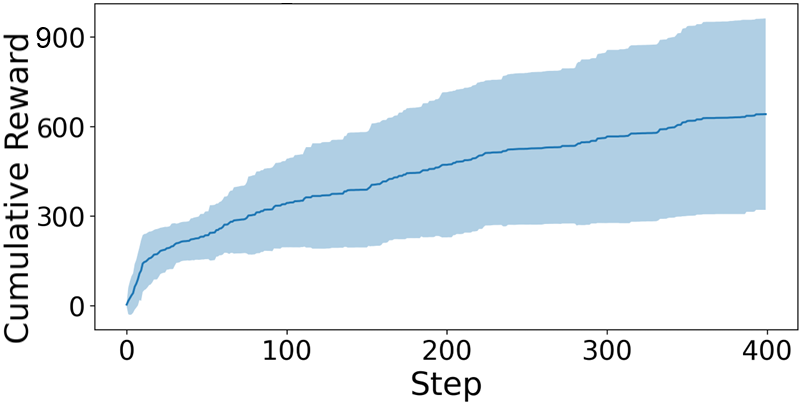}
 \captionsetup{font=scriptsize}
 \caption{CC500: Late Phase}
 \label{fig:cc500}
 \end{subfigure}
\caption{\small{Cumulative rewards comparison for the PPO algorithm in early vs. late attack phases across environments. Results show that PPO agent achieves higher rewards in the late phase, indicating a progression from exploratory behaviour to exploitation-focused strategies as training advances.}}

\label{fig:PPO_early_late}
\end{figure*}

\begin{figure*}[t!]
\centering

\begin{subfigure}[b]{0.48\textwidth}
\centering
\includegraphics[width=\textwidth]{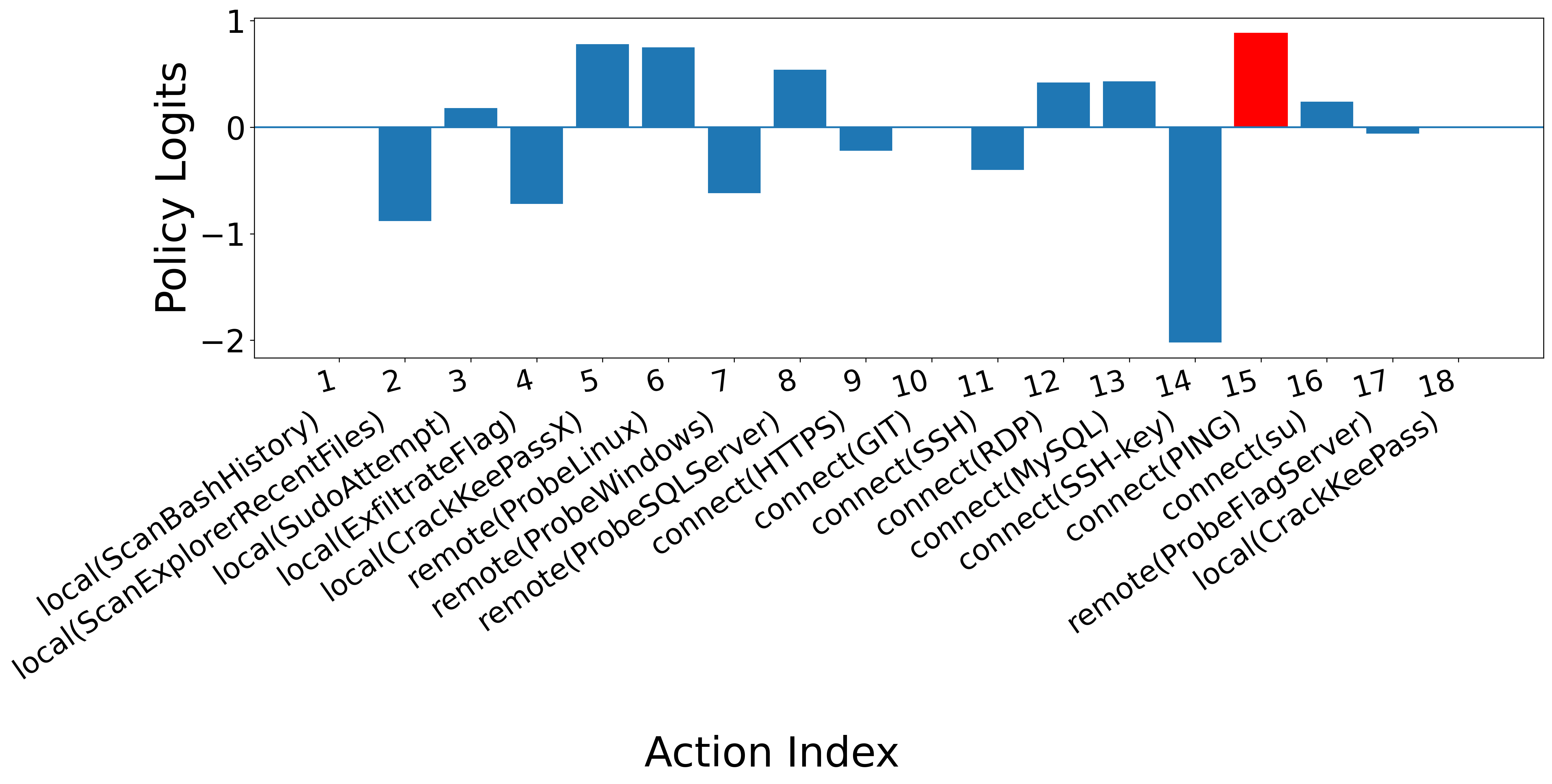}
\captionsetup{font=scriptsize}
\caption{CTF: Episode 10}
\label{fig2:ctf1}
\end{subfigure}
\hfill
\begin{subfigure}[b]{0.48\textwidth}
\centering
\includegraphics[width=\textwidth]{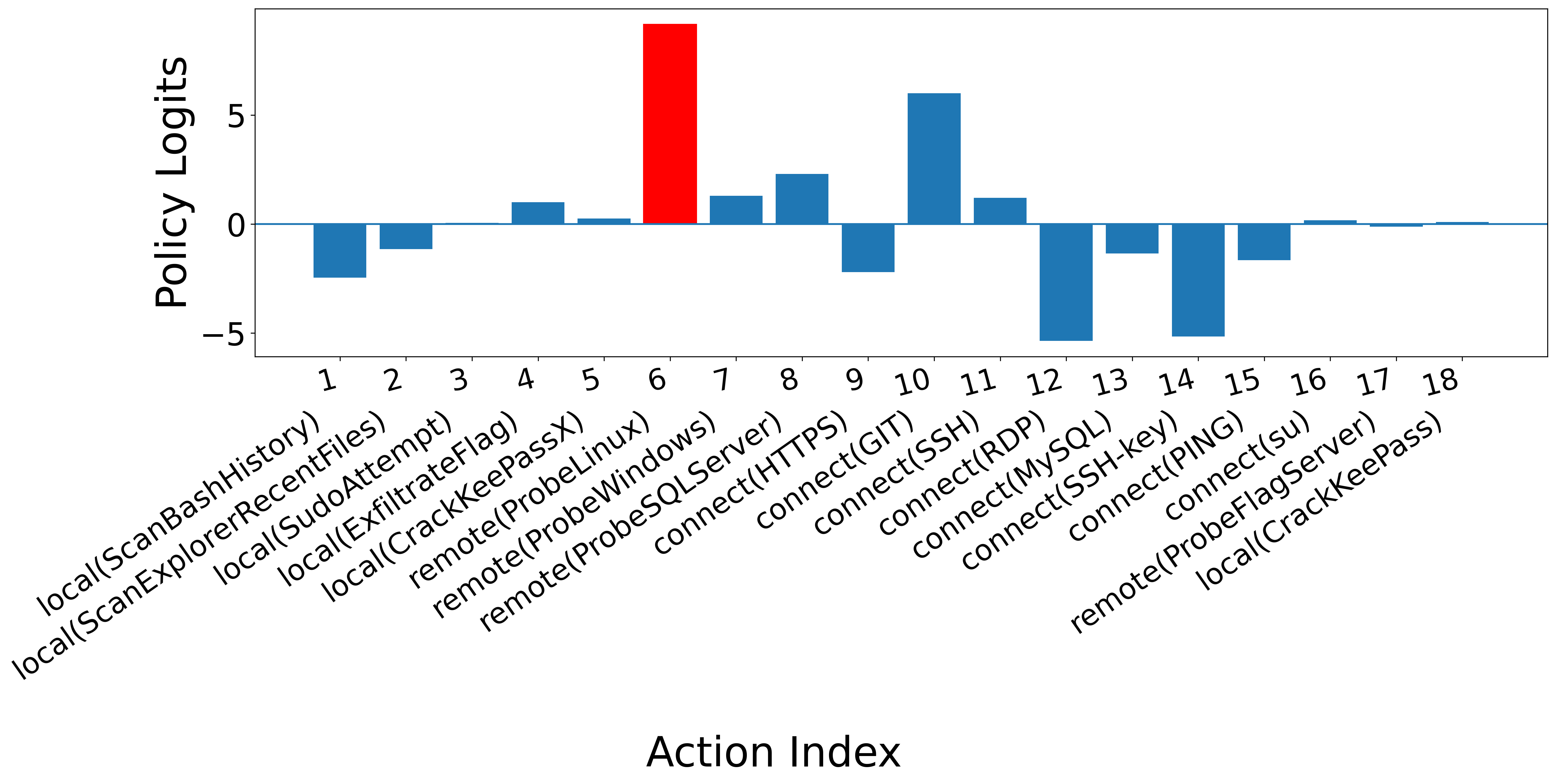}
\captionsetup{font=scriptsize}
\caption{CTF: Episode 25}
\label{fig2:ctf2}
\end{subfigure}

\vspace{2mm}

\begin{subfigure}[b]{0.48\textwidth}
\centering
\includegraphics[width=\textwidth]{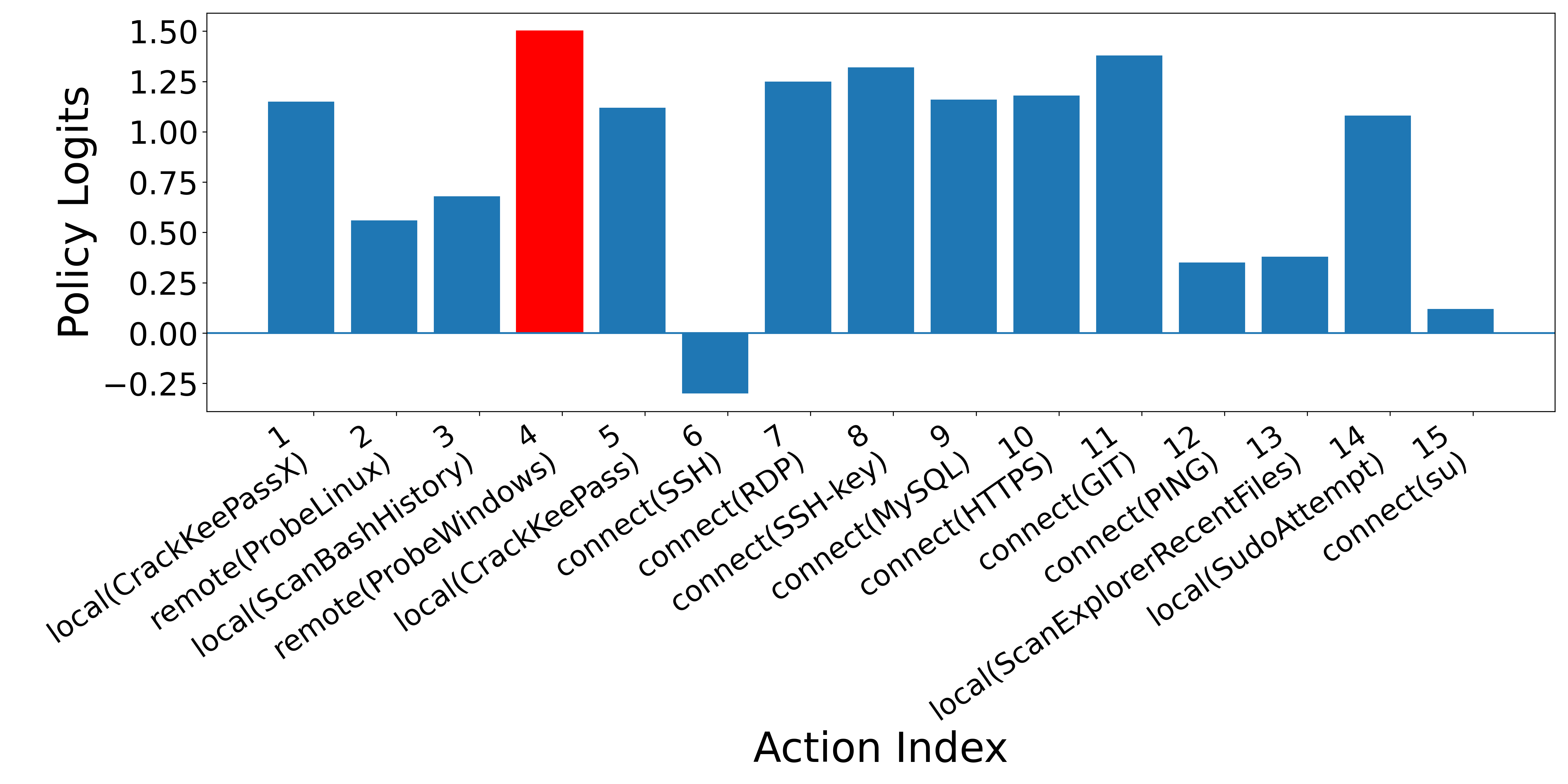}
\captionsetup{font=scriptsize}
\caption{CC22: Episode 20}
\label{fig2:cc22_1}
\end{subfigure}
\hfill
\begin{subfigure}[b]{0.48\textwidth}
\centering
\includegraphics[width=\textwidth]{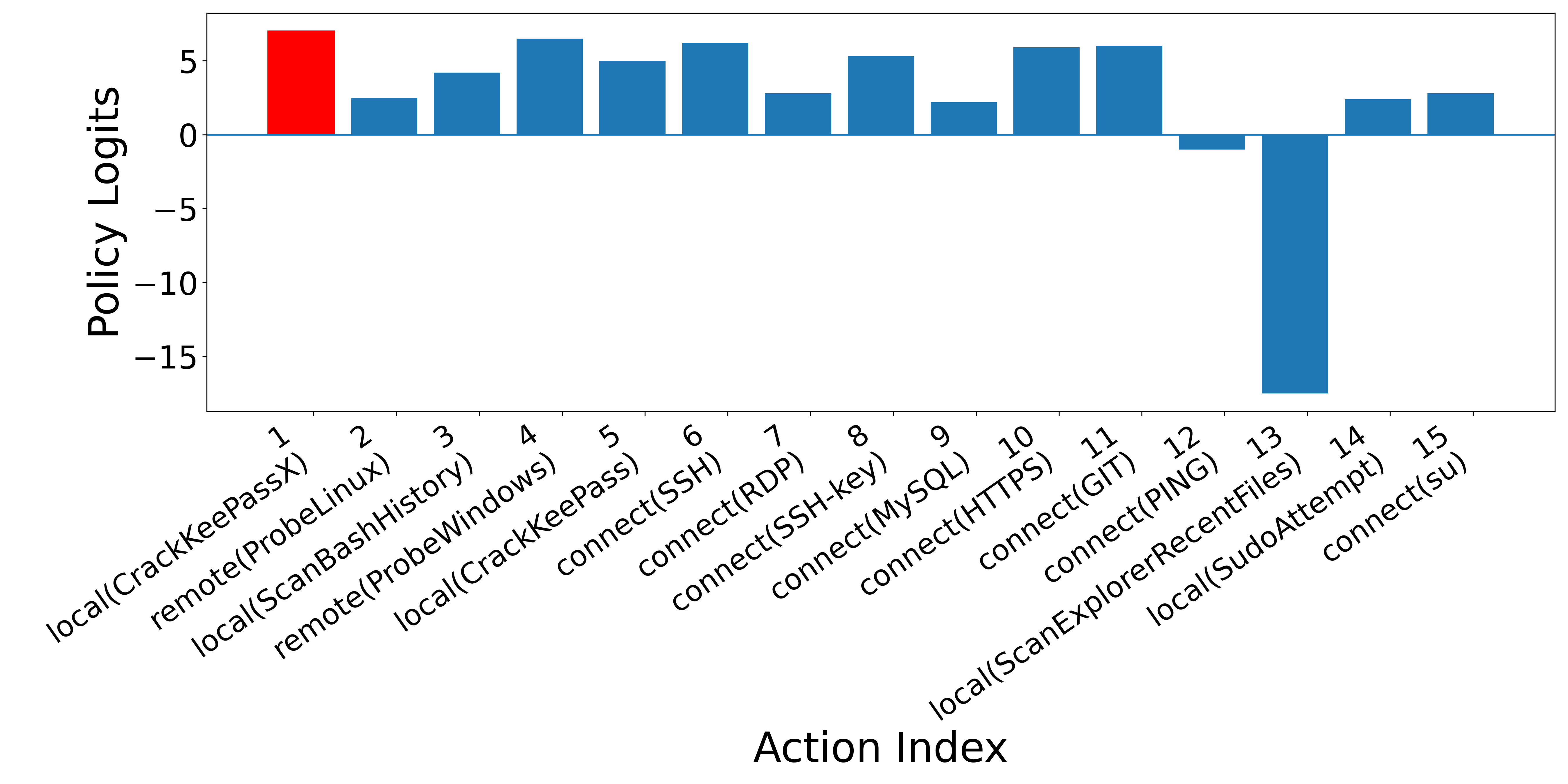}
\captionsetup{font=scriptsize}
\caption{CC22: Episode 35}
\label{fig2:cc22_2}
\end{subfigure}

\vspace{2mm}

\begin{subfigure}[b]{0.48\textwidth}
\centering
\includegraphics[width=\textwidth]{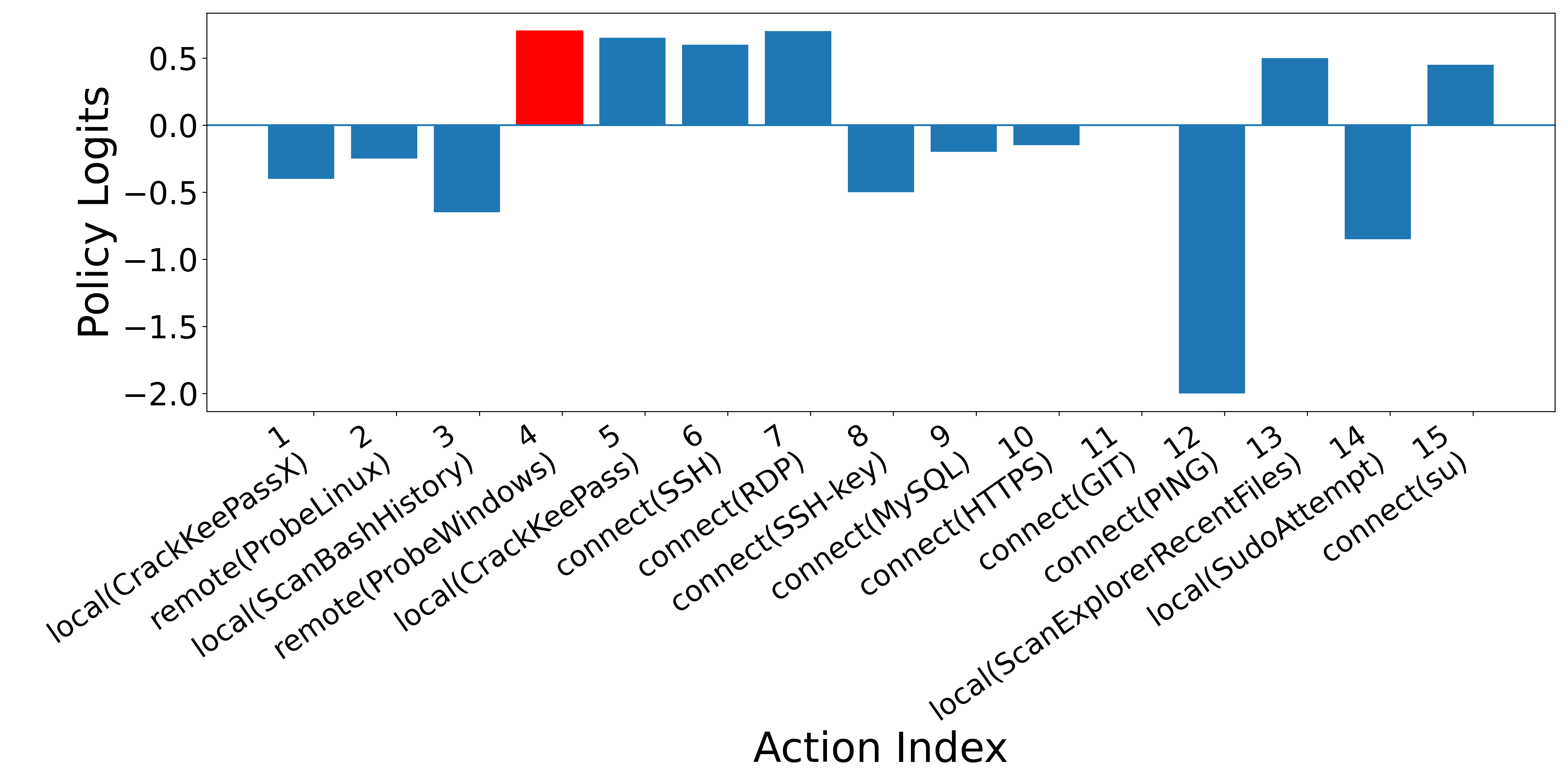}
\captionsetup{font=scriptsize}
\caption{CC100: Episode 10}
\label{fig2:cc100_1}
\end{subfigure}
\hfill
\begin{subfigure}[b]{0.48\textwidth}
\centering
\includegraphics[width=\textwidth]{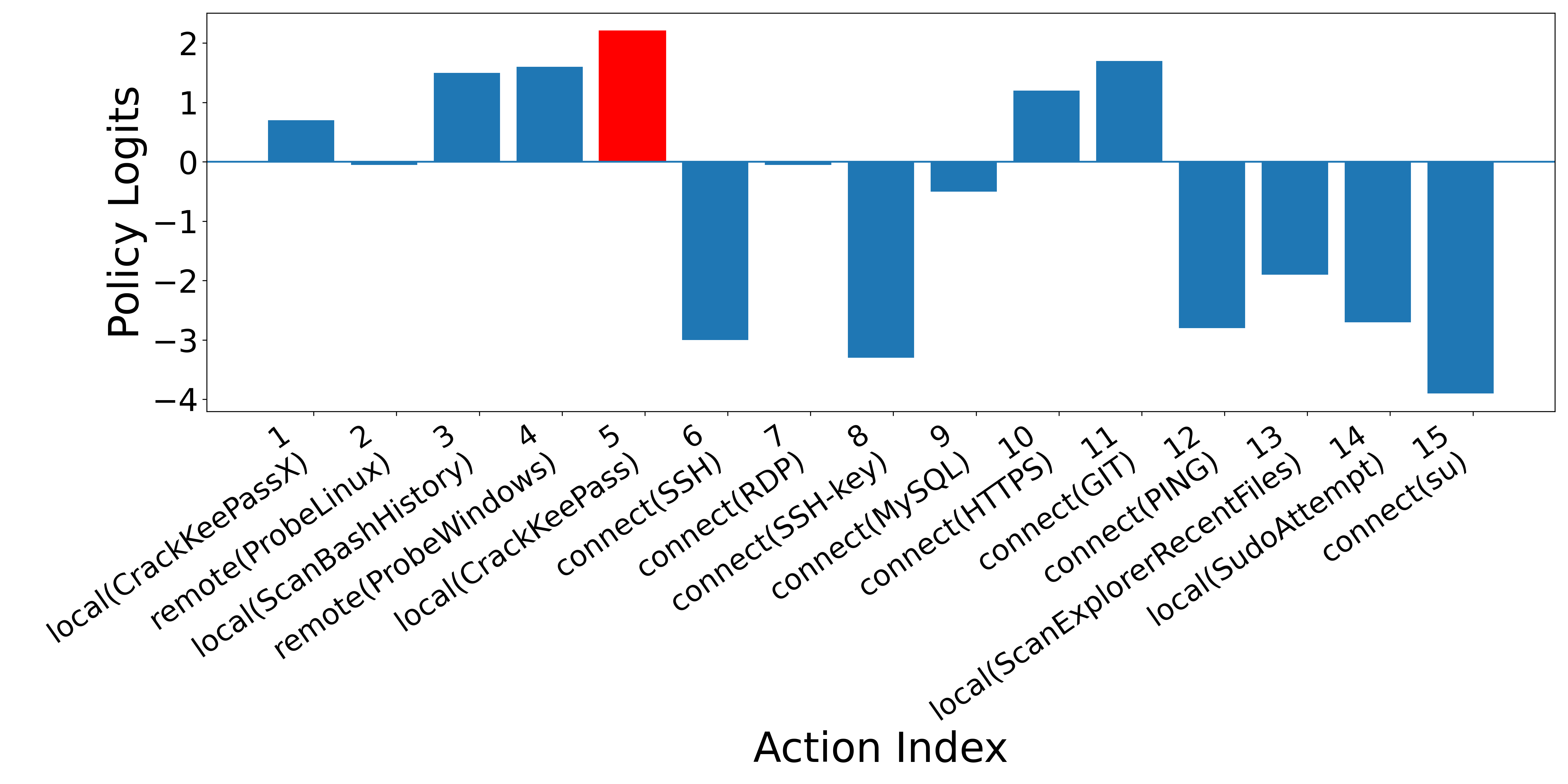}
\captionsetup{font=scriptsize}
\caption{CC100: Episode 35}
\label{fig2:cc100_2}
\end{subfigure}

\vspace{2mm}

\begin{subfigure}[b]{0.48\textwidth}
\centering
\includegraphics[width=\textwidth]{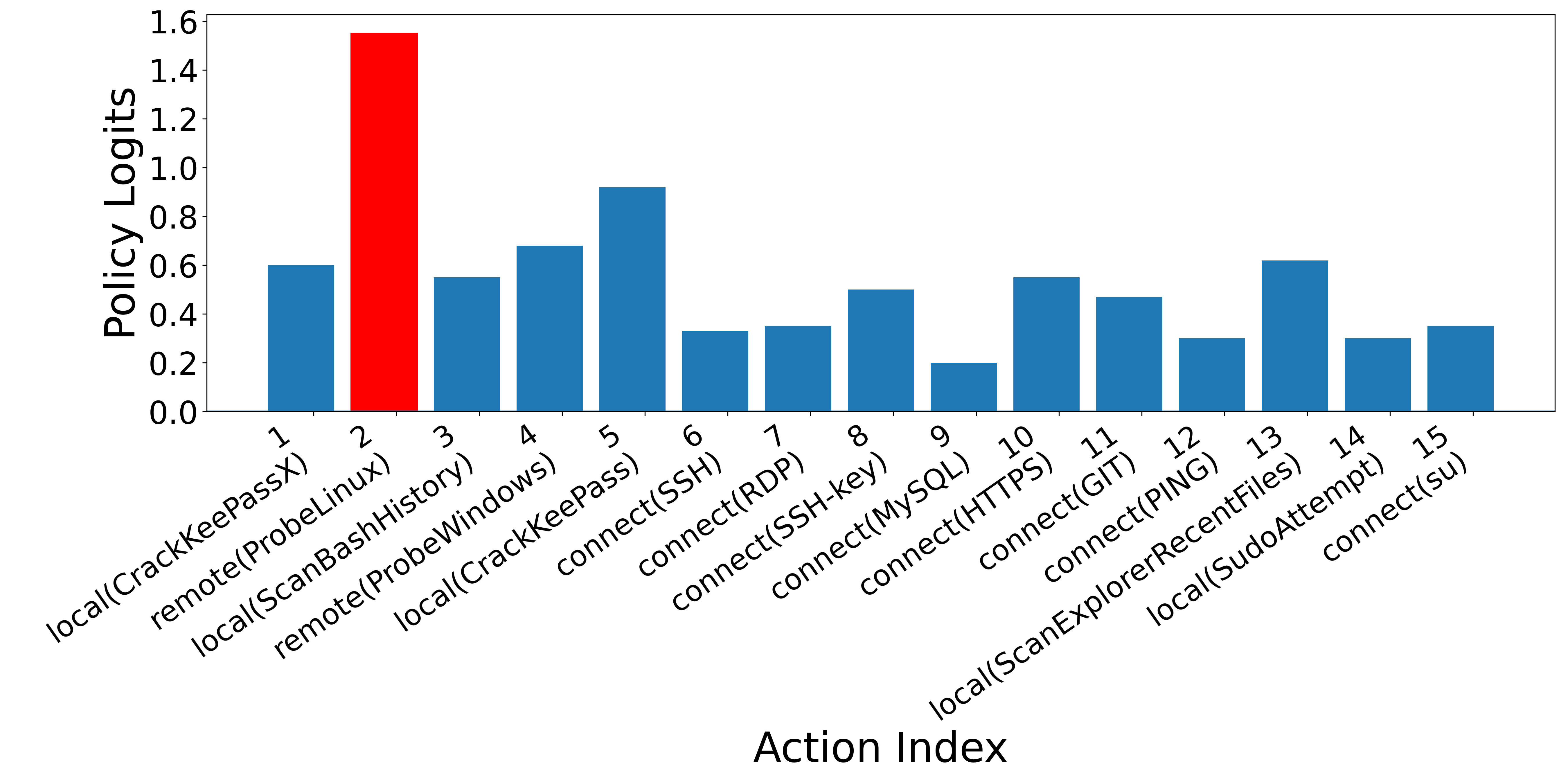}
\captionsetup{font=scriptsize}
\caption{CC500: Episode 5}
\label{fig2:cc500_1}
\end{subfigure}
\hfill
\begin{subfigure}[b]{0.48\textwidth}
\centering
\includegraphics[width=\textwidth]{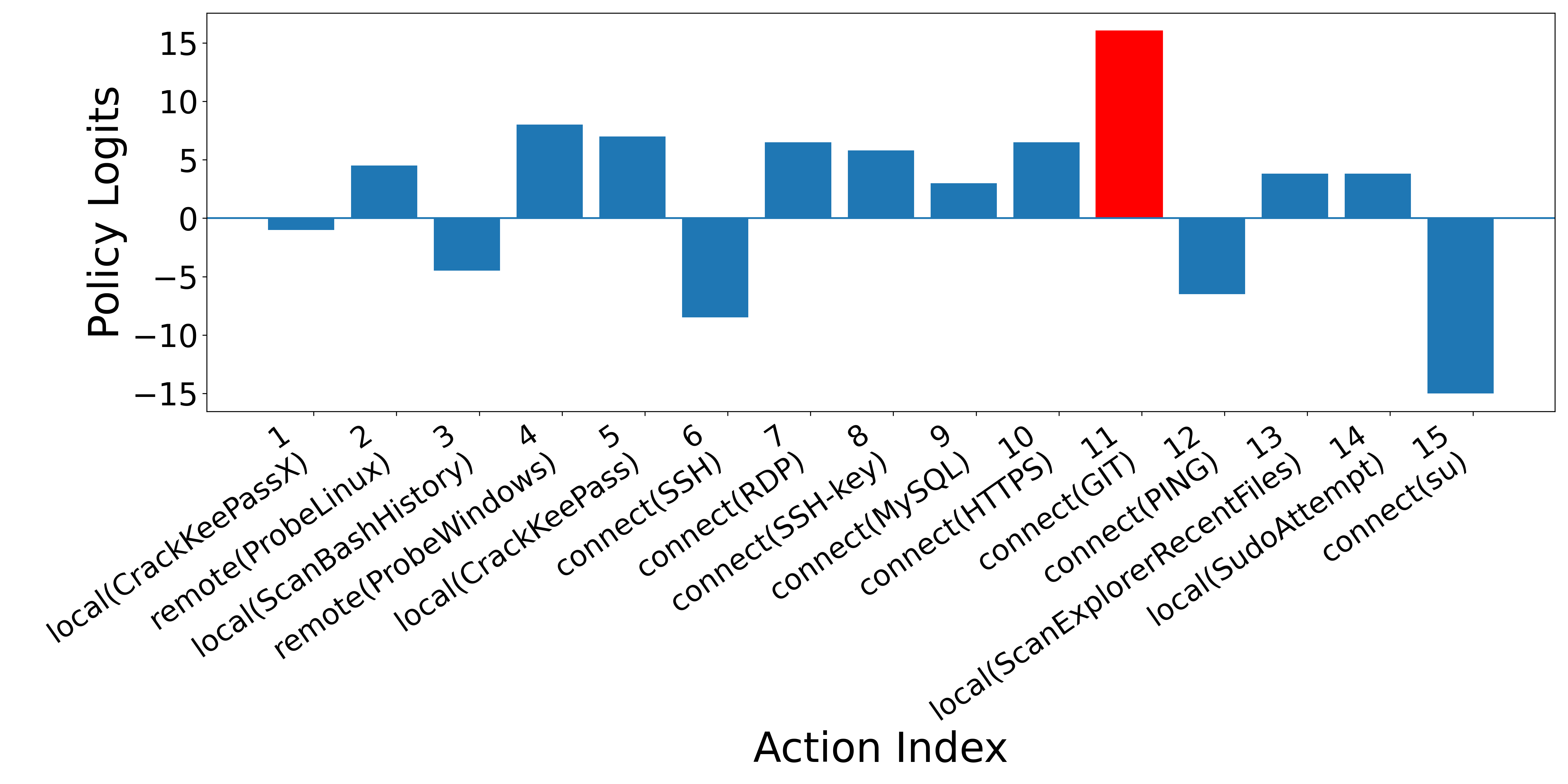}
\captionsetup{font=scriptsize}
\caption{CC500: Episode 40}
\label{fig2:cc500_2}
\end{subfigure}
\caption{\small{{Emergence of action preferences through state-aggregated policy logits across episodes. The highlighted dominant action marks the shift from exploratory to high-value, environment-specific tactics. Compared with DQL, PPO’s clipped updates yield smoother yet consistent preference consolidation. Action indices for CyberBattleChain (1–15) and CTF (1–18) correspond to the x-axis and are listed in Table \ref{tab:action_ladder} and Table \ref{tab:abstract-actions}, respectively.}}}
 \label{fig:PPO_ACTIONS}
\end{figure*}

\subsection{Cross-Algorithm Generalisability of the Explainability Framework}\label{s7}
While Sections \ref{s1} to \ref{s5} establish detailed interpretability results under Deep Q-Learning, a fundamental question remains: do these findings reflect intrinsic properties of adversarial learning, or are they artefacts of a particular optimisation strategy? Our framework is explicitly \emph{model-agnostic}, designed to extract explainability signals from any RL agent that exposes either value estimates or policy outputs. To substantiate this claim, we replicate two representative experimental setups, \emph{Phase-aware behavioural evolution} (Setup 3) and \emph{Temporal value dynamics} (Setup 4), using the policy-gradient \textbf{Proximal Policy Optimisation (PPO)} algorithm with matched network architectures. This cross-paradigm evaluation probes whether the strategic and tactical explainability signatures identified earlier, such as phase transitions, action consolidation, and learning stability, remain invariant under fundamentally different optimisation mechanisms, thereby establishing the framework’s generality across learning paradigms rather than dependence on any single algorithm.

\subsubsection{Strategic-Level Generalisability: Phase Transitions and Behaviour Evolution}

Figures \ref{fig:MDPlate} (DQL) and \ref{fig:PPO_early_late} (PPO) show highly consistent phase-aware reward trajectories across both learning families and all evaluated environments. Despite their fundamentally different optimisation mechanisms, bootstrapped value updates in DQL versus clipped policy-gradient optimisation in PPO, both agents exhibit the same macro-level behavioural progression: an exploratory reconnaissance phase with sparse rewards, followed by a decisive exploitation phase marked by rapid reward growth and stable convergence. This pattern mirrors the phase structure identified for DQL in Section \ref{s3} and persists under PPO, demonstrating that the exploration–exploitation transition is not an artefact of value-based learning but a stable structural property of adversarial training dynamics. The figures further show algorithm-specific differences superimposed on this shared structure: DQL displays sharper reward inflection points as critical vulnerabilities are discovered, whereas PPO produces smoother, more gradual transitions with narrower confidence bands, consistent with its variance-controlled updates and entropy regularisation.

The results show that the timing and qualitative form of the phase transition remain consistent across both algorithms and network scales, indicating that the strategic dynamics revealed by our framework arise from the underlying structure of adversarial learning itself, rather than from the particulars of any individual optimisation algorithm.

\subsubsection{Tactical-Level Generalisability: Temporal Value Evolution and Action Preferences}

At the tactical level, Figures \ref{fig:PL1_actions} (DQL) and \ref{fig:PPO_ACTIONS} (PPO) reveal a consistent pattern in the temporal emergence of dominant actions, a principal indicator of tactical consolidation. Across both value-based and policy-gradient agents, early training is characterised by diffuse exploratory preferences, followed by progressive concentration onto a small subset of high-impact, environment-specific actions. In CTF environment, both algorithms initially prioritise reconnaissance-oriented actions (e.g., \texttt{connect(PING)}, \texttt{connect(MySQL)}), before converging on \texttt{remote(ProbeLinux)} as the dominant exploit, marking the transition from exploration to exploitation. In the more complex CC500 environment, both DQL and PPO ultimately favour critical pivoting actions such as \texttt{remote(ProbeWindows)} and \texttt{connect(GIT)}, despite substantial differences in their underlying optimisation dynamics. While DQL’s Q-values exhibit sharper episodic fluctuations and discrete shifts in dominant actions, PPO’s policy logits evolve more smoothly with broader preference distributions, reflecting their contrasting update mechanisms: DQL’s bootstrapped value updates induce higher variance, whereas PPO’s clipped-gradient formulation promotes gradual policy refinement. Occasionally, PPO exhibited locally over-confident policy behaviour, where one action’s preference sharply dominated others late in training. Although this reflects expected policy-gradient saturation under sparse-reward conditions, it also demonstrates the framework’s ability to surface such optimisation dynamics for inspection. These parallel dominance trajectories reveal that the same subset of actions preserves relative importance across algorithms, even as value scales and convergence rates differ. Collectively, this demonstrates that the \textit{hierarchy of action importance} remains structurally aligned between DQL and PPO, confirming that the framework captures algorithm-independent tactical regularities in adversarial RL.

\begin{figure*}[t!]
\centering
\begin{subfigure}[b]{0.49\textwidth}
\centering
\includegraphics[width=\textwidth]{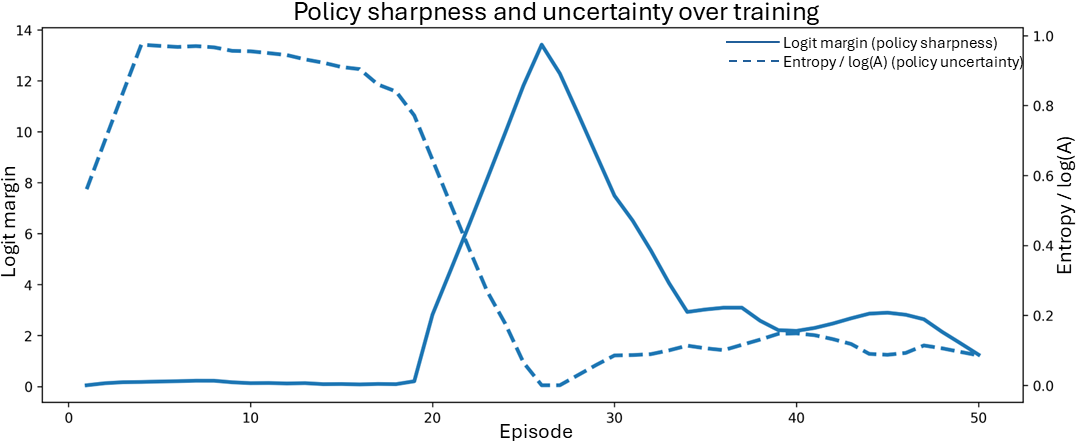}
\caption{\small Evolution of policy confidence and uncertainty.}
\label{fig:policysharpness}
\end{subfigure}
\hspace{0.06em}
\begin{subfigure}[b]{0.49\textwidth}
\centering
\includegraphics[width=\textwidth]{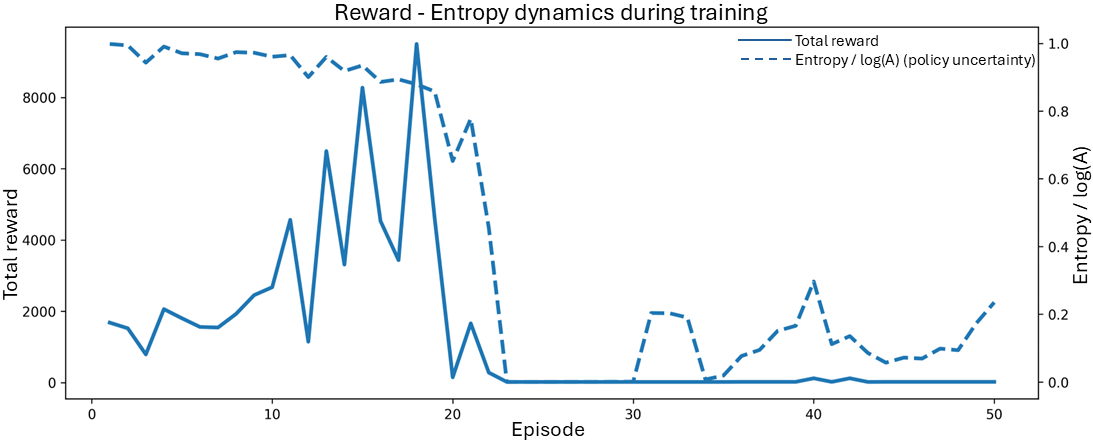}
\caption{\small Reward and policy entropy dynamics over training.}
\label{fig:reward-entropy}
\end{subfigure}

\caption{\small {Explainable signatures of policy collapse.} 
(a) Evolution of policy confidence (logit margin) and uncertainty (normalised entropy) over training. A sharp rise in confidence and collapse in entropy mark the transition to an over-confident, low-exploration regime.
(b) Reward–entropy dynamics during training. Following the entropy collapse, reward stagnates, indicating behavioural failure despite high confidence.}
\label{fig:policy-collapse-xai}
\end{figure*}

\begin{tcolorbox}
Across both value-based (DQL) and policy-gradient (PPO) agents, the same strategic and tactical explainability signatures consistently emerge. Phase transitions from exploration to exploitation, consolidation of dominant actions, and stability patterns persist despite fundamentally different optimisation mechanisms. This demonstrates that the signals extracted by our framework reflect intrinsic properties of adversarial learning dynamics rather than artefacts of a particular RL algorithm, confirming the model-agnostic nature of the proposed explainability approach.
\end{tcolorbox}

\vspace{0.9in}

\subsection{Policy-Level Explainability: Internal Confidence--Uncertainty Dynamics}\label{s8}

To evaluate the diagnostic power of our explainability framework under the most challenging conditions, where learning appears stable under reward-based evaluation, we analyse PPO trained in the CC500 environment. CC500 represents a large-scale, more complex, and highly partially observable setting, while PPO is widely regarded as a robust, entropy-regularised optimisation method. Failure observed in this regime therefore cannot be dismissed as an artefact of value-based learning or small environments.


This experiment isolates the internal dynamics of the PPO policy in the CC500 environment to determine whether observed learning failures originate within the policy itself. We deliberately exclude task-level performance signals in order to identify structural failure mechanisms before they become observable in reward trajectories. At each training episode, we instrument two complementary policy-level signals: \emph{policy entropy}, capturing decision uncertainty and exploratory capacity, and the \emph{logit margin} (the difference between the highest and second-highest action logits), capturing instantaneous decision confidence. Their temporal evolution is shown in Figure~\ref{fig:policy-collapse-xai}(a). During early training (Episodes~1–18), the policy operates in a flexible regime: entropy remains high, logit margins are near zero, and action preferences are weakly differentiated. This configuration reflects sustained exploration and effective uncertainty management. Between Episodes~19–25, a sharp structural transition occurs. Policy entropy collapses rapidly toward zero while the logit margin rises abruptly, forcing the policy into a highly concentrated, near-deterministic configuration. From this point onward, the agent exhibits persistent over-confidence, vanishing uncertainty, and a near-complete loss of exploratory behaviour. Crucially, this internal policy collapse occurs \emph{before} any corresponding degradation in task-level performance, indicating that failure emerges within the policy rather than as a response to declining reward.

Figure~\ref{fig:policy-collapse-xai}(b) contextualises this internal degeneration by relating entropy dynamics to reward evolution. Prior to collapse, high entropy coincides with volatile but improving reward. Once entropy contracts, reward becomes unstable and subsequently deteriorates, revealing a growing misalignment between internal decision commitment and effective behaviour. These structural breakdowns are not detectable under reward-only evaluation.

By exposing the contraction of the policy’s decision space during training, this analysis establishes the internal origin of learning failure and provides an early-warning signal that precedes performance collapse. The following section examines how this internal degeneration propagates to produce global performance failure.

\begin{tcolorbox}
Policy entropy and confidence dynamics reveal that learning failure originates inside the policy rather than in task performance. The agent enters an over-confident, low-uncertainty regime while rewards are still improving, demonstrating that policy-level explainability can surface latent failure mechanisms that reward-centric evaluation cannot detect.
\end{tcolorbox}

\begin{figure}[t]
\centering
\includegraphics[width=\columnwidth, trim=0 0 0 0, clip]{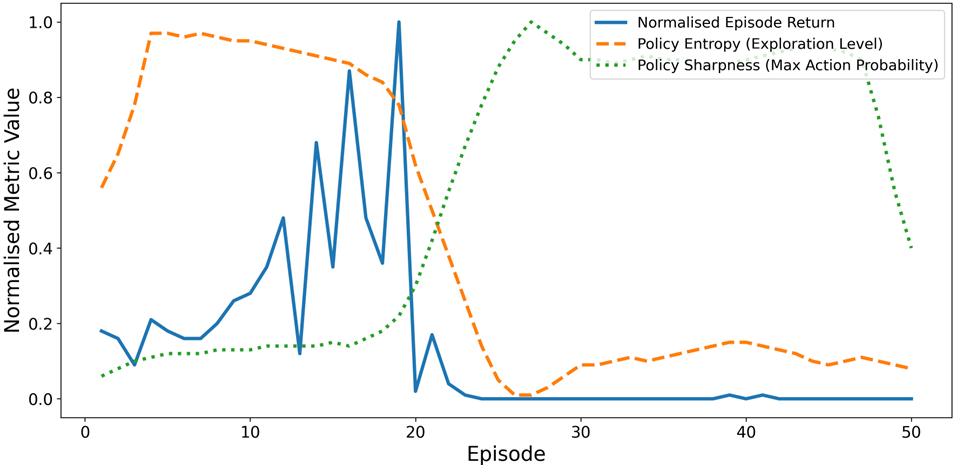}
\caption{\small{Joint evolution of task performance (normalised return), exploration (policy entropy), and decision confidence (policy sharpness) over training.}}
\label{fig:threecurves}
\end{figure}

\subsection{Joint Explainability of Policy and Performance Collapse}\label{s9}

This experimental setup examines the temporal relationship between internal policy degradation and task-level performance collapse. Figure~\ref{fig:threecurves} jointly visualises policy entropy, policy sharpness, and normalised return, enabling direct comparison of internal learning dynamics with external behavioural outcomes. During early training (Episodes~1–18), the agent operates in a stable learning regime: high policy entropy sustains exploration, policy sharpness remains low, and task performance improves steadily. These signals evolve coherently, indicating effective optimisation. Between Episodes~19–26, a critical transition occurs. Policy entropy contracts rapidly while policy sharpness saturates, forcing the policy into an over-confident, low-uncertainty regime. Importantly, this internal collapse precedes any degradation in task-level performance. Only after the policy has entered this degenerate state does task performance collapse and fail to recover, marking the onset of persistent learning failure. This temporal ordering establishes internal policy dynamics as the primary locus of failure and motivates the behavioural analysis in the following section.

\begin{tcolorbox}
 Joint inspection of policy entropy, policy sharpness, and task performance reveals a consistent causal ordering: internal policy degeneration occurs \emph{before} reward collapse. The agent enters an over-confident, low-uncertainty regime while performance is still improving, and only subsequently does task performance deteriorate irreversibly. This demonstrates that reward collapse is a downstream symptom rather than the root cause of learning failure, and that training-time explainability is required to identify failure mechanisms before they become externally observable.
\end{tcolorbox}

\subsection{Behavioural Lock-In and Dominant Action Dynamics}\label{s10}

\begin{figure}[t]
\centering
\includegraphics[width=\columnwidth]{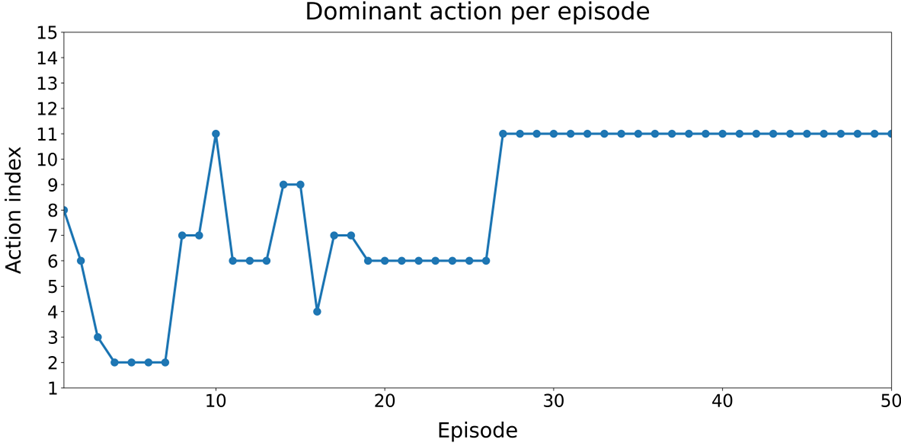}
\caption{\small Temporal evolution of the policy’s dominant action (argmax of episode-mean action probabilities), illustrating early exploratory instability followed by persistent behavioural lock-in.}
\label{fig:actionperep}
\end{figure}

Figure~\ref{fig:actionperep} grounds the internal failure mechanisms identified in Sections~\ref{s8} to \ref{s9} in observable agent behaviour. Whereas prior analyses revealed collapse in exploration and confidence dynamics within the policy, this figure demonstrates how those internal pathologies manifest externally as persistent action repetition.

\begin{table*}[t!]
\centering
\caption{Comparison of post-hoc explainability methods with our training-time framework in adversarial RL.}
\label{tab:xai_comparison}
\small
\setlength{\tabcolsep}{4pt}
\renewcommand{\arraystretch}{1.15}
\begin{tabular}{p{2cm}p{3cm}p{3cm}p{4cm}p{2cm}}
\toprule

{\textbf{Method}} & 
{\textbf{Phase Transition}} & 
{\textbf{Instability}} & 
{\textbf{Temporal}} & 
{\textbf{Diagnostic}} \\

{} & 
{\textbf{Detectable}} & 
{\textbf{Detectable}} & 
{\textbf{Insight}} & 
{\textbf{Score}} \\

\midrule

{\textbf{SHAP}} 
& {$\times$} 
& {$\times$} 
& {$\times$ (snapshot only)} 
& {Low} \\

{\textbf{LIME}} 
& {$\times$} 
& {$\times$} 
& {$\times$ (local approximation only)} 
& {Low} \\

{\textbf{CODEX}} 
& {$\checkmark$ (coarse regimes)} 
& {$\times$} 
& {Limited (episode-level)} 
& {Moderate} \\

{\textbf{Ours}} 
& {$\checkmark$ (phase-aware)} 
& {$\checkmark$ (TD-error, PER)} 
& {$\checkmark$ (continuous training-time)} 
& {High} \\

\bottomrule
\end{tabular}
\end{table*}

To characterise behavioural evolution over training, we compute episode-wise state-mean policy logits and record the dominant action for each episode. During early training (Episodes~1–18), the dominant action fluctuates rapidly, reflecting sustained exploration and the absence of stable behavioural commitment. Around Episodes~18–19, the policy enters a transient fixation phase, repeatedly selecting Action~6 for several consecutive episodes (Episodes~18–27). This marks a sharp reduction in behavioural diversity and coincides with the initial entropy collapse and confidence amplification identified in Section~\ref{s8}. At Episode~28, a second and terminal transition occurs. The policy switches to Action~11 (\texttt{connect(GIT)}) and remains persistently committed for the remainder of training (Episodes~28–50). From this point onward, behavioural adaptation ceases entirely, marking a practically irreversible stage of policy degeneration. This persistent lock-in aligns with the joint explainability analysis in Section~\ref{s9}, where internal policy collapse precedes and precipitates task-level performance failure.

This behavioural freezing is independently corroborated by PPO action-distribution snapshots (e.g., Figure~\ref{fig:PPO_ACTIONS}, CC500 Episode~40), which show the contraction of an initially broad action distribution into a single dominant choice. Together, these results demonstrate that internal policy collapse translates directly into concrete behavioural failure, completing the explainability chain from internal dynamics to external outcomes.

\begin{tcolorbox}
Explainability signals do not merely describe internal learning states; they predict concrete behavioural outcomes. Once policy entropy collapses and decision confidence saturates, the agent transitions into persistent action repetition from which recovery does not occur within training. Behavioural lock-in therefore serves as the observable endpoint of internal policy failure, validating the operational relevance of training-time explainability.
\end{tcolorbox}

\begin{figure*}[t!]
    \centering
    \includegraphics[width=\textwidth,height=0.35\textheight, keepaspectratio]{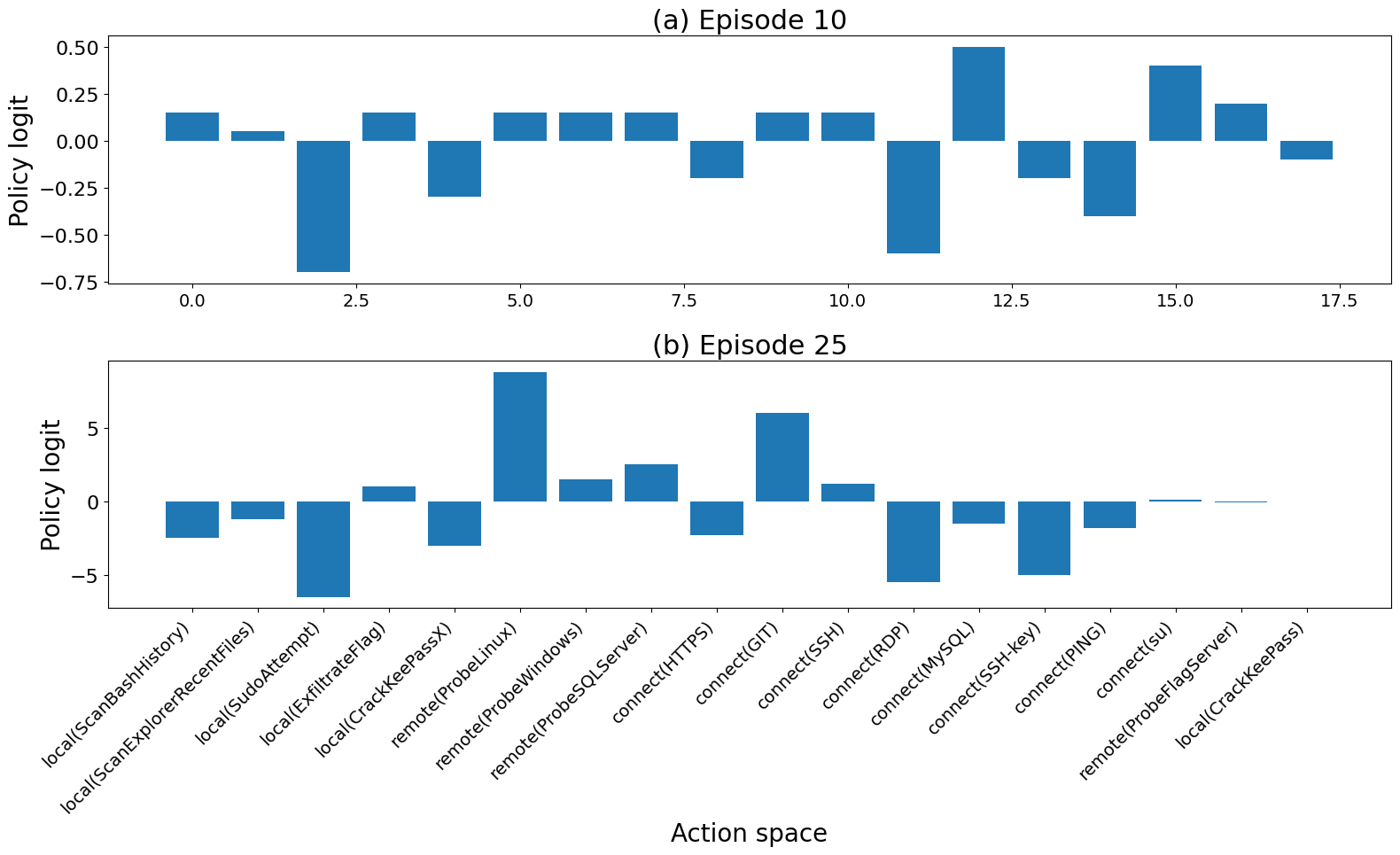}
\caption{{Policy logits across actions at Episodes 10 and 25. Early-stage behaviour is diffuse with low-magnitude logits, while later-stage behaviour shows a clear action hierarchy with dominant and suppressed actions.}}
    \label{fig:ourframework_compare}
\end{figure*}

\begin{figure}[t!]
    \centering
        \includegraphics[width=\columnwidth]{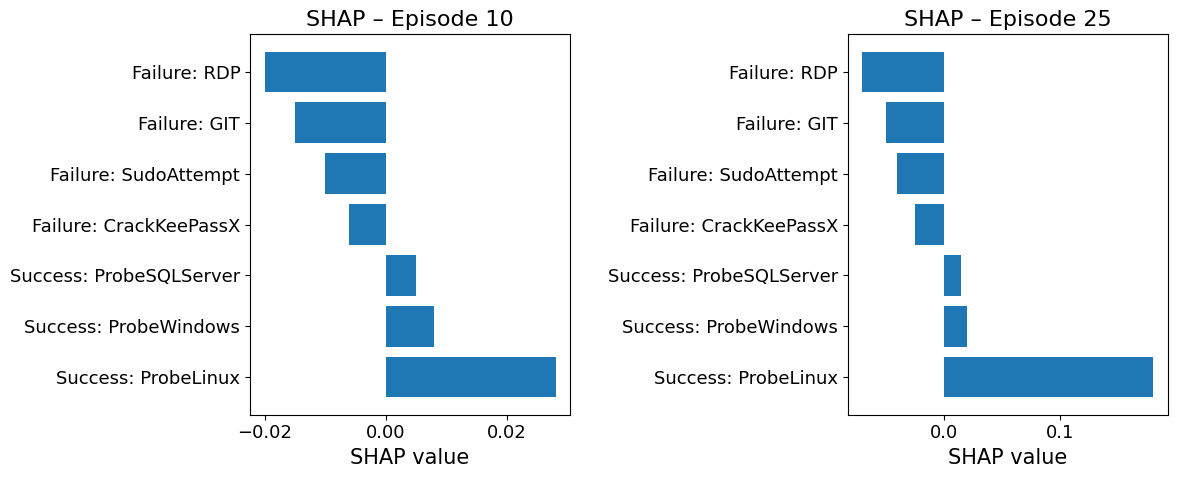}
\caption{{SHAP attribution for \texttt{remote(ProbeLinux)} at Episodes 10 and 25. Early attributions are small and distributed, while later attributions concentrate on successful probing, indicating increased decision confidence.}}
    \label{fig:shap_compare}
\end{figure}

{
\section{Comparative Analysis with Post-hoc 
Explainability Methods}

Explainability methods for reinforcement learning 
operate at fundamentally different levels of 
abstraction. Local post-hoc techniques such as 
SHAP and LIME provide instance-level feature 
attribution, explaining \textit{what} drives 
individual decisions. Behavioural abstraction 
approaches such as CODEX summarise attribution 
patterns across episodes to characterise 
\textit{what} behavioural regimes emerge. Our 
framework, by contrast, operates at the 
training-time level, providing a mechanistic 
account of \textit{how} and \textit{why} 
policies evolve through underlying optimisation 
dynamics, including Q-value updates, TD-error 
propagation, and replay prioritisation. As these 
approaches address different aspects of 
explainability, they are not directly comparable 
in a like-for-like manner but instead provide 
complementary perspectives across different 
levels of analysis. To enable a structured and 
consistent comparison, all methods are applied to 
the same representative training checkpoints 
(Episodes~10 and~25). Table \ref{tab:xai_comparison} compares the diagnostic capabilities of post-hoc explainability methods with our training-time framework.\\

\noindent \textbf{Our Framework, Policy Logit Evolution 
(Figure~\ref{fig:ourframework_compare}).} We begin with our framework, as it establishes 
the behavioural ground truth against which 
post-hoc explanations are interpreted. At 
Episode~10 (panel~a), policy logits span a 
narrow range of approximately $[-0.75,~+0.5]$ 
and are distributed across the action space 
with no clearly dominant action, reflecting 
high uncertainty and exploratory behaviour 
characteristic of early training. By 
Episode~25 (panel~b), the logit distribution 
becomes highly structured, expanding 
substantially to approximately $[-7,~+7.5]$. 
A clear action hierarchy emerges, with 
\texttt{remote(ProbeLinux)} consolidating as 
the dominant action (~+7.5), followed by 
\texttt{connect(GIT)} (~+6), while ineffective 
actions, such as \texttt{connect(RDP)}, 
\texttt{connect(SSH-key)}, and 
\texttt{local(SudoAttempt)}, are strongly 
suppressed. This transformation is strongly 
indicative of policy consolidation and 
convergence toward goal-directed exploitation.

Beyond these snapshots, our framework captures 
the underlying learning dynamics driving this 
transition, including Q-value evolution, 
TD-error propagation, replay prioritisation, 
and phase-aware behavioural changes. This 
provides a mechanistic account of how and why 
policy preferences emerge and stabilise over 
training. The identification of 
\texttt{remote(ProbeLinux)} as the dominant 
action provides a natural anchor for the 
post-hoc analyses that follow.\\

\begin{figure}[t!]
    \centering
    \includegraphics[width=\columnwidth]{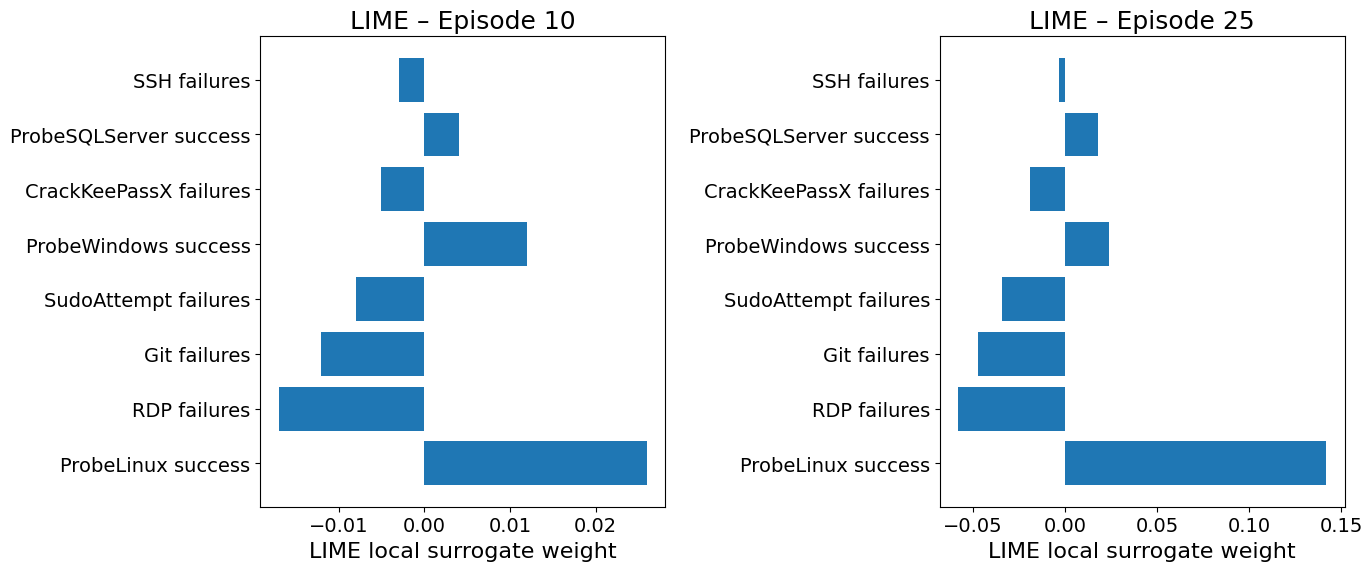}
\caption{{LIME attribution for \texttt{remote(ProbeLinux)} at Episodes 10 and 25. Local surrogate weights show diffuse contributions during exploration and concentrated positive signals for successful probing at convergence.}}
    \label{fig:lime_compare}
\end{figure}

\noindent \textbf{SHAP 
(Figure~\ref{fig:shap_compare}).} SHAP operates at the local, instance level, 
providing feature attributions for individual 
decisions rather than modelling global or 
temporal policy behaviour. Consequently, it 
cannot analyse the evolution of action 
preferences over training in the same manner 
as our framework. To enable a meaningful 
comparison, we anchor the analysis to the 
dominant action \texttt{remote(ProbeLinux)} 
identified by our framework.

At Episode~10, SHAP attributions are small and 
distributed across features, reflecting weak 
and competing signals consistent with early 
exploration. By Episode~25, the attribution 
landscape becomes more concentrated, with 
\texttt{Success: ProbeLinux} showing a strong 
positive contribution, while failure-related 
features remain consistently negative. The 
attribution scale also expands, indicating 
increased confidence aligned with the policy 
consolidation observed in our framework.

However, SHAP remains inherently local and 
static. It explains \textit{what} drives a 
decision at a given point in time but does not 
capture \textit{how} preferences evolve or the 
learning dynamics underlying their emergence. 
In contrast, our framework explicitly models 
this temporal evolution, linking changes in 
action preference to the underlying optimisation 
processes during training.\\

\noindent \textbf{LIME 
(Figure~\ref{fig:lime_compare}).} LIME provides local explanations by fitting a 
surrogate model to approximate the policy’s 
behaviour in the neighbourhood of a specific 
decision. While it differs from SHAP in its 
use of local approximation, it similarly does 
not model global or temporal policy dynamics. 
We therefore apply LIME to the same dominant 
action \texttt{remote(ProbeLinux)} to examine 
how local approximations reflect policy 
behaviour across training stages.  At Episode~10, LIME produces diffuse, 
low-magnitude contributions across features, 
indicating weak and competing signals 
consistent with exploratory behaviour. By 
Episode~25, the attributions become more 
concentrated, with \texttt{Success: ProbeLinux} 
emerging as the dominant positive contributor, 
while failure-related features remain negative. 
Notably, LIME also reveals an additional 
early-stage signal (e.g., \texttt{ProbeWindows 
success}), suggesting a broader exploratory 
tendency captured by the surrogate model.

Despite this nuance, LIME remains inherently 
local and dependent on approximation quality. 
It explains decision behaviour within a limited 
neighbourhood but does not capture the temporal 
progression or learning dynamics that shape 
policy evolution. In contrast, our framework 
provides a temporally grounded view, linking 
changes in action preference to underlying 
optimisation processes throughout training.\\

\begin{figure}[t!]
    \centering
    \includegraphics[width=\columnwidth,height=0.2\textheight]{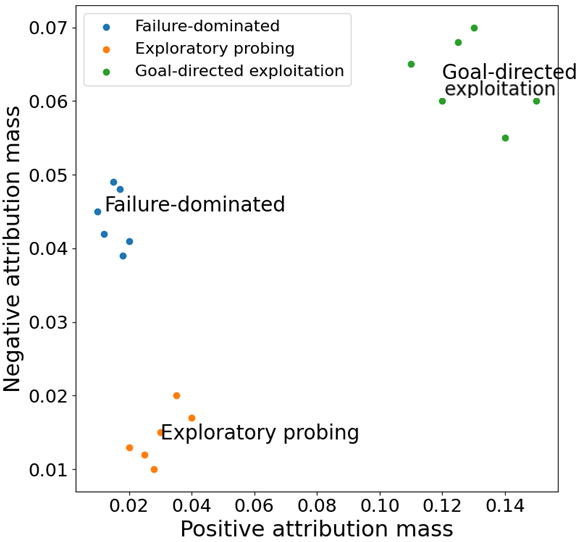}
\caption{{CODEX-style clustering in attribution space (positive vs. negative mass). Episodes separate into failure-dominated, exploratory, and goal-directed regimes, reflecting progression from exploration to exploitation.}}
\label{fig:codex_compare}
\end{figure}

\noindent \textbf{CODEX-style Behavioural 
Abstraction (Figure~\ref{fig:codex_compare}).} 
Unlike SHAP and LIME, which operate at the 
level of individual decisions, CODEX provides 
a higher-level view by aggregating attribution 
statistics across episodes and clustering 
similar behavioural patterns into regimes. 
Three distinct regimes emerge: 
\textit{failure-dominated} (low positive, high 
negative attribution mass), \textit{exploratory 
probing} (low attribution mass on both axes), 
and \textit{goal-directed exploitation} (high 
positive and high negative attribution mass 
simultaneously). Particularly noteworthy is 
the goal-directed cluster, which exhibits high 
attribution mass on both axes simultaneously, 
indicating that the agent makes strongly 
differentiated, high-confidence decisions with 
clear winners and losers in the action space, consistent with the logit consolidation 
identified by our framework. The separation 
of these clusters reflects a clear progression 
from unsuccessful behaviour, through 
exploration, to confident exploitation, 
consistent with the exploration--exploitation 
dynamics observed in our framework.

While this abstraction provides a compact and 
interpretable summary of behavioural regimes, 
it remains fundamentally descriptive. It 
characterises \textit{what} behavioural states 
exist but does not explain the mechanisms 
governing transitions between them or the 
underlying learning dynamics that drive this 
progression. In contrast, our framework 
captures these temporal and mechanistic 
processes, linking behavioural transitions 
to the optimisation dynamics of the agent.

Taken together, these results provide a layered 
understanding of explainability. SHAP and LIME 
explain \textit{what} drives individual 
decisions, while CODEX summarises \textit{what} 
behavioural regimes emerge. Our framework, in 
contrast, explains \textit{how} and \textit{why} 
these behaviours develop by exposing the 
underlying optimisation dynamics during 
training. This distinction is particularly 
important in sequential adversarial settings, 
where understanding the evolution of strategies 
is essential for debugging, intervention, and 
trustworthy evaluation of autonomous cyber 
agents.
}

{

\section{Policy Adaptation Under Evolving Defensive Dynamics}

Understanding how RL agents adapt when previously learned strategies become unreliable is critical for analysing robustness and behavioural flexibility. While earlier experiments focused on policy formation under static conditions, we extend the analysis to settings where defensive mechanisms actively modify the environment during training. We introduce two classes of defenders operating at different temporal scales: continuous system recovery and targeted mitigation of frequently exploited pathways. These mechanisms disrupt the underlying structure, enabling analysis of how estimates, action preferences, and learning signals reorganise as the agent adapts to evolving conditions.

\noindent \paragraph{Experimental Design}
We evaluate three training regimes in the CTF environment using identical configurations (DQL, 20 episodes, 1500 iterations per episode):
\begin{itemize}
    \item \textbf{Baseline:} Standard CTF environment without modification.
    
    \item \textbf{Stochastic Re-hardening (Defender 1):} A persistent defender induces non-stationarity by stochastically reverting compromised nodes via automated recovery (e.g., resets, credential invalidation), with each node restored independently with probability $p = 0.3$, yielding sustained disruption while preserving learnability. 
    \item \textbf{Targeted Patch Interventions (Defender 2):} A reactive defender introduces structural perturbations by disabling the dominant exploit pathways at two intervention points, $P_1$ (batch $\approx 15{,}000$) and $P_2$ (batch $\approx 21{,}000$), inducing discrete shifts in the action-value landscape\footnote{Empirically, the highest-performing action at $P_1$ is \texttt{remote(ProbeLinux)}, while at $P_2$ it is \texttt{connect(GIT)}.}.

\end{itemize}

These regimes introduce two complementary forms of defensive influence: persistent disruption of state persistence and targeted removal of high-value exploit pathways.\\

\paragraph{TD-Error Dynamics.}
Figure~\ref{fig:td_error_ctf} shows the evolution of average TD-error (PER priority), revealing how defensive mechanisms reshape learning dynamics across three regimes: stable convergence (baseline), stochastic re-hardening (Defender 1), and targeted patch interventions (Defender 2).

In the baseline, TD-error follows a canonical trajectory, with an initial spike ($\approx 0.013$) rapidly decaying to a stable floor ($\approx 0.001$) by batch $\approx 3{,}000$, followed by a mild increase ($\approx 0.002$--$0.003$), indicating stable value propagation and convergence. Under stochastic re-hardening, TD-error remains persistently elevated ($\approx 0.003$--$0.006$) with sustained oscillations, reflecting repeated invalidation of learned transitions and preventing stable convergence. Under patch interventions, TD-error tracks the baseline until $P_1$, where a sharp spike reflects abrupt removal of a dominant action, followed by a smaller spike at $P_2$ , indicating faster adaptation due to prior diversification. In both cases, TD-error rapidly returns to baseline, indicating effective policy reorganisation.

{Importantly, these dynamics are quantitatively captured by TD-error magnitude and temporal variance, enabling systematic differentiation between convergence, disruption, and adaptation. For example, TD-error increases by approximately 7$\times$ at $P_1$ under patching, signalling substantial policy disruption not reflected in reward, which shows only minor degradation.}

These TD-error patterns characterise how learning dynamics evolve under each defensive condition.\\

\paragraph{Reward Behaviour.}
Figure~\ref{fig:reward_ctf} shows cumulative reward across episodes, reflecting the performance impact of defensive dynamics. In the baseline, reward rapidly plateaus at $\approx 470$ by episode 5, indicating stable exploitation and convergence. Under stochastic re-hardening, reward is reduced to $\approx 60$--$70\%$ of baseline with increased variability, as repeated node resets limit sustained accumulation; however, the smooth trajectory suggests continued local progress despite lack of global stability, consistent with elevated TD-error. Under patch interventions, reward closely follows the baseline, with brief dips at $P_1$ and $P_2$ followed by rapid recovery, indicating effective adaptation via alternative actions.

Importantly, while TD-error reveals substantial internal changes, including sharp spikes and reorganisation of value estimates, these changes are only weakly reflected in reward. This highlights that reward captures outcomes, but not the underlying adaptation process.

\textit{\textbf{Key Insight and Interpretation.}}
Across conditions, the baseline exhibits stable convergence, Defender~2 (patching) achieves near-baseline performance with efficient recovery, while Defender~1 (re-hardening) induces the most persistent disruption. Specifically, patch interventions trigger sharp TD-error spikes followed by rapid stabilisation, indicating effective policy reorganisation, whereas re-hardening sustains elevated TD-error, reflecting continuous invalidation of learned transitions and preventing convergence. Notably, despite substantial internal restructuring (6--8$\times$ TD-error increase under patching), reward remains largely stable, while under re-hardening, moderate reward degradation corresponds to persistent learning instability.

These observations reveal that Defender~2 is less detrimental to long-term performance but induces discrete structural adaptation, whereas Defender~1 enforces sustained non-stationarity, leading to continuous adaptation without stabilisation. More broadly, this comparison highlights a key limitation of reward: it reflects outcomes but fails to capture underlying learning dynamics. In contrast, TD-error provides a mechanistic signal of when and how policies are disrupted and reorganised, enabling precise analysis of adaptation under different defensive regimes.

\begin{figure}[t]
\centering
\includegraphics[width=\linewidth]{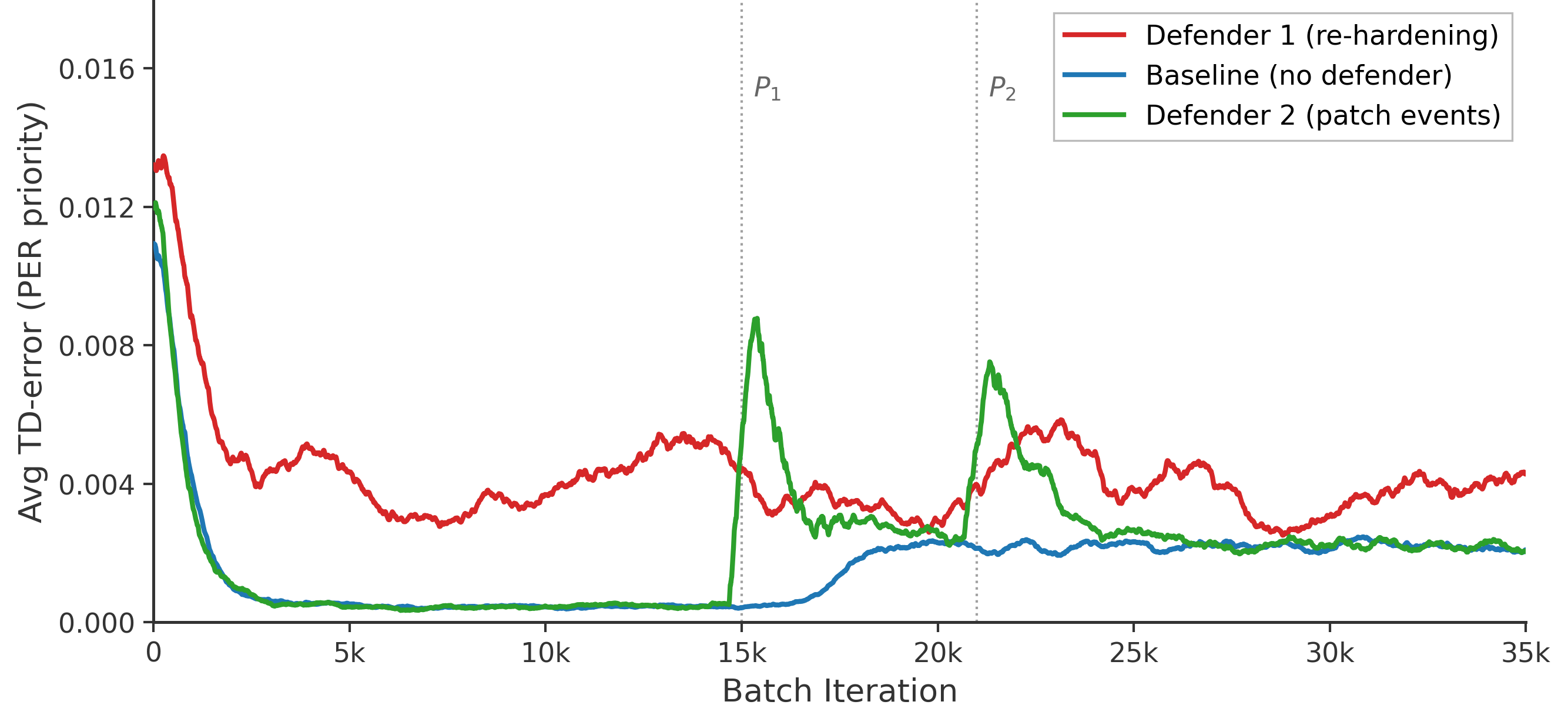}
\caption{{
TD-error under defensive dynamics.
Baseline converges; re-hardening maintains elevated TD-error (continuous disruption); patching causes spikes at $P_1$, $P_2$ followed by recovery (policy reorganisation).
}}
\label{fig:td_error_ctf}
\end{figure}

\begin{figure}[t]
\centering
\includegraphics[width=\linewidth]{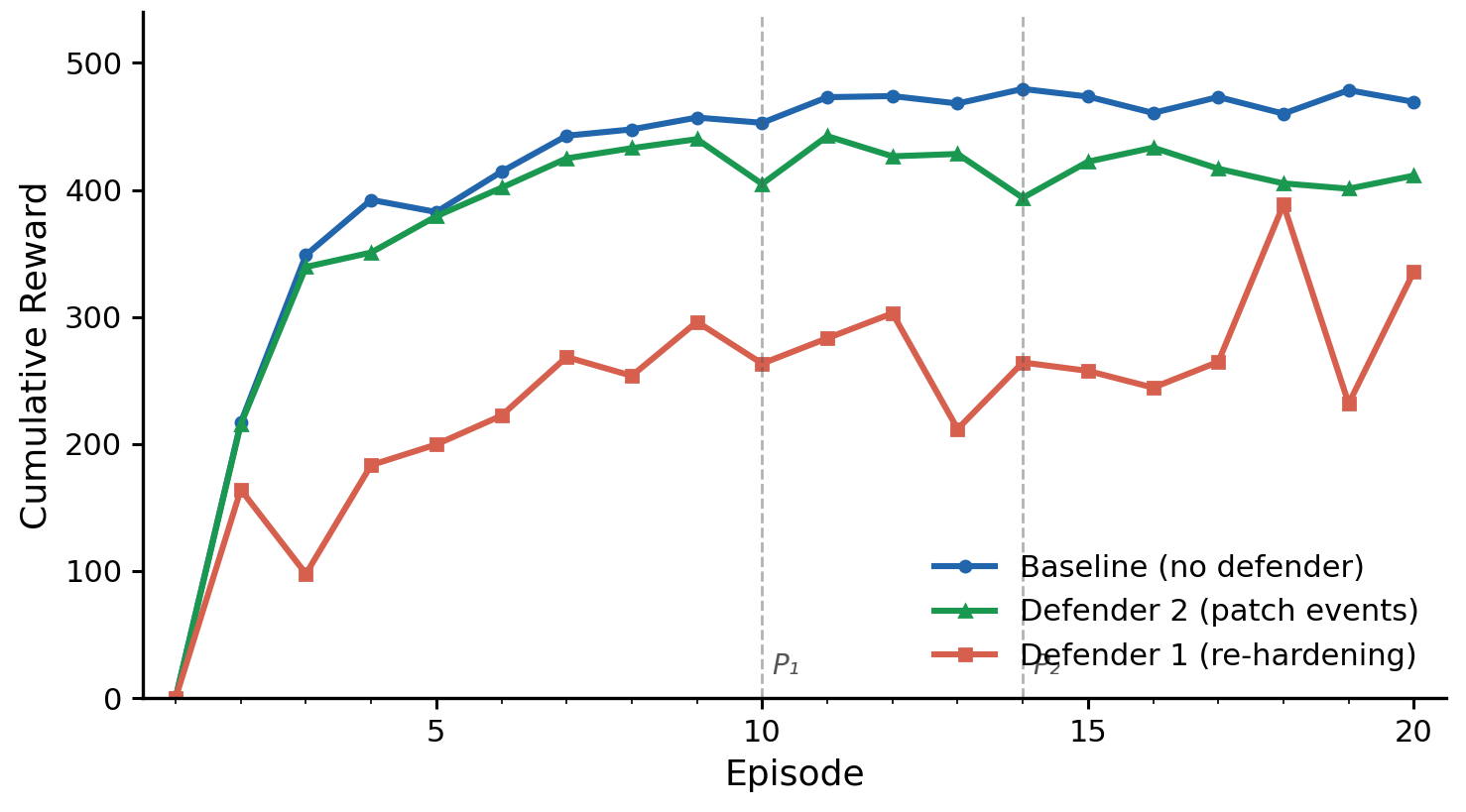}
\caption{{
Cumulative reward under defensive dynamics.
Baseline plateaus; re-hardening reduces reward; patching causes small, transient drops despite significant internal changes.
}}
\label{fig:reward_ctf}
\end{figure}

}

\section{Discussion}

\noindent \textbf{Key Insights.}
Across all experiments, three consistent insights emerge. 
First, training-time signals expose phase transitions, instability, and behavioural collapse that remain invisible under reward-only evaluation, demonstrating that cumulative reward is an unreliable indicator of learning health in adversarial cyber environments. 
Second, commonly used reinforcement learning enhancements, most notably Prioritised Experience Replay (PER) and entropy-driven exploration in policy-gradient methods, can exhibit pathological behaviour in sparse-reward, partially observable settings. While PER is often assumed to accelerate learning or promote earlier lateral movement, our analysis shows that it instead amplifies transient, high-TD-error transitions, introducing replay bias that distorts value propagation and destabilises strategy formation in large environments. Similar sensitivity arises in policy-gradient methods, where aggressive entropy decay can induce premature determinism rather than sustained exploration.  More broadly, our analysis suggests that richer or denser reward signals do not prevent premature escalation, but can instead obscure early over-commitment by maintaining stable reward trajectories despite collapsing exploration. Third, these failure modes emerge early during training, often well before any degradation in task-level performance is observable. Together, these findings establish training-time explainability as a critical diagnostic capability for understanding, debugging, and tuning autonomous cyber attacker agents.\\

\noindent\textbf{Practical Use for ML Model Developers.}
The primary beneficiaries of this framework are developers of RL-based autonomous attacker agents. The framework serves as a training-time diagnostic tool for understanding, debugging, and validating learning behaviour. It enables:
\begin{itemize}
 \item \textit{Visualisation of Q-value trajectories} to trace how action preferences emerge, stabilise, or shift during training, distinguishing genuine strategy formation from early bias.
 \item \textit{Identification of critical learning transitions} via TD-error spikes and PER prioritisation, highlighting experiences that disproportionately influence policy updates.
 \item \textit{Detection of brittle convergence} or excessive determinism through internal confidence and entropy signals, exposing premature policy collapse even when rewards appear stable.
 \item \textit{Evaluation of exploration strategies} based on behavioural richness and adaptability rather than cumulative reward alone.
\end{itemize}
\vspace{0.02in}

These capabilities are especially valuable in partially observable environments, where incomplete visibility make reward curves a poor proxy for learning health. By exposing internal learning dynamics, the framework supports the development of more interpretable and resilient attacker models, clarifying not only \textit{what} the agent learns, but \textit{how} and \textit{when} that learning occurs.\\

\noindent\textbf{Implications for Defensive Simulation.}
While our framework is primarily intended for developers of autonomous attacker agents, we emphasise that its core contributions lie in attacker modelling and training-time diagnosis. Nevertheless, the same diagnostic signals may provide useful context for defensive simulation during offline training and scenario development. As agentic attackers become more sophisticated, understanding not only \textit{what} strategies emerge but also \textit{how} and \textit{when} they adapt can support more informed threat modelling and defensive preparedness.

\textit{Example:} In a simulated environment such as CC500, defenders may observe recurring spikes in transition priority at specific stages of training, indicating moments where the attacker model shifts from reconnaissance-driven behaviour to credential reuse or lateral movement. Although such internal learning signals are not observable in live attacks, they can inform defensive simulation by highlighting which phases of an attack lifecycle are most sensitive to adaptation. This insight can be used to design targeted defensive stress tests, evaluate the robustness of segmentation or access-control policies, and assess how defensive assumptions hold under adaptive attacker behaviour.\\

\noindent\textbf{Limitations and Future Work.}
While CyberBattleSim provides a controlled and extensible 
environment for systematic experimentation, it abstracts 
several real-world complexities. In particular, it does 
not fully capture concurrent adversaries, deception-aware 
agents, or sophisticated evasion tactics, and its network 
architectures lack the heterogeneity of operational 
enterprise systems, such as layered access controls, 
segmented subnets, and diverse user behaviours. These 
simplifications enable controlled analysis but limit 
direct extrapolation to real-world deployments. Moreover, 
our current study focuses on a single autonomous attacker 
operating in isolation. Future work will extend the 
framework to multi-agent settings to investigate 
co-evolving attacker--defender dynamics and emergent 
behaviours under strategic interaction.

{In this work, we focus on establishing \textit{technical and behavioural} interpretability of RL-based cyber agents by exposing training-time dynamics that are not observable through reward-based evaluation. These signals provide actionable insights that RL developers can directly use during training—for example, to detect policy collapse, identify instability, and recognise behavioural shifts, enabling earlier and more targeted debugging.

Building on this foundation, further exploration of how these signals integrate into practitioner workflows represents a natural next step, including task-based and usability-oriented evaluations to refine their effectiveness in real-world settings.}\\

\noindent \textbf{Key Takeaways.}
Understanding autonomous cyber attackers requires visibility beyond final performance metrics. Our analysis shows that reward curves alone provide an incomplete, and in some cases misleading, view of learning behaviour in adversarial cyber environments. Effective evaluation therefore requires insight into learning dynamics at multiple levels of abstraction. In particular, strategic signals reveal phase transitions from reconnaissance to exploitation as environments evolve under uncertainty, while policy-level signals expose how action preferences emerge, stabilise, or collapse during training. Together, these perspectives clarify not only \textit{what} strategies are learned, but critically \textit{how} and \textit{when} they form.

Across large-scale environments, we find that learning failure does not originate in declining reward, but emerges internally during training as exploration capacity progressively contracts. As policy entropy decreases and decision confidence sharpens, agents become locked into narrow, brittle behaviours, even while task-level rewards remain stable or improve. This decoupling demonstrates that reward curves can actively mask underlying learning pathologies rather than reveal them.

Within this context, our results expose a concrete replay-induced failure mode. Contrary to common assumptions, Prioritised Experience Replay does not reliably accelerate learning or promote earlier lateral movement in sparse, partially observable cyber environments. Instead, it oversamples transient, high-TD-error transitions, introducing replay bias that distorts value propagation and destabilises convergence in large environments. These effects are detectable only through training-time diagnostic signals, well before they manifest as observable performance collapse.

Taken together, these findings suggest that evaluating autonomous cyber agents solely on final performance is insufficient. Training-time explainability must be treated as a first-class diagnostic tool when developing, benchmarking, and validating reinforcement learning–based attacker models, particularly in sparse, partially observable cybersecurity settings.

\section{Conclusion}

In this paper, we introduced a unified, multi-layer explainability framework for demystifying the decision-making of RL-based attacker agents in cyber environments. By combining MDP- and policy-level analysis, the framework offers fine-grained, temporally grounded insights into how adversaries explore networks, adapt, and exploit network vulnerabilities. Through experiments in CyberBattleSim, we demonstrated how the framework reveals strategy shifts, learning patterns, and decision inflection points often hidden in standard RL evaluations. Unlike prior post-hoc only or domain-specific work, our agent-agnostic approach generalises to defender agents, enabling unified interpretation of both offensive and defensive strategies. It equips red teams with visibility into how RL agents develop adversarial tactics and offers blue teams a powerful tool to interpret and counter evolving threats. This work lays the foundation for transparent, trustworthy, and operationally useful RL in cybersecurity. {While focused on adversarial cyber operations, our framework is readily applicable to other sequential decision-making domains, such as autonomous drone navigation, financial fraud detection, and critical infrastructure protection, where explainability and behavioural traceability are crucial.} Future directions include validating the usability and actionability of these explanations for human stakeholders (e.g., security analysts and RL developers),  multi-agent extensions, and application in high-stakes AI safety contexts.

\bibliographystyle{IEEEtran}
\bibliography{ref}

@inproceedings{Goel2022defending,
title = {Defending Active Directory by Combining Neural Network Based Dynamic Program and Evolutionary Diversity Optimisation},
author = {Goel, Diksha and Ward-Graham, Max Hector and Neumann, Aneta and Neumann, Frank and Nguyen, Hung and Guo, Mingyu},
booktitle = {Proceedings of the Genetic and Evolutionary Computation Conference},
year = {2022},
pages = {1191–1199},
numpages = {9},
location = {Boston, Massachusetts},
series = {GECCO '22}
}

@inproceedings{goel2023evolving,
  title={Evolving Reinforcement Learning Environment to Minimize Learner's Achievable Reward: An Application on Hardening Active Directory Systems},
  author={Goel, Diksha and Neumann, Aneta and Neumann, Frank and Nguyen, Hung and Guo, Mingyu},
booktitle = {Proceedings of the Genetic and Evolutionary Computation Conference},
year = {2023},
pages = {1348–1356},
numpages = {9},
location = {Lisbon, Portugal},
series = {GECCO '23}
}

@misc{schwartz2019nasim,
title={NASim: Network Attack Simulator},
author={Schwartz, Jonathon and Kurniawatti, Hanna},
year={2019},
howpublished={\url{https://networkattacksimulator.readthedocs.io/}},
}

@inproceedings{schaul2016prioritized,
  author = {Schaul, Tom and Quan, John and Antonoglou, Ioannis and Silver, David},
  title = {Prioritized Experience Replay},
  booktitle = {Proceedings of the 4th International Conference on Learning Representations (ICLR)},
  year = {2016},
  note = {arXiv:1511.05952},
  url = {https://arxiv.org/abs/1511.05952}
}

@inproceedings{tokic2010adaptive,
  author = {Tokic, Michel},
  title = {Adaptive $\epsilon$-greedy Exploration in Reinforcement Learning},
  booktitle = {Proceedings of the 22nd International Conference on Tools with Artificial Intelligence (ICTAI)},
  year = {2010},
  pages = {243--250},
  doi = {10.1109/ICTAI.2010.41},
  publisher = {IEEE}
}

@inproceedings{thompson2024entity,
  title={Entity-based reinforcement learning for autonomous cyber defence},
  author={Thompson, Isaac Symes and Caron, Alberto and Hicks, Chris and Mavroudis, Vasilios},
  booktitle={Proceedings of the Workshop on Autonomous Cybersecurity},
  pages={56--67},
  year={2024}
}

@article{wiebe2023learning,
  title={Learning cyber defence tactics from scratch with multi-agent reinforcement learning},
  author={Wiebe, Jacob and Mallah, Ranwa Al and Li, Li},
  journal={arXiv preprint arXiv:2310.05939},
  year={2023}
}

@article{andrew2022developing,
  title={Developing optimal causal cyber-defence agents via cyber security simulation},
  author={Andrew, Alex and Spillard, Sam and Collyer, Joshua and Dhir, Neil},
  journal={arXiv preprint arXiv:2207.12355},
  year={2022}
}

@inproceedings{bierbrauer2023autonomous,
  title={Autonomous cyber warfare agents: dynamic reinforcement learning for defensive cyber operations},
  author={Bierbrauer, David A and Schabinger, Robert M and Carlin, Caleb and Mullin, Jonathan and Pavlik, John A and Bastian, Nathaniel D},
  booktitle={Artificial Intelligence and Machine Learning for Multi-Domain Operations Applications V},
  volume={12538},
  pages={42--56},
  year={2023},
  organization={SPIE}
}

@inproceedings{sultana2021autonomous,
  title={Autonomous network cyber offence strategy through deep reinforcement learning},
  author={Sultana, Madeena and Taylor, Adrian and Li, Li},
  booktitle={Artificial Intelligence and Machine Learning for Multi-Domain Operations Applications III},
  volume={11746},
  pages={490--502},
  year={2021},
  organization={SPIE}
}

@article{watkins1992qlearning,
  author = {Watkins, Christopher J. C. H. and Dayan, Peter},
  title = {Q-learning},
  journal = {Machine Learning},
  volume = {8},
  number = {3-4},
  year = {1992},
  pages = {279--292},
  doi = {10.1007/BF00992698},
  publisher = {Springer}
}

@book{bellman1957dynamic,
  author = {Bellman, Richard E.},
  title = {Dynamic Programming},
  year = {1957},
  publisher = {Princeton University Press},
  address = {Princeton, NJ}
}

@misc{msft:cyberbattlesim,
  Author = {{Microsoft Defender Research Team}},
  Note = {{Created by Christian Seifert, Michael Betser, William Blum, James Bono, Kate Farris, Emily Goren, Justin Grana, Kristian Holsheimer, Brandon Marken, Joshua Neil, Nicole Nichols, Jugal Parikh, Haoran Wei}},
  Publisher = {GitHub},
  Howpublished = {\url{https://github.com/microsoft/cyberbattlesim}},
  Title = {CyberBattleSim},
  Year = {2021}
}

@misc{cage_cyborg_2022, 
  Title = {Cyber Operations Research Gym}, 
  Note = {Created by Maxwell Standen, David Bowman, Son Hoang, Toby Richer, Martin Lucas, Richard Van Tassel, Phillip Vu, Mitchell Kiely, KC C., Natalie Konschnik, Joshua Collyer}, 
  Publisher = {GitHub}, 
  Howpublished = {\url{https://github.com/cage-challenge/CybORG}}, 
  Year = {2022} 
}

@article{molina2021network,
  title={Network defense is not a game},
  author={Molina-Markham, Andres and Winder, Ransom K and Ridley, Ahmad},
  journal={arXiv preprint arXiv:2104.10262},
  year={2021}
}

@article{foley2023inroads,
  title={Inroads into autonomous network defence using explained reinforcement learning},
  author={Foley, Myles and Wang, Mia and Hicks, Chris and Mavroudis, Vasilios and others},
  journal={arXiv preprint arXiv:2306.09318},
  year={2023}
}

@article{gyevnar2023causal,
  title={Causal explanations for sequential decision-making in multi-agent systems},
  author={Gyevnar, Balint and Wang, Cheng and Lucas, Christopher G and Cohen, Shay B and Albrecht, Stefano V},
  journal={arXiv preprint arXiv:2302.10809},
  year={2023}
}

@article{mathes2023codex,
  title={CODEX: A Cluster-Based Method for Explainable Reinforcement Learning},
  author={Mathes, Timothy K and Inman, Jessica and Col{\'o}n, Andr{\'e}s and Khan, Simon},
  journal={arXiv preprint arXiv:2312.04216},
  year={2023}
}

@article{sharma2022explainable,
  title={Explainable artificial intelligence for cybersecurity},
  author={Sharma, Deepak Kumar and Mishra, Jahanavi and Singh, Aeshit and Govil, Raghav and Srivastava, Gautam and Lin, Jerry Chun-Wei},
  journal={Computers and Electrical Engineering},
  volume={103},
  pages={108356},
  year={2022},
  publisher={Elsevier}
}

@inproceedings{yu2023airs,
  title={{AIRS}: Explanation for Deep Reinforcement Learning-Based Security Applications},
  author={Yu, Jiahao and Guo, Wenbo and Qin, Qi and Wang, Gang and Wang, Ting and Xing, Xinyu},
  booktitle={32nd USENIX Security Symposium (USENIX Security 23)},
  pages={7375--7392},
  year={2023}
}

@article{alabdulkarim2022experiential,
  title={Experiential explanations for reinforcement learning},
  author={Alabdulkarim, Amal and Singh, Madhuri and Mansi, Gennie and Hall, Kaely and Riedl, Mark O},
  journal={arXiv preprint arXiv:2210.04723},
  year={2022}
}

@inproceedings{goel2024optimizing,
  title={Optimizing Cyber Defense in Dynamic Active Directories through Reinforcement Learning},
  author={Goel, Diksha and Moore, Kristen and Guo, Mingyu and Wang, Derui and Kim, Minjune and Camtepe, Seyit},
  booktitle={European Symposium on Research in Computer Security},
  pages={332--352},
  year={2024},
  organization={Springer}
}

@inproceedings{wei2021non,
  title={Non-stationary reinforcement learning without prior knowledge: An optimal black-box approach},
  author={Wei, Chen-Yu and Luo, Haipeng},
  booktitle={Conference on learning theory},
  pages={4300--4354},
  year={2021},
  organization={PMLR}
}

@inproceedings{madumal2020explainable,
  title={Explainable reinforcement learning through a causal lens},
  author={Madumal, Prashan and Miller, Tim and Sonenberg, Liz and Vetere, Frank},
  booktitle={Proceedings of the AAAI conference on artificial intelligence},
  volume={34},
  number={03},
  pages={2493--2500},
  year={2020}
}

@article{nashed2025causal,
  title={Causal Explanations for Sequential Decision Making},
  author={Nashed, Samer B and Mahmud, Saaduddin and Goldman, Claudia V and Zilberstein, Shlomo},
  journal={Journal of Artificial Intelligence Research},
  volume={83},
  year={2025}
}

@article{dwivedi2023explainable,
  title={Explainable AI (XAI): Core ideas, techniques, and solutions},
  author={Dwivedi, Rudresh and Dave, Devam and Naik, Het and Singhal, Smiti and Omer, Rana and Patel, Pankesh and Qian, Bin and Wen, Zhenyu and Shah, Tejal and Morgan, Graham and others},
  journal={ACM computing surveys},
  volume={55},
  number={9},
  pages={1--33},
  year={2023},
  publisher={ACM New York, NY}
}

@inproceedings{nguyen2021evaluation,
  title={Evaluation of explainable artificial intelligence: Shap, lime, and cam},
  author={Nguyen, Hung Truong Thanh and Cao, Hung Quoc and Nguyen, Khang Vo Thanh and Pham, Nguyen Dinh Khoi},
  booktitle={Proceedings of the FPT AI Conference},
  pages={1--6},
  year={2021}
}

@inproceedings{alenezi2021explainability,
  title={Explainability of cybersecurity threats data using shap},
  author={Alenezi, Rafa and Ludwig, Simone A},
  booktitle={2021 IEEE symposium series on computational intelligence (SSCI)},
  pages={01--10},
  year={2021},
  organization={IEEE}
}

@article{claypoole2025interpreting,
  title={Interpreting Agent Behaviors in Reinforcement-Learning-Based Cyber-Battle Simulation Platforms},
  author={Claypoole, Jared and Cheung, Steven and Gehani, Ashish and Yegneswaran, Vinod and Ridley, Ahmad},
  journal={arXiv preprint arXiv:2506.08192},
  year={2025}
}

@inproceedings{terranova:hal-05182437,
  TITLE = {{Scalable and Generalizable RL Agents for Attack Path Discovery via Continuous Invariant Spaces}},
  AUTHOR = {Terranova, Franco and Lahmadi, Abdelkader and Chrisment, Isabelle},
  URL = {https://hal.science/hal-05182437},
  BOOKTITLE = {{2025 28th International Symposium on Research in Attacks, Intrusions and Defenses (RAID)}},
  ADDRESS = {Gold Coast, Australia},
  PAGES = {18},
  YEAR = {2025},
  MONTH = Oct,
  HAL_ID = {hal-05182437},
  HAL_VERSION = {v1},
}

@article{goel2023enhancing,
  title={Enhancing network resilience through machine learning-powered graph combinatorial optimization: Applications in cyber defense and information diffusion},
  author={Goel, Diksha},
  journal={arXiv preprint arXiv:2310.10667},
  year={2023}
}

@inproceedings{slack2020fooling,
  title={Fooling lime and shap: Adversarial attacks on post hoc explanation methods},
  author={Slack, Dylan and Hilgard, Sophie and Jia, Emily and Singh, Sameer and Lakkaraju, Himabindu},
  booktitle={Proceedings of the AAAI/ACM Conference on AI, Ethics, and Society},
  pages={180--186},
  year={2020}
}

@article{bello2025level,
  title={The level of strength of an explanation: A quantitative evaluation technique for post-hoc XAI methods},
  author={Bello, Marilyn and Amador, Rosalis and Garcia, Maria-Matilde and Del Ser, Javier and Mesejo, Pablo and Cord{\'o}n, {\'O}scar},
  journal={Pattern Recognition},
  volume={161},
  pages={111221},
  year={2025},
  publisher={Elsevier}
}

@article{guo2021edge,
  title={Edge: Explaining deep reinforcement learning policies},
  author={Guo, Wenbo and Wu, Xian and Khan, Usmann and Xing, Xinyu},
  journal={Advances in Neural Information Processing Systems},
  volume={34},
  pages={12222--12236},
  year={2021}
}

@inproceedings{topin2019generation,
  title={Generation of policy-level explanations for reinforcement learning},
  author={Topin, Nicholay and Veloso, Manuela},
  booktitle={Proceedings of the AAAI Conference on Artificial Intelligence},
  volume={33},
  number={01},
  pages={2514--2521},
  year={2019}
}

@article{li2017deep,
  title={Deep reinforcement learning: An overview},
  author={Li, Yuxi},
  journal={arXiv preprint arXiv:1701.07274},
  year={2017}
}

@article{deshmukh2023explaining,
  title={Explaining rl decisions with trajectories},
  author={Deshmukh, Shripad Vilasrao and Dasgupta, Arpan and Krishnamurthy, Balaji and Jiang, Nan and Agarwal, Chirag and Theocharous, Georgios and Subramanian, Jayakumar},
  journal={arXiv preprint arXiv:2305.04073},
  year={2023}
}

@inproceedings{takagi2024abstracted,
  title={Abstracted trajectory visualization for explainability in reinforcement learning},
  author={Takagi, Yoshiki and Tabalba, Roderick and Kirshenbaum, Nurit and Leigh, Jason},
  booktitle={2024 IEEE Conference on Artificial Intelligence (CAI)},
  pages={75--82},
  year={2024},
  organization={IEEE}
}

\begin{center}
\Large\bfseries{Appendix}
\end{center}

\begin{table*}[H]
\centering
\caption{Action indices and descriptions for CTF environment. These indices (1-18) correspond to the x-axis in Figure \ref{fig:PL1_actions} and are used to interpret the emergence of action preferences.}
\renewcommand{\arraystretch}{0.9}
\begin{tabular}{cll}
\hline
\textbf{Index} & \textbf{Abstract Action} & \textbf{Description} \\
\hline
1 & local(ScanBashHistory) & Scan bash history for host references \\
2 & local(ScanExplorerRecentFiles) & Scan Windows recent-files list \\
3 & local(SudoAttempt) & Attempt sudo privilege escalation \\
4 & local(ExfiltrateFlag) & Read local flag file \\
5 & local(CrackKeePassX) & Crack KeepPassX vault (Linux) \\
6 & remote(ProbeLinux) & Probe to confirm Linux OS \\
7 & remote(ProbeWindows) & Probe to confirm Windows OS \\
8 & remote(ProbeSQLServer) & Probe for SQL-Server service \\
9 & connect(HTTPS) & Connect to HTTPS service \\
10 & connect(GIT) & Connect to Git service \\
11 & connect(SSH) & SSH connection with cached creds \\
12 & connect(RDP) & RDP connection with cached creds \\
13 & connect(MySQL) & Connect to MySQL service \\
14 & connect(SSH-key) & SSH using leaked key \\
15 & connect(PING) & ICMP ping \\
16 & connect(su) & Command/connection via su creds \\
17 & remote(ProbeFlagServer) & Probe the flag server \\
18 & local(CrackKeePass) & Crack KeepPass vault (Windows) \\
\hline
\end{tabular}
\label{tab:abstract-actions}
\end{table*}

\begin{table*}[t!]
\centering
\small
\caption{Agent state at Episode 20, Step 956, with five owned nodes and no discovered nodes, reflecting an early stage of exploration under partial observability.}
\label{956}
\begin{tabular}{lllll}
\hline
\textbf{id} & \textbf{status} & \textbf{properties} & \textbf{local\_attacks} & \textbf{remote\_attacks} \\
\hline
client & owned & [] & [SearchEdgeHistory] & [] \\
Website & owned & [MySql, Ubuntu, nginx/1.10.3] & [CredScanBashHistory] & [ScanPageSource, ScanPageContent] \\
Website[user=monitor] & owned & [MySql, Ubuntu, nginx/1.10.3] & [CredScan-HomeDirectory] & [] \\
AzureStorage & owned & [CTFFLAG:LeakedCustomerData] & [] & [AccessDataWithSASToken] \\
AzureResourceManager & owned & [CTFFLAG:LeakedCustomerData2] & [] & [ListAzureResources] \\
\hline
\end{tabular}
\end{table*}

\begin{table*}[t!]
\centering
\small
\caption{Agent state at Episode 20, Step 957. The node GitHubProject is discovered (discovered) through a remote exploit from a previously owned node, demonstrating initial expansion of the visible attack surface.}
\label{957}
\begin{tabular}{lllll}
\hline
\textbf{id} & \textbf{status} & \textbf{properties} & \textbf{local\_attacks} & \textbf{remote\_attacks} \\
\hline
client & owned & [] & [SearchEdgeHistory] & [] \\
Website & owned & [MySql, Ubuntu, nginx/1.10.3] & [CredScanBashHistory] & [ScanPageSource, ScanPageContent] \\
Website[user=monitor] & owned & [MySql, Ubuntu, nginx/1.10.3] & [CredScan-HomeDirectory] & [] \\
AzureStorage & owned & [CTFFLAG:LeakedCustomerData] & [] & [AccessDataWithSASToken] \\
AzureResourceManager & owned & [CTFFLAG:LeakedCustomerData2] & [] & [ListAzureResources] \\
GitHubProject & discovered & -- & -- & [CredScanGitHistory] \\
\hline
\end{tabular}
\end{table*}

\begin{table*}[t!]
\centering
\footnotesize
\renewcommand{\arraystretch}{0.95}
\caption{Agent state at Episode 20, Step 958. The agent reveals Website.Directory via directory traversal, highlighting application-layer reconnaissance and continued exploration beyond direct exploitation.}
\label{958}
\begin{tabular}{lllll}
\hline
\textbf{id} & \textbf{status} & \textbf{properties} & \textbf{local\_attacks} & \textbf{remote\_attacks} \\
\hline
client & owned & [] & [SearchEdgeHistory] & [] \\
Website & owned & [MySql, Ubuntu, nginx/1.10.3] & [CredScanBashHistory] & [ScanPageSource, ScanPageContent] \\
Website[user=monitor] & owned & [MySql, Ubuntu, nginx/1.10.3] & [CredScan-HomeDirectory] & [] \\
AzureStorage & owned & [CTFFLAG:LeakedCustomerData] & [] & [AccessDataWithSASToken] \\
AzureResourceManager & owned & [CTFFLAG:LeakedCustomerData2] & [] & [ListAzureResources] \\
GitHubProject & discovered & -- & -- & [CredScanGitHistory] \\
Website.Directory & discovered & -- & -- & [NavigateWebDirectory, NavigateWebDirectoryFurther] \\
\hline
\end{tabular}
\end{table*}

\begin{table*}[t!]
\centering
\footnotesize
\renewcommand{\arraystretch}{0.95}
\caption{Agent state at Episode 20, Step 959. The node Sharepoint is discovered through sequential reconnaissance actions, evidencing the agent’s multi-step lateral movement and expanding visibility into deeper network segments.}
\label{959}
\begin{tabular}{lllll}
\hline
\textbf{id} & \textbf{status} & \textbf{properties} & \textbf{local\_attacks} & \textbf{remote\_attacks} \\
\hline
client & owned & [] & [SearchEdgeHistory] & [] \\
Website & owned & [MySql, Ubuntu, nginx/1.10.3] & [CredScanBashHistory] & [ScanPageSource, ScanPageContent] \\
Website[user=monitor] & owned & [MySql, Ubuntu, nginx/1.10.3] & [CredScan-HomeDirectory] & [] \\
AzureStorage & owned & [CTFFLAG:LeakedCustomerData] & [] & [AccessDataWithSASToken] \\
AzureResourceManager & owned & [CTFFLAG:LeakedCustomerData2] & [] & [ListAzureResources] \\
GitHubProject & discovered & -- & -- & [CredScanGitHistory] \\
Website.Directory & discovered & -- & -- & [NavigateWebDirectory, NavigateWebDirectoryFurther] \\
Sharepoint & discovered & -- & -- & [ScanSharepointParentDirectory] \\
\hline
\end{tabular}
\end{table*}

\noindent \textbf{Intermediate Exploration Trajectory of the RL Attacker Under Partial Observability}

To illustrate the agent’s behaviour in \textbf{\textit{CTF environment}} under partial observability, Tables \ref{956}–\ref{959} present four consecutive snapshots of its internal state at Episode 20, Steps 956 to 959. Each row corresponds to a node known to the agent at that step, and the columns represent key aspects of the agent’s perception and action space:

\vspace{0.5em}
\begin{itemize}
 \item \textbf{\texttt{id}}: Unique identifier of a node (e.g., \texttt{Website}, \texttt{GitHubProject}).
 \item \textbf{\texttt{status}}: Agent’s access level to the node (\texttt{owned}, \texttt{discovered}, or unknown).
 \item \textbf{\texttt{properties}}: Node attributes such as OS, services, or flags (e.g., \texttt{Ubuntu}, \texttt{nginx}, \texttt{CTFFLAG}).
 \item \textbf{\texttt{local\_attacks}}: Exploits executable from within the node (e.g., \texttt{CredScanBashHistory}, \texttt{SearchEdgeHistory}).
 \item \textbf{\texttt{remote\_attacks}}: Actions targeting other nodes from the current node (e.g., \texttt{ScanPageSource}, \texttt{NavigateWebDirectory}).
\end{itemize}

These tables illustrate how the agent progressively uncovers the network topology through a sequence of exploratory actions involving local exploits and remote scans.

At Step 956 (Table \ref{956}), the agent has compromised five nodes, \texttt{Website}, \texttt{Website[user=monitor]}, \texttt{AzureStorage}, \texttt{AzureResourceManager}, and \texttt{client}, each marked as \texttt{owned}. The table lists the local and remote actions currently available from these nodes, including \texttt{CredScanBashHistory}, \texttt{SearchEdgeHistory}, \texttt{ScanPageSource} and others. Now, the agent will execute additional actions from the owned nodes to discover new nodes or compromise further targets, gradually expanding its visibility into the environment.

At Step 957 (Table \ref{957}), the agent expands its knowledge of the environment by discovering a new node, \texttt{GitHubProject}, which is now marked as \texttt{discovered}. This node has not yet been compromised, but its appearance indicates that it has been identified as a reachable target through action executed from one of the owned nodes.

At Step 958 (Table \ref{958}), the agent continues its exploration by discovering a new node, \texttt{Website.Directory}, which is marked as \texttt{discovered}. This node is revealed through actions executed from an owned node. While the node has not been compromised, its discovery indicates that the agent is actively probing the structure of the environment to uncover additional entry points.

At Step 959 (Table \ref{959}), the agent continues its exploration by discovering a new node, \texttt{Sharepoint}, which is now marked as \texttt{discovered}. Although the node remains uncompromised, its appearance in the state table indicates that the agent is systematically expanding its reach by chaining reconnaissance actions across accessible paths. At this point, the agent has visibility over eight nodes in total, five marked as \texttt{owned} and three as \texttt{discovered}, demonstrating a steady progression in environmental awareness and exploration capabilities.

These four steps represent an intermediate segment within the agent’s broader training trajectory. Over the course of training, the agent incrementally learns to navigate the environment under partial observability by discovering new nodes, exploiting available actions, and updating its internal state representation. This process, combining exploration and exploitation, enables the agent to gradually expand its understanding of the network and identify strategic paths for lateral movement. Examining such intermediate snapshots highlights the underlying mechanisms by which the agent refines and adapts its attack policy over time.

\end{document}